\documentclass{LMCS}
\usepackage{amssymb,amsfonts,graphicx,graphics}
\usepackage[arrow,curve,frame,ps,dvips,pdftex]{xy}
\def\<{\langle}
\def\>{\rangle}
\def\mths{\mathsurround=0 pt}
\def\eqalign#1{\null\,\vcenter{\openup\jot\mths
  \ialign{\strut\hfil$\displaystyle{##}$&$\displaystyle{{}##}$\hfil
      \crcr#1\crcr}}\,}
\def\(#1{\relax\hbox to1.4em{\rm(\hss#1\hss)}}
\def\mA{\mathcal{A}}
\def\mB{\mathcal{B}}
\def\mC{\mathcal{C}}
\def\mD{\mathcal{D}}
\def\mF{\mathcal{F}}
\def\mM{\mathcal{M}}
\def\mP{\mathcal{P}}
\def\mR{\mathcal{R}}
\def\mS{\mathcal{S}}
\def\mV{\mathcal{V}}
\def\mBS{\mathcal{BS}}
\def\mFE{\mathcal{FE}}
\def\mSFE{\mathcal{SFE}}
\def\mSTR{\mathcal{STR}}

\def\add{\mathop{\mathit{add}}\nolimits}
\def\cwd{\mathop{\mathit{cwd}}\nolimits}
\def\dEf{\mathop{\mathit{def}}\nolimits}
\def\edg{\mathop{\mathit{edg}}\nolimits}
\def\lab{\mathop{\mathit{lab}}\nolimits}
\def\ren{\mathop{\mathit{ren}}\nolimits}
\def\rwd{\mathop{\mathit{rwd}}\nolimits}
\def\val{\mathop{\mathit{val}}\nolimits}
\def\Sdg{\mathop{\mathit{Sdg}}\nolimits}
\def\Rep{\mathop{\mathit{Rep}}\nolimits}
\def\Sep{\mathop{\mathit{Sep}}\nolimits}
\def\Inc{\mathop{\mathit{Inc}}\nolimits}
\def\Dec{\mathop{\mathit{Dec}}\nolimits}
\def\Elim{\mathop{\mathit{Elim}}\nolimits}
\def\Eval{\mathop{\mathit{Eval}}\nolimits}
\def\Good{\mathop{\mathit{Good}}\nolimits}
\def\Split{\mathop{\mathit{Split}}\nolimits}
\def\doi{2 (2:2) 2006}
\lmcsheading%
{\doi}
{46}
{}
{}
{Jun.~24, 2005}
{Mar.~23, 2006}
{}

\begin{document}

\title[Graph decompositions]{The monadic second-order logic of graphs XVI: \\
Canonical graph decompositions}
\author[B.~Courcelle]{Bruno Courcelle}
\address{LaBRI, Bordeaux 1 University, 33405 Talence, France}
\email{courcell@labri.fr}
\keywords{Monadic second-order logic, split decomposition, modular
decomposition, clique-width}
\subjclass{F.4.1}

\begin{abstract}
  \noindent This article establishes that the \emph{split
  decomposition} of graphs introduced by Cunnigham, is definable in
  Monadic Second-Order Logic.This result is actually an instance of a
  more general result covering canonical graph decompositions like the
  modular decomposition and the Tutte decomposition of 2-connected
  graphs into 3-connected components.  As an application, we prove
  that the set of graphs having the same cycle matroid as a given
  2-connected graph can be defined from this graph by Monadic
  Second-Order formulas.
\end{abstract}

\maketitle

\section{Introduction}\label{S:intro}\label{S:1}

  Hierarchical graph decompositions are useful for the construction of
  efficient algorithms, and also because they give structural
  descriptions of the considered graphs. Cunningham and Edmonds have
  proposed in \cite{CuEd80} a general framework for defining
  decompositions of graphs, hypergraphs and matroids.  This framework
  covers many types of decompositions. Of particular interest is the
  \emph{split decomposition} of directed and undirected graphs defined
  by Cunningham in \cite{cunningham82}.

  A hierarchical decomposition of a certain type is \emph{canonical}
  if, up to technical details like vertex labellings, there is a
  unique decomposition of a given graph (or hypergraph, or matroid) of
  this type. To take well-known examples concerning graphs, the
  \emph{modular decomposition} is canonical, whereas, except in
  particular cases, there is no useful canonical notion of
  \emph{tree-decomposition} of minimal tree-width. The general results
  of \cite{CuEd80} define canonical decompositions.

  The present article shows that many of these canonical
  decompositions can be defined by \emph{monadic second-order (MS)
  formulas} "inside" the considered graphs or hypergraphs (we do not
  consider decompositions of matroids in this article).  More
  precisely, we prove that under the quite natural and generally
  satisfied hypothesis that the elementary decomposition steps are
  definable by an MS formula, the mapping from a graph or a hypergraph
  to the \emph{tree} representing its canonical decomposition (of the
  type under consideration) is a\emph{ monadic second-order (MS)
  transduction}, i.e., a transformation of relational structures
  defined by MS formulas.\ Furthermore, in many concrete cases, a
  certain relational structure based on this tree represents the
  considered decomposition, in such a way that the decomposed graph
  can be reconstructed from it.  We call it a\emph{ graph
  representation of the decomposition} in the case where it uses
  relations of arity at most 2.  Otherwise, we call it a\emph{
  hypergraph representation} (because relational structures can be
  viewed as directed labelled ranked hypergraphs). These
  representations can be constructed from the graphs (equipped with
  arbitrary linear orderings of their sets of vertices or edges) by MS
  transductions.  Roughly speaking, we obtain that, from the point of
  view of MS logic, a graph is equivalent to the graph or hypergraph
  representation of its canonical decomposition, which means that an
  MS property of the canonical decomposition of a graph $G$ is
  (equivalent to) an MS property of $G$ and conversely, that an MS
  property of $G$ is (equivalent to) an MS property of the
  (hyper)graph representation of its canonical decomposition.

  This article contributes to the understanding of the power of MS
  logic for representing graph properties and graph theoretical
  notions like canonical graph decompositions and equivalences on
  graphs. When a graph property is expressible in MS logic, it can be
  checked in polynomial time on graphs of bounded tree-width or
  clique-width. When a graph transformation is expressible in MS
  logic, it preserves the property that a set has bounded tree-width
  or clique-width. We refer the reader to \cite{courcelle97} and
  \cite{CoMaRo00} for detailed expositions of these consequences.

%\subsection*{Why canonical decompositions are interesting?}
\subsection*{Why are canonical decompositions interesting?}

  Canonical decompositions and their (hyper)graph representations are
  interesting for several reasons.

  First they contain useful informations on the structure of the
  graphs.  This structural information has two forms: the tree, and
  the elementary graphs from which the considered graph is built. In
  most cases, hierarchical decompositions can be viewed as
  constructions of graphs or hypergraphs by means of particular
  composition operations (like \emph{graph substitution} in the case
  of modular decomposition) in terms of \emph{prime} graphs or
  hypergraphs, i.e., of those which are undecomposable. We will
  discuss this "algebraic" aspect whenever relevant.

  Second, hierarchical graph decompositions are useful for the
  construction of polynomial algorithms. For example, the first step
  of the polynomial algorithm recognizing circle graphs by Bouchet
  \cite{bouchet87} consists in constructing the split decomposition of
  the given graph. It uses the fact that a graph is a circle graph if
  and only if each component of its split decomposition is a circle
  graph. The planarity testing algorithm by Hopcroft and Tarjan
  \cite{HoTa72} begins with the decomposition of a graph into
  3-connected components.  Hence a good understanding of hierarchical
  graph structure is useful for constructing low degree polynomial
  algorithms.

  Third the (hyper)graph representation of the canonical decomposition
  of a graph requires in many cases less space to be stored than the
  given graph.

  Finally, canonical decompositions are also useful for establishing
  logical properties. For example, it is proved in \cite{courcelle04}
  that Seese's Conjecture holds for \emph{interval graphs}, and that
  it holds in general if and only if it holds for \emph{comparability
  graphs}.  The proof makes an essential use of modular
  decompositions.\ (Seese's Conjecture says that if a set of graphs
  has a decidable satisfiability problem for MS logic, then it has
  bounded clique-width.  A slight weakening of this conjecture is
  established in \cite{CoOu04}).

  The companion article \cite{cour05} develops an application of split
  decomposition to \emph{circle graphs} that we review briefly.  A
  circle graph is the \emph{intersection graph} of a \emph{set of
  chords of a circle. }A graph is a circle graph if and only if all
  components of its split decomposition are circle graphs. Those
  components which are prime are \emph{uniquely representable} as
  intersection graphs of sets of chords. It is proved in \cite{cour05}
  that the unique representation of a prime circle graph can be
  defined by MS formulas (one uses a description of sets of chords by
  finite relational structures). From the split decomposition of a
  circle graph $G$ and the chord representations of its prime
  components, one can define all chord representations of $G$. And
  this can be done by MS formulas, since the split decomposition and
  the chord representations of the prime components of $G$ can be
  defined by MS formulas. Hence, from a given circle graph, one can
  define by MS formulas (using also linear orders of the sets of its
  vertices), all chord representations of this graph. (To be precise,
  this construction rests upon a result by Courcelle and Oum
  \cite{CoOu04} which uses MS formula with set predicates of the form
  $Even(X)$ expressing that a set $X$ has even cardinality.)

  In the present article, we prove a result of the same general form:
  all 2-connected graphs \emph{equivalent} to a given graph $G$, in
  the sense that they have the same cycle matroid, can be defined from
  $G$ and the linear orderings of its vertices by a fixed MS\
  transduction. As for circle graphs, the proof uses a canonical
  decomposition of the considered graph, constructed by MS formulas,
  from which can be defined all the equivalent graphs.  This
  construction is based on \emph{Whitney's 2-isomorphism theorem},
  which characterizes the graphs equivalent to $G$ as those derived
  from $G$ by a sequences of transformations called \emph{twistings}.

\subsection*{Main results and overview of the article}

  First, we give a general set theoretical and logical setting in
  which decompositions of combinatorial structures can be
  defined. This framework covers actually two cases. In the first
  case, studied in Section \ref{S:2}, the decomposition tree is rooted
  and directed.  The fundamental example is the very well-known
  \emph{modular decomposition}. The decompositions of this type
  correspond to definitions of graphs by algebraic expressions based
  on graph operations. In Section \ref{S:3}, we consider the second
  case, where the decomposition tree is unrooted and undirected. In
  both cases, and under easily applicable conditions, we prove that
  the decomposition tree is definable by MS formulas, which
  generalizes the MS definability result of the modular decomposition
  of \cite{courcelle96}.  In Section \ref{S:4} we develop the
  application to the \emph{split decomposition} of Cunnigham
  \cite{cunningham82} and we prove its MS definability, which is our
  second main result.  We do not assume the reader familiar with this
  notion and this section presents it in detail. We prove in Section
  \ref{S:2} the "logically effective" version of the above mentioned
  theorem by Whitney.  Appendices 1 and 2 review definitions, basic
  properties and technical lemmas on MS logic, MS\ transductions and
  clique-with. This work has been presented at the International
  Conference on Graph Theory, Hy\`{e}res, France, in September 2005.

\section{Partitive families of sets}\label{S:2}

  Trees, graphs and relational structures are finite. Two sets
  \emph{meet} if they have a nonempty intersection.  They
  \emph{overlap} if they meet and are incomparable for inclusion. We
  write $A\perp B$ if and only if $A$ and $B$ overlap. The terms
  \emph{minimal, least,} and \emph{maximal} applied to sets refer,
  unless otherwise specified, to inclusion.

\subsection{Rooted trees and families of sets}\label{SS:2.1}

  We define the family of sets associated with a partition of a set
  $V$, the blocks of which form a rooted tree.  This is a
  generalization of the modular decomposition of a graph where $V$ is
  its set of vertices.

\begin{defi}\label{D:2.1}
%\textbf{Definitions 2.1\: }
\textit{Set families and trees.}  A \emph{rooted tree} $T$ has its
  edges directed so that every node is accessible from the root by a
  unique directed path. The \emph{leaves} are the nodes of outdegree
  0. The other nodes are the \emph{internal nodes}. The set of nodes
  is denoted by $N_{T}$ and the set of internal nodes by
  $N_{T}^{int}$. Although a tree is a graph, we will use the term
  "nodes" for the vertices of a tree (or a forest). This particular
  terminology will be useful for clarity in situations where we
  discuss simultaneously a graph and a tree representing it.  A rooted
  tree is \emph{proper} if no node has \emph{outdegree} 1, hence if
  every node is either a leaf, or has at least two \emph{sons}. We
  denote the son relation by $son_{T}$.
\end{defi}

  Let $T$ be a rooted tree and $\mV=(V(u))_{u\in N_{T}} $ be a
  partition of a nonempty set $V$ such that $V(u)$ is nonempty if $u$
  is a leaf (but is possibly empty otherwise).  For each node $u$, we
  let $T(u)$ be the rooted subtree consisting of $u$ (its root) and
  the nodes reachable from $u$ by a directed path.  We let
  $\overline{V}(u)$ be the union of the sets $V(z)$ where $z$ is a
  node of $T(u)$. Hence, $\overline{V}(u)=V$ if $u$ is the root. The
  family $\mF=\mF(T,\mV)$ of sets $\overline{V}(u)$ has the following
  properties:
\begin{itemize}
\item[(P0)] $V\in\mF$, $\varnothing\notin\mF$,
\item[(P1)] no two elements of $\mF$ overlap.
\end{itemize}

  An element of a family $\mF$ of subsets of a set $V$ will be called
  an $\mF$-\emph{module}. Every family $\mF$ satisfying properties P0
  and P1 is associated as above with a rooted tree $T_{\mF}$ that is
  defined as follows. We take $\mF$ as set of nodes, $V$ as root, the
  inverse of inclusion as ancestor relation. The leaves are the
  minimal $\mF$-modules. For a node $N$, we let $V(N)=N-\bigcup\{M\in
  \mF\mid M\subset N\}$ and we denote this family of sets by
  $\mV_{\mF}$. We have $\overline{V_{\mF}}(N)=N$.  Hence
  $\mF(T_{\mF},\mV_{\mF})=\mF$.  We have $V_{\mF}(N)\neq\varnothing$
  for every node $N$ of outdegree 1.  Every pair $(T,\mV)$ such that
  $\mF(T,\mV)=\mF$ and $V_{\mF}(N)$ is nonempty for every node $N$ of
  outdegree 1 is isomorphic to $(T_{\mF},\mV_{\mF})$ (this means that
  there exists an isomorphism $h$ of $T$ onto $T_{\mF}$ such that
  $V_{\mF}(h(u))=V(u)$ for every node $u$ of $T$).

  If $M,P\in\mF$ their least common ancestor in $T_{\mF}$ is the least
  $\mF$-module $N$ containing $M\cup P$. We define a binary relation
  $box_{T_{\mF}}(x,N)$ holding if and only if $x$ belongs to
  $V_{\mF}(N)$. We also define a binary relation
  \emph{mod}$_{T_{\mF}}(x,N)$ holding if and only if $x$ belongs to
  $\overline {V_{\mF}}(N)$.  This relation is membership since the
  nodes of $T_{\mF}$ are the $\mF$-modules.  However, it will be
  useful later when we construct $T_{\mF}$ as an abstract tree, and
  not as a set of sets ordered by inclusion. The relations\
  $box_{T_{\mF}}$ and \emph{mod}$_{T_{\mF}}$ are definable from each
  other with the help of the son relation of the tree $T_{\mF}$.

  If the family $\mF$ satisfies the stronger property:

\begin{itemize}
\item[(P'0)] $V\in\mF$, $\varnothing\notin\mF$, $\{v\}\in
\mF$ for each $v\in V$,
\end{itemize}
  then the leaves of $T_{\mF}$ are the singletons $\{v\}$, $T_{\mF}$
  is a proper tree and $V_{\mF}(u)$ is empty if $u$ is an internal
  node. If a family $\mF$ satisfies only P0 and P1, then the family
  $\mF^{+}=\mathcal{F\cup }\{\{v\}\mid v\in V\}$ satisfies P'0 and
  P1. The corresponding tree $T_{\mF^{+}}$ is obtained from $T_{\mF}$
  as follows: for each $v$ such that $\{v\}\notin \mF$, we add $\{v\}$
  as a new leaf with father the least $\mF$-module containing $v$.

  Let $\mC$ be a class of relational structures (see Appendix 1 for
  definitions). For each $S\in\mC$, we let $\mF(S)$ be a family of
  subsets of its domain $D_{S}$. We say that $\mF$ is
  \emph{MS-definable} if there exists an MS formula $\varphi(X)$ such
  that for every $S$ in $\mC$, $\mF(S)=\{A\mid A\subseteq
  D_{S},S\models\varphi(A)\}$. With these definitions:

\begin{prop}\label{P:2.1}
  Let $\mC$ be a set of $\mR$-structures and
  $\mathcal{F(}S\mathcal{)}$ be an MS-definable family of subsets of
  $D_{S}$ which satisfies P0 and P1 for every $S\in\mC$.  There exists
  a domain extending MS-transduction that associates with
  $(S,\preccurlyeq)$, where $S=\<D_{S},(R_{S})_{R\in\mR}\>\in\mC$ and
  $D_{S}$ is linearly ordered by $\preccurlyeq$, the structure:
\[\Dec(S)=\<D_{S}\cup N_{T},(R_{S})_{R\in\mR}, son_{T},box_{T}\>\]
  where $T=T_{\mF(S)}$ and $box_{T}=box_{T_{\mF(S)}}$.
\end{prop}

  We will give a proof of this proposition adapted from that of
  \cite{courcelle96}, Section 5. In the structure $\Dec(S)$ the domain
  consists of elements of different natures.  If we are given a
  structure $\<D_{U},(R_{U})_{R\in\mR}$, $son_{U},box_{U}\>$ \
  intended to be isomorphic to $\<D_{S}\cup N_{T},(R_{S})_{R\in\mR}$,
  $son_{T},box_{T}\>$ we can identify the nodes of $T$ as the elements
  $x$ of $D_{U}$ such that $son_{U}(x,y)$ or $son_{U}(y,x)$ or
  $box_{T}(y,x)$ holds for some $y$. (We assume $D_{S}$ nonempty; $T$
  may have a single node).

  \emph{Monadic Second-order logic} (\emph{MS logic} in short) and
  \emph{Monadic Second-order transductions} (\emph{MS transductions})
  are defined in Appendix 1.  We only recall here that an MS\
  transduction (also called sometimes an \emph{MS interpretation}) is
  a transformation of relational structures that is specified by MS\
  formulas forming its \emph{definition scheme}.  It transforms a
  structure $S$ into a structure $T$ (possibly over a different set of
  relations) such that the domain $D_{T}$ of $T$ is a subset of
  $D_{S}\times\{1,\dots,k\}$.  The numbers $1,\dots,k$ are just a
  convenience for the formal definition; we are actually interested
  by relational structures up to isomorphism. In many cases, this
  transformation involves a bijection of $D_{S}$ onto a subset of
  $D_{T}$, and the definition scheme can be constructed in such a way
  that this bijection is the mapping: $x\mapsto(x,1)$. Hence, in this
  case $D_{T}$ contains $D_{S}\times\{1\}$, an isomorphic copy of
  $D_{S}$ and we will say that the MS transduction is \emph{domain
  extending}, because it defines the domain of $T$ as an extension of
  that of $S$. This does not imply that the relations of $T$ extend
  those of $S$. An \emph{FO transduction} is a transduction defined by
  a first-order definition scheme.

\begin{defi}\label{D:2.2}
  \emph{The leaves of a tree}.  Let $T$ be a proper rooted tree. We
  write $x\leq y$ if $x$ is below $y$ and we denote by $y\vee z$ the
  least upper bound of two elements $y$ and $z$. The root is thus the
  unique maximal element of $T$ for this order.  We let
  $\lambda(T)=\<Leaves(T),R_{T}\>$ where $Leaves(T)$ denotes the set
  of leaves of $T$ and $R_{T}(x,y,z)$ holds if and only if $x\leq
  y\vee z$. The next lemma shows that if $Leaves(T)$ is linearly
  ordered by some auxiliary order $\preccurlyeq$, then $T$ is
  definable from $(\lambda(T),\preccurlyeq)$ by a domain extending
  MS-transduction. The resulting tree $T$ does not depend on the
  linear order $\preccurlyeq$.
\end{defi}

\begin{lem}\label{L:2.1}
  There exists a domain extending FO transduction that maps
  $(\lambda(T),\preccurlyeq)$ to $T$, whenever $T$ is a proper rooted
  tree and $\preccurlyeq$ is a linear order on $Leaves(T)$.
\end{lem}

\proof Let $T$ be a proper rooted tree and $\preccurlyeq$ be a linear
  order on its leaves. For every internal node $x$ of $T$ we let:
\begin{itemize}
\item $fl(x)$ be the $\preccurlyeq$-smallest leaf below $x$, called
  the \emph{first leaf} \emph{below} $x$, and we let $fs(x)$ be the
  unique son $y$ of $x$ such that $fl(x)\leq y$;
\item $rep(x)$ be the $\preccurlyeq$-smallest leaf below $x$ and not
  below $fs(x)$ (this is well-defined because in a proper tree, every
  internal node has at least two sons).
\end{itemize}
  We call $rep(x)$ the \emph{leaf representing} $x$.  We have
  $fl(x)<x$, $rep(x)<x$, and $fl(x)\prec rep(x)$.

\medskip\noindent\textbf{Claim 1}: Let $x,y$ be two internal nodes. If
  $rep(x)=rep(y)$ then $x=y$.

\medskip\noindent\emph{Proof of the claim.}
  By contradiction. Let $x,y$ be distinct internal nodes such
  that $u=rep(x)= rep(y)$. Since $u$ is below $x$ and $y$, $x$ and $y$
  are comparable.  We can assume that $x<y$. By the definitions, $u$
  is not below $fs(x)$. Hence $fl(x)\prec$ $u$ and $fl(x)\leq
  fs(x)<x$. Since $x<y$, $u$ and $fl(x)$ are below the same son of
  $y$, call it $z$; we may have $x=z$. We have $fl(y)\leq fs(y)<y$,
  where $fs(y)\neq z$ since $u=rep(y)$.  Hence, since $u=rep(y)$, $u$
  is the $\preccurlyeq$-smallest leaf among the set of leaves below
  $y$ and not below $fs(y)$, and this set contains $fl(x)$. Hence $u$
  $\preccurlyeq$ $fl(x)$, contradicting the above observation that
  $fl(x)\prec$ $u$.\qed

  One can define a bijection of the nodes of $T$ onto a subset of
  $\lambda(T)\times\{1,2\}$. Each leaf $u$ is mapped to $(u,1)$,
  hence the transduction we are constructing will be domain
  extending. Each internal node $u$ is mapped to $(rep(u),2)$.

\medskip\noindent\textbf{Claim 2:}
  One can write a first-order formula $\alpha(x,y,z)$ such that:
\[(\lambda(T),\preccurlyeq)\models\alpha(x,y,z)
  \qquad\hbox{if and only if}\qquad
  x\neq y
  \quad\hbox{and}\quad
  z=rep(x\vee y)\;.
\]

\medskip\noindent\emph{Proof of the claim.}
  We recall that $R(u,v,w)$ means: $u\leq v\vee w$ for leaves
  $u,v,w$. The relation $\leq$ denotes the ancestor relation in $T$
  and should not be confused with the linear order $\preccurlyeq$ on
  the set of leaves of $T$. Using $R$, one can construct an FO formula
  $\beta(x,y,z)$ expressing that $x\neq y$ and $z=fl(x\vee y)$. An FO
  formula $\gamma (x,y,u,v)$ can be constructed to express that:
\[x\neq y,\ u\neq v,\ u\leq x\vee y,\ v\leq x\vee y,\ x\leq u\vee v,
\hbox{\ and\ } y\leq u\vee v\ ,
\] 
  which means that for leaves $u,v\neq u,x,y\neq x$, $u$ and $v$ are
  below $x\vee y$ but are not below the same son of this node. We can
  construct $\alpha (x,y,z)$ so as to express the following:
\[\exists u[\beta(x,y,u)\wedge \hbox{"$z$ is the $\preccurlyeq$-smallest
  element such that $\gamma(x,y,u,z)$ holds"}]\;.
\]
  This completes the proof Claim 2.\qed

  We let $N =(\lambda(T)\times\{1\})\cup(REP_{T}\times\{2\})$, where
  $REP_{T}$ is the set of leaves of the form $rep(x\vee y)$ for some
  leaves $x,y\neq x$.  We order $N$ by letting:
\[\eqalign{
(x,1)\leq(y,1)&\hbox{\ if and only if $x=y$,}\cr
(x,2)\leq(y,1)&\hbox{\ never holds,}\cr
(x,1)\leq(y,2)&\hbox{\ if and only if $ $there exist leaves $u,v$
  such that}\cr
              &\hbox{\ $y=rep(u\vee v)$ and $R_{T}(x,u,v)$ holds,}\cr
(x,2)\leq(y,2)&\hbox{\ if and only if $ $there exist leaves
                       $u,v,w,z$ such that}\cr
              &\hbox{\ $x=rep(u\vee v)$, $y=rep(w\vee z)$, $R_{T}(u,w,z)$ 
                       and $R_{T}(v,w,z)$ hold.}\cr
  }
\]

\medskip\noindent\textbf{Claim 3 }: The tree $(T,\leq)$ is isomorphic to
  $(N,\leq)$ under the bijection which maps a leaf $u$ of $T$ to $(u,1)$
  and an internal node $u$ to $(rep(u),2)$.

\medskip\noindent\emph{Proof of the claim.}
  The four clauses above correspond to the facts that two different
  leaves are incomparable, that an internal node cannot be below a
  leaf, that a leaf $x$ is below an internal node $u\vee v$ if and
  only if $R_{T}(x,u,v)$ holds, and that an internal node $u\vee v$ is
  below $w\vee z$ if and only if $u$ and $v$ are both below $w\vee
  z$.\qed

  These claims give the desired result because the set $REP_{T}$ is
  FO\ definable in the structure $(\lambda(T),\preccurlyeq)$ by Claim
  2.  The ancestor relation defined by the formulas before Claim 3 is
  also FO\ definable.  From it, one can obtain an FO definition of the
  $son$ relation.  Hence, we have an FO transduction as
  claimed.\qed

\medskip\noindent\textbf{Remark.}
%\begin{rem}\label{R:curly} 
  \emph{On the role of $\preccurlyeq$.}  The role of the linear order
  $\preccurlyeq $ is to make possible the construction of a set
  $REP_{T}$ so that FO formulas can specify in a unique way the
  element of $REP_{T}$ intended to represent an internal node, and to
  express in terms of this encoding, the $son$ relation of $T$. The
  tree $T$ is uniquely defined for every structure $\lambda (T)$.
  \emph{Uniquely} means here that if $T$ and $T^{\prime }$ are such
  that $\lambda (T)=\lambda (T^{\prime })$, there is a unique
  isomorphism of $T$ onto $T^{\prime }$ that is the identity on
  leaves.
\medskip%\end{rem}

\medskip\noindent\emph{Proof of Proposition \ref{P:2.1}.} 
  We will use $\mF(S)^{+}$ instead of $\mF(S)$. (We
  have $\mF(S)^{+}=\mF(S)$ if $\mF(S)$
  satisfies P'0).  It is clear that $\mF(S)^{+}$ is MS
  definable. We construct a structure with domain a subset of
  $D_{S}\times\{1,2,3\}$. Its domain is the union of three sets:
\begin{itemize}
\item[-] the set $D_{S}\times\{1\}$, a copy of $D_{S}$,
\item[-] the set $D_{S}\times\{2\}$ which is the set of leaves of
  $T=T_{\mathcal{F(S)}^{+}}$ (the pair $(v,2)$ represents the leaf
  $\{v\}$ for each $v$ in $D_{S}$),
\item[-] and of a subset of $D_{S}\times\{3\}$, namely $REP_{T}\times\{3\}$, \
  (cf.\ the proof of Lemma \ref{L:2.1}) in bijection with the set of
  internal nodes of $T$.
\end{itemize}
  The relation $R_{T}(z,x,y)$ is "$z\in N$ where $N$ is the least set
  in $\mF(S)$ that contains $x$ and $y$".  This is expressible
  by an MS formula. Hence the structure $\lambda(T)$ is definable from
  $S$ by an MS-transduction.  Since the set of leaves of $T$ is
  linearly ordered by $\preccurlyeq$ (because $D_{S}$ is, and is in
  bijection with $D_{S}$) we can obtain $T$ from
  $(\lambda(T),\preccurlyeq)$ by a domain extending MS-transduction.

  Then we reduce $T=T_{\mathcal{F(S)}^{+}}$ into $T_{\mathcal{F(S)}}$
  by eliminating the leaves $(v,2)$ such that
  $\{v\}\notin\mF(S)$.  The relation
  $box_{\mathcal{T}_{\mathcal{F(}S\mathcal{)}}}$ is also MS\ definable
  since \emph{mod}$_{\mathcal{T}_{\mathcal{F(}S\mathcal{)}}}(x,u)$
  holds (cf.\ Definition \ref{D:2.1}) if and only if in the tree
  $T_{\mathcal{F(S)}^{+}}$, the singleton $\{x\}$ is a leaf below
  $u$. We obtain thus an MS\ transduction. A definition scheme for it
  can be written from the above description.\qed

\medskip\noindent\textbf{Remark.}
%\begin{rem}\label{R:alter}
  One could alternatively define the domain of the constructed
  structure as a subset of $ D_{S}\times \{1,2\}$ by letting $(v,1)$
  represent simultaneously the element $v$ of $D_{S}$ and the leaf
  $\{v\}$ of $T_{\mathcal{F(S)}}$, in the case where $\{v\}\in
  \mF(S)$. In this case, the internal nodes of
  $T_{\mathcal{F(S)}}$ are pairs $(v,2)$. However, in most cases, we
  will keep separated the domain $D_{S}$ of the structure and the set
  of nodes of its decomposition tree, even if there is a "natural"
  bijection between a subset of $D_{S}$ and a set of nodes of the
  tree.
\medskip%\end{rem}

  An MS property is \emph{order-invariant} if it is expressible by an
  MS\ formula using an auxiliary linear order $\preccurlyeq$ of the
  domain of the considered structure, that can be chosen arbitrarily.
  See Appendix 1 for a more precise definition.

\begin{cor}\label{C:2.1}
  Under the hypotheses of Proposition \ref{P:2.1}, every MS property
  of the structures $\Dec(S)$ for $S\in\mC$ is equivalent to an
  order-invariant MS property of the structures $S$.
\end{cor}

\proof Let $\mP$ be an MS property of the structures
  $\Dec(S)$. By Proposition A.1.2 (in Appendix 1), applied to the
  transduction of Proposition \ref{P:2.1}, $\mP(\Dec(S))$ is
  equivalent to an MS property $\mathcal{Q}$ of $(S,\preccurlyeq)$,
  where $\preccurlyeq$ is \emph{any} linear order of $D_{S}$.  For any
  two linear orders $\preccurlyeq$ and $\preccurlyeq\prime$ on
  $D_{S}$, one obtains isomorphic structures $\Dec(S)$ by the remark
  before the proof of Proposition \ref{P:2.1}. Hence $\mathcal{Q}$ is
  an order-invariant MS property.\qed

\begin{defi}\label{D:2.3}
  \emph{Partitive families of sets} Let $V$ be a nonempty set. A
  family $\mF$ of subsets of $V$ is \emph{weakly partitive} if
  it satisfies the following properties:
\begin{itemize}
\item[(P0)] $V\in\mF$, $\varnothing\notin\mF$.
\item[(P2)] For every two overlapping $\mF$-modules $A$ and
  $B$ we have $A\cup B$, $A\cap B$, $A-B\in\mF$.
\end{itemize}
  It is \emph{partitive} if, in addition, it satisfies the following property:
\begin{itemize}
\item[(P3)] For every two overlapping $\mF$-modules $A$ and $B$ we
  have $A\Delta B\in\mF$, (where $ A\Delta B=(A-B)\cup(B-A))$.
\end{itemize}
\end{defi}

  The \emph{strong} $\mF$-modules are the $\mF$-modules which do not
  overlap any other $\mF$-module. The family $Strong(\mF)$ of strong
  $\mF$-modules satisfies P0 and P1.  The corresponding rooted tree
  $T_{Strong(\mF)}$ is called the \emph{decomposition tree} of $\mF$
  or of the structure $S$, if $\mF=\mF(S)$ is known from the context.
  Its leaves are the minimal $\mF$-modules (they are strong).  They
  are the singletons $\{v\}$ for all elements $v$ of $V$ if $\mF$
  satisfies P'0. Since a singleton does not overlap any set, if $\mF$
  is weakly partitive or partitive, then $\mF^{+} $ is weakly
  partitive or partitive respectively.

  The conditions of partitivity and weak partitivity on a family $\mF$
  imply some particular structure associated with the nodes of
  $T_{Strong(\mF)}$. They are formulated in an easier way in terms of
  the tree $T_{Strong(\mF^{+})}$ rather than in terms of
  $T_{Strong(\mF)}$. We recall that the nodes of $T_{Strong(\mF^{+})}$
  are subsets of $V$.

\begin{thm}\label{T:2.1}
   Let $\mF$ be a partitive family.
%\begin{itemize}
%\item[(1)]

  \noindent\emph{(1)}\ 
  Every internal node $N$ of the tree $T_{Strong(\mF^{+})}$ \
  satisfies one of the following two properties:
\begin{itemize}
\item[T1:] $N$ has $k$ sons, $N_{1},\dots,N_{k},k\geq2$, and for every
  nonempty subset $I$ of $\{1,\dots,k\}$, the set $\bigcup\{N_{i}\mid i\in
  I\}$ belongs to $\mF$.
\item[T2:] $N$ has $k$ sons, $N_{1},\dots,N_{k},k\geq2$, and for every
  subset $I$ of $\{1,\dots,k\}$, the set $\bigcup\{N_{i}\mid i\in I\}$
  belongs to $\mF$ if and only if $I$ is $\{1,\dots,k\}$ or singleton.
\end{itemize}

%\item[(2)]
  \noindent\emph{(2)}\ 
  If an $\mF$-module is not strong, it is of the form
  $\bigcup\{N_{i}\mid i\in I\}$ for some node $N$ satisfying \emph{T1}
  and a non singleton set $I$ $\subset\{1,\dots,k\}$.

%\end{itemize}
  Let $\mF$ be weakly partitive.
%\begin{itemize}
%\item[(3)]

  \noindent\emph{(3)}\ 
  Every internal node $N$ of the tree $T_{Strong(\mF^{+})}$ \
  satisfies one of properties \emph{T1}, \emph{T2} or %T3 where T3 is the following property:
\begin{itemize}
\item[T3:] The sons of $N$ can be numbered $N_{1},\dots,N_{k}$,
  $k\geq2$, in such a way that for every subset $I$ of $\{1,\dots,k\}$,
  the set $\bigcup\{N_{i}\mid i\in I\}$ belongs to $\mF$ if and only if
  $I$ is an interval $[m,n] $ for some $m,n$ with $1\leq m\leq n\leq k$.
\end{itemize}

%\item[(4)]
  \noindent\emph{(4)}\
  If an $\mF$-module is not strong, it is of the form
  $\bigcup\{N_{i}\mid i\in I\}$ for some node $N$ satisfying \emph{T1}
  and a non singleton set $I$ $\subset\{1,\dots,k\}$, or of the form
  $\bigcup\{N_{i}\mid i\in\lbrack m,n]\}$ for a node $N$ satisfying
  \emph{T3} and $m<n$.\qed
%\end{itemize}
\end{thm}

  See \cite{ChHaMa81,CuEd80,habib81,montgolfier03,MoRa84} for the
  proof. The nodes of types T1, T2, T3 are called respectively the
  \emph{complete nodes}, the \emph{prime nodes} and the \emph{linear
  nodes}. In this theorem, one could require $k\geq3$ in conditions
  T1\ and T3 because the nodes with two sons satisfy T2, and then
  properties T1, T2 and T3 would be mutually exclusive. However, in
  the application of this theorem to the modular decomposition,
  properties T1, T2, T3 and the notions of complete, prime or linear
  nodes correspond to three different graph operations, and those
  corresponding to T1 and T3 may have two arguments only.

  This theorem will be used for classes $\mC$ of relational
  structures, where for each $S$ in $\mC$:
\begin{itemize}
\item[(i)] we have a partitive or weakly partitive MS definable family
  $\mF(S)$ of subsets of its domain $D_{S}$,
\item[(ii)] for each node $N$ of the decomposition tree of $S$, one can
  express $S[N]$, the substructure of $S$ induced by $N$, as a
  composition of the substructures $S[N_{1}],S[N_{2}],\dots,S[N_{k}]$ by
  an operation $f$ (such operations can be seen as generalized
  concatenations) where $N_{1},\dots,N_{k}$ are the sons of $N$,
\item[(iii)] the nature of this operation $f$ can be determined by an
  MS formula with free variables which take $N_{1},N_{2},\dots,N_{k}$ as
  values.
\end{itemize}

  In this case, Proposition \ref{P:2.1} can be improved, and one can
  define an MS transduction that takes as input
  $S=\<D_{S},(R_{S})_{R\in\mR}\>$, together with an arbitrary linear
  order $\preccurlyeq$ of $D_{S}$ and produces a structure $\Rep(S)$
  consisting of the decomposition tree of $S$ augmented with some
  relations which encode the operations $f$, and from which $S$ can be
  reconstructed by an MS transduction. Such a structure contains
  information on the hierarchical construction of $S$, and it is, in
  some cases, a space efficient representation of $S$.  (See the book
  by Spinrad \cite{spinrad03} on efficient graph representations in a
  very general sense).

  Hence our method consists in doing the following steps:
\begin{itemize}
\item[(i)] first, we construct from $(S,\preccurlyeq)$ a structure:
\[\Dec(S)=\<D_{S}\cup N_{T},(R_{S})_{R\in\mR},son_{T},box_{T}\>\]
  which includes $S$ \emph{and} the decomposition tree $T$ together
  with the relation $box_{T}$ which links both; this structure is
  independent of $\preccurlyeq$ up to isomorphism;

\item[(ii)] second, we construct, \emph{if possible}, a structure
  $\Rep(S)$ with domain $D_{S}\cup N_{T}$ and relations
  $son_{T},box_{T}$ together with some relations encoding the
  operations $f$.  The objective is here to have a space efficient
  representation of $S$, from which $S$ can be reconstructed by an MS
  transduction.
\end{itemize}

  In some cases, the structure $\Rep(S)$ encodes a term over a
  signature of operations on graphs or, more generally, on relational
  structures, the value of which is $S$.  If these constructions can
  be done with MS\ transductions then Corollary \ref{C:2.1} applies to
  $\Rep(S)$ in place of $\Dec(S)$.  At this point it is not appropriate
  to formalize more this notion in the general setting.  We rather
  show its application in two important examples, the \emph{modular
  decomposition} based on the family of \emph{nonempty modules of a
  graph}, and the decomposition in \emph{blocks} of certain directed
  acyclic graphs called \emph{inheritance graphs}. Some new results
  are also established.

\subsection{The modular decomposition}\label{SS:2.2}

  Graphs are simple, directed, loop-free.  \emph{Simple} means that
  there is at most one edge from a vertex $x$ to a vertex $y$. Graphs
  are finite, as already indicated. We denote by $x\longrightarrow y$
  the existence of an edge from $x$ to $y$. The undirected graphs are
  those where each edge $x\longrightarrow y$ has an \emph{opposite
  edge} $y\longrightarrow x$.  We write $x-y$ if $x\longrightarrow y$
  and $y\longrightarrow x$. We denote by $V_{G} $the set of vertices
  of a graph $G$. If $X$ is a set of vertices of $G$, we denote by
  $G[X]$ its \emph{induced subgraph} consisting of $X$ and all the
  edges, the two ends of which are in $X$.  If $E$ is a set of edges,
  we denote by $G[E]$ its subgraph consisting the edges of $E$ and
  their end vertices.

\begin{defi}\label{D:2.4} 
  \emph{Modules and graph substitution.}  A \emph{module} of a graph
  $G$ is a subset $M$ of $V_{G}$ such that for every vertices $x,y$ in
  $M$ and every vertex $z$ not in $M$: $x\longrightarrow z$ implies
  $y\longrightarrow z$ and $z\longrightarrow x$ implies
  $z\longrightarrow y$. In words this means that every vertex not in
  $M$ "sees" all vertices of $M$ in the same way. This frequently
  rediscovered notion is surveyed in \cite{MoRa84} (see also
  \cite{EhMc94} for numerous references using various names for the
  same notion). The book by Spinrad \cite{spinrad03} contains also
  many definitions, results, algorithms and references.
\end{defi}

  We denote by $\mM(G)$ the family of nonempty modules of a graph $G$.
  It satisfies Property P'0 (each singleton is a module) and is weakly
  partitive.  It is partitive if $G$ is undirected.  We denote by
  $\mS(G)$ the corresponding family of \emph{strong modules}: they are
  the nonempty modules that do not overlap any module. The tree of
  strong modules is called the \emph{modular decomposition}, and its
  leaves are the vertices of the considered graph ("are" means that we
  identify $v$ and $\{v\}$). The relevant operations that combine
  substructures are \emph{vertex-substitutions}, that we now review.

  If $G$ and $H$ are graphs with disjoint sets of vertices, and $u$ is
  a vertex of $G$, we denote by $G[H/u]$ the graph such that:
\begin{itemize}
\item[\(a]  its set of vertices is $V_{G} \cup V_{H}-\{u\}$,
\item[\(b] its edges are those of $H$, those of $G$ that are not
  incident with $u$, the edges $x\longrightarrow y$ whenever $x\in
  V_{G}-\{u\}$, $x\longrightarrow u$ in $G$, $y\in V_{H}$, and the edges
  $y\longrightarrow x$ whenever $x\in V_{G}-\{u\}$, $u\longrightarrow x$
  in $G$, $y\in V_{H}$.
\end{itemize}

  If $G$ and $H$ are not disjoint, we replace $H$ by an isomorphic
  copy disjoint with $G$.  When we write: let $K$ be a graph of the
  form $G[H/u]$ we assume, unless otherwise specified, that $G$ and
  $H$ are disjoint. This graph is called the result of the
  \emph{substitution of H for u in G}. It is undirected if $G$ and $H$
  are.

  If $u_{1},\dots,u_{n}$ are vertices of $G$ and $H_{1},\dots,H_{n}$ are
  graphs, we define $G[H_{1}/u_{1}$, $\dots,H_{n}/u_{n}]$ as
  $G[H_{1}/u_{1}]\dots[H_{n}/u_{n}]$.  The order in which substitutions
  are done is irrelevant, hence we can consider they are done
  simultaneously.

  A graph is \emph{prime} if it has at least 3 vertices and is not of
  the form $G[H/u]$, except in a trivial way with $G$ or $H$ reduced
  to a single vertex. The paths $a\longrightarrow b\longrightarrow c$
  and $a-b-c-d$ are examples of small prime graphs.

  We will also use the graph operations $\oplus,\otimes$ and
  $\overrightarrow {\otimes}$: $G\oplus H$ is the disjoint union of
  $G$ and $H$, $G\overrightarrow{\otimes}H$ is $G\oplus H$ augmented
  with edges from each vertex of $G$ to each vertex of $H$, and
  $G\otimes H$ is $G\overrightarrow {\otimes}H$ augmented with edges
  from each vertex of $H$ to each vertex of $G$.  In all cases, we
  replace if necessary $H$ by an isomorphic copy disjoint with
  $G$. These operations can be defined by $K[G/u,H/v]$ for graphs $K$
  with two vertices $u$ and $v$, and, respectively, no edge, an edge
  from $u$ to $v$, edges between $u$ and $v$ in both directions. They
  are associative.  We will consider them as operations of variable
  arity in the usual way. The operations $\oplus$ and $\otimes$ are
  also commutative. They transform undirected graphs into undirected
  graphs. More generally, every graph $G$ can be turned as follows
  into a graph operation.  We enumerate its vertices as
  $v_{1},\dots,v_{n}$, and we define an $n$-ary graph operation
  $\sigma_{G}$ (where $\sigma$ stands for \textit{substitution}) by
  $\sigma_{G}(H_{1},\dots,H_{n})$ = $G[H_{1}/v_{1},\dots,H_{n}/v_{n}]$.

  Let us go back to modular decomposition.  The complete nodes are of
  two possible types, $\oplus$ or $\otimes$, because if $N$ is a
  "complete" strong module with $n$ sons $N_{1},\dots,N_{n}$, then
  either $G[N]=G[N_{1}]\oplus\dots\oplus G[N_{n}]$ or
  $G[N]=G[N_{1}]\otimes\dots\otimes G[N_{n}]$.

  If $N$ is "linear" with $n$ sons $N_{1},\dots,N_{n}$ ordered in this
  way (cf.\ T3 in Theorem \ref{T:2.1}), then either
  $G[N]=G[N_{1}]\overrightarrow{\otimes}\dots
  \overrightarrow{\otimes}G[N_{n}]$ or $G[N]=G[N_{n}]\overrightarrow
  {\otimes}\dots\overrightarrow{\otimes}G[N_{1}]$.

  If $ N$ is "prime", with $n$ sons $N_{1},\dots,N_{n}$ then
  $G[N]=\sigma_{K}(G[N_{1}],\dots$, $G[N_{n}])$ for some prime graph
  $K$. We have $n\geq3$; the operations corresponding to the graphs
  $K$ with 2 vertices are $\oplus,\otimes$ and
  $\overrightarrow{\otimes}$.

  The terms \emph{complete, linear} and \emph{prime} are defined after
  Theorem \ref{T:2.1}. If the given graph is a \emph{dag}, i.e., a directed
  graph without circuits, then no node is of type $\otimes$. If it is
  undirected, no node is of type $\overrightarrow{\otimes}$.

  An MS formula $\varphi_{\oplus}(X,Y)$ can express that $X$ is a
  complete strong module of type $\oplus$ and $Y$ is one of its sons.
  An MS formula $\varphi_{\otimes}(X,Y)$ can do the same for
  $\otimes$.  An MS formula
  $\varphi_{\overrightarrow{\otimes}}(X,Y,Z)$ can express that $X$ is
  a linear strong module, $Y$ and $Z$ are two sons such that
  $G[X]=\dots\overrightarrow
  {\otimes}G[Y]\overrightarrow{\otimes}G[Z]\overrightarrow{\otimes}\dots$. An
  MS\ formula $\varphi_{\Pr}(X,Y,Z)$ can express that $X$ is a prime
  strong module, $Y$ and $Z$ are two sons such that
  $G[X]=K[\dots,G[Y]/u_{i},\dots,G[Z]/u_{j},\dots.]$ where
  $u_{i}\longrightarrow u_{j}$ in the prime graph $K$.

  By using these formulas, one can build the \emph{graph
  representation of the modular decomposition} of a graph $G$, denoted
  by $Gdec(G)$.  This is a binary relational structure consisting of
  the rooted tree $T_{\mS(G)} $ enriched with the following
  informations:
\begin{itemize}
\item[\(a] its nodes of types $\oplus,\otimes$ and
  $\overrightarrow{\otimes}$ are labelled by their respective types,
\item[\(b] if $N$ is a "linear" node and
  $G[N]=G[N_{1}]\overrightarrow{\otimes}\dots
  \overrightarrow{\otimes}G[N_{n}]$, we set an edge
  $N_{i}\longrightarrow N_{i+1}$ for each $i=1,\dots,n-1$.
\item[\(c] If $N$ is a "prime" strong module, $N_{i}$ and $N_{j}$ are two sons
  and
\[G[N]=K[\dots,G[N_{i}]/u_{i},\dots,G[N_{j}]/u_{j},\dots]\ ,\]
  we set an edge $N_{i}\longrightarrow N_{j}$ whenever
  $u_{i}\longrightarrow u_{j}$ in $K$.
\end{itemize}

  This representation is in certain cases space efficient.  Consider
  the graph $G$ of a strict linear order on $n$ elements (in other
  words, a\emph{\ transitive tournament}). This graph has $n$ vertices
  and $n(n-1)/2$ edges.\ The graph $Gdec(G)$ has $n+1$ vertices and
  $2n-1$ edges.

  The tree $T_{\mS(G)}$ and the structure $Gdec(G)$ can be constructed
  by MS transductions using an arbitrary linear order of the vertices
  of the given graph as auxiliary information. From the relational
  structure $Gdec(G), $ that is actually a vertex- and edge-labelled
  directed graph, one can reconstruct $G$ by an MS\ transduction. We
  refer the reader to \cite{courcelle96} for illustrated examples and
  further developments, and to \cite{CoDe05} for the extension of
  these constructions to countable graphs.

  There are two distinct extensions of modular decomposition to
  hypergraphs: the decomposition into \emph{committees} of undirected
  unranked hypergraphs (see \cite{ChHaMa81,CuEd80}) and the modular
  decomposition of $k$-\emph{structures} which are $k$-ary relational
  structures, hence are labelled directed hypergraphs of rank $k$ (see
  \cite{EhMc94}). In both cases the families of sets are MS definable
  and Proposition \ref{P:2.1} is applicable.

\subsection{Factors in directed acyclic graphs}\label{SS:2.3}

  We review some results of Courcelle \cite{courcelle99}, Capelle
  \cite{capelle97}, Habib et al. \cite{HaHuSp95} concerning directed
  acyclic graphs.  We show that they can be reformulated in the
  framework of this section and slightly improved.

\begin{defi}\label{D:2.5}
  \emph{2-graphs and 2-dags} In this subsection, we consider directed
  graphs, possibly with\emph{\ multiple edges} (hence, not necessarily
  simple, as in the previous section).  Those without circuits (and
  loops) are called \emph{dags} (for \emph{directed acyclic
  graphs}). We denote by $E_{G}$ the set of edges of a graph $G$.  A
  \emph{2-graph} is a graph with two distinct distinguished vertices
  denoted by $s_{1}(G)$ and $s_{2}(G)$ called its \emph{sources}. We
  denote by $G^{0}$ the underlying graph, i.e., the same graph without
  distinguished vertices (the sources are turned into "ordinary"
  vertices). We use "2-graph" as an abreviation of "graph with 2
  sources" also called sometimes "2-terminal graph"; 2-graphs
  \emph{are not} particular $C$-graphs in the sense of the definition
  of clique-width, recalled in Appendix 2.

  A \emph{2-dag} is a 2-graph without circuits such that $s_{1}(G)$ is
  the unique vertex of indegree 0, $s_{2}(G)$ is the unique vertex of
  outdegree 0 and every vertex is on a directed path from $s_{1}(G)$
  to $s_{2}(G)$. We denote by $V_{G}^{0}$ the set
  $V_{G}-\{s_{1}(G),s_{2}(G)\}$, called the set of \emph{internal
  vertices} of $G$. An orientation of a graph making it into a 2-dag
  is also called a \emph{bipolar orientation}.

  For example, the graph of the "Wheatstone bridge" consisting of the
  directed path $a\longrightarrow b\longrightarrow c\longrightarrow d$
  with additional edges $a\longrightarrow c$ and $b\longrightarrow d$
  is a 2-dag if its two sources are $a$ and $d$ (in this order) and is
  a 2-graph and a dag but is not a 2-dag if its two sources are $a$
  and $b$.

  A \emph{factor} of a 2-dag $G$ is a 2-dag $H$ such that $H^{0}$ is a
  subgraph of $G^{0}$ and if an edge of $G$ has one end in
  $V_{H}^{0}$, then it is in $H$. An edge is a factor (its two ends
  being the sources) and a 2-dag is one of its own factors. We let
  $\mFE(G)$ denote the set of edge sets of the factors of
  $G$.
\end{defi}

  The following proposition is proved in \cite{courcelle99}, Lemma 3.5
  and Corollary 3.6.  Its first assertion is also proved in
  \cite{capelle97} and \cite{HaHuSp95}.

\begin{prop}\label{P:2.2}
  For every 2-dag $G$, the family $\mFE(G)$ is weakly
  partitive and MS definable.\qed
\end{prop}

  Note that $\mFE(G)$ satisfies Property P'0.  The family of strong
  $\mFE(G)$-modules is denoted by $\mSFE(G)$.  In order to define the
  tree $T_{\mSFE(G)}$ of a 2-dag $G$ by an MS\ transduction we need to
  quantify over edge sets.  Hence, we represent graphs by their
  \emph{incidence structures}.  For a graph $G$, we let $\Inc(G)=$
  $\<V_{G}\cup E_{G},inc_{G}\>$ where $V_{G}$ is the set of vertices,
  $E_{G}$ is the set of edges and $inc_{G}$ is the ternary relation
  such that $inc_{G}(e,x,y)$ holds if and only if $e:x\longrightarrow
  y$ in $G$. By the general definitions, the binary relation
  \emph{mod}$_{\mSFE(G)}$ defines for every node $x$ of $T_{\mSFE(G)}$
  the set of edges of the corresponding factor, that we will denote by
  $G(x)$; hence $G(x)=G[N]$ if $N $ is the $\mFE(G)$-module
  represented by the node $x$. The following proposition is Theorem
  3.12 of \cite{courcelle99}.

\begin{prop}\label{P:2.3}
  There exists an MS transduction that transforms $\Inc(G)$ into
  $\Dec(G)=\<V_{G}\cup E_{G}\cup
  N_{T_{\mSFE(G)}},inc_{G},son_{T_{\mSFE(G)}},\emph{mod}_{\mSFE(G)}\>$
  for every 2-dag $G$.\qed
\end{prop}

  Although the leaves of $T_{\mSFE(G)}$ are (or correspond to) the
  edges of $G$, we keep $N_{T_{\mSFE(G)}}$ and $E_{G}$ disjoint in the
  structure $\Dec(G)$.  Proposition \ref{P:2.1} could be used here
  because the family $\mSFE(G)$ is MS definable (since graphs are
  represented by their incidence structures), but it would give a
  weaker result than Proposition \ref{P:2.3}, because we would need an
  auxiliary linear ordering of the edges of $G$ as input of the
  transduction, which is not the case in Proposition \ref{P:2.3}.

  The MS transduction of Proposition \ref{P:2.3} uses edge set
  quantifications. In the case of simple graphs, one can do the same
  \emph{without edge set quantifications}.

\begin{cor}\label{C:2.2}
  There exists a monadic second-order transduction that transforms the
  structure $\<V_{G},\edg_{G}\>$ into:
\[\<V_{G}\cup N_{T_{\mSFE(G)}},\edg_{G},son_{T_{\mSFE(G)}},\emph{fact}_{\mSFE(G)}\>\]
  for every simple 2-dag $G$, where $fact_{\mSFE(G)}(v,x)$ is defined
  to hold if and only if $v$ is a vertex of $G(x)$.
\end{cor}

\proof The proof of Theorem 3.12 in \cite{courcelle99} defines $\<V_{G}\cup
  E_{G}\cup N_{T},inc_{G},son_{T}$, \emph{mod}$_{T}\>$ from
  $\<V_{G}\cup E_{G},inc_{G}\>$ by an MS transduction that specifies
  the set of nodes of $T=T_{\mSFE(G)}$ as follows: Its internal nodes
  are pairs $(v,i)$ where $i=2$ or 3 and $v$ is a vertex, or pairs
  $(e,3)$ where $e$ is an edge.  The latter case corresponds to
  factors which are sets of at least two parallel edges.  In the case
  of simple graphs, there are no such factors, hence these pairs are
  not needed. A leaf of $T$ corresponds to a factor of $G $ reduced to
  a single edge $e$ and is defined as the pair $(e,4)$.  However, from
  the above remark, its father is an internal node defined as a pair
  $(v,i)$ where $i=2$ or 3 for a vertex $v$. Hence, this leaf can be
  defined as the pair $(v,4)$ in the former case and $(v,5)$ in the
  latter. It follows that $T$ can be specified with a set of nodes
  defined as a subset of $V_{G}\times \{2,3,4,5\}$.\qed

  We now review the graph operations associated with the various types
  of nodes of $T_{\mSFE(G)}$ where $G$ is a 2-dag. We define them
  actually for 2-graphs.

\begin{defi}\label{D:2.6}
  \emph{Operations on 2-graphs} The main operation is
  \emph{edge-substitution}. Two other operations will be defined as
  particular instances of it. For a 2-graph $K$ with directed edges
  $e_{1},\dots,e_{k}$, we denote by
  $K[G_{1}/e_{1},\dots,G_{k}/e_{k}]$ the result $H$ of the
  substitution of the 2-graphs $G_{1},\dots,G_{k}$ for the edges
  $e_{1},\dots,e_{k}$ respectively. For defining $H$, we assume what
  follows:
\begin{itemize}
\item[(i)] $K,G_{1},\dots,G_{k}$ have pairwise disjoint sets of edges,
\item[(ii)] $e_{i}:s_{1}(G_{i})\longrightarrow s_{2}(G_{i})$ for each $i$,
\item[(iii)] $K,G_{1},\dots,G_{k}$ have no vertices in common other
  than the ends of the edges $e_{i}$ and the sources of $G_{i}$, as
  required by (ii).
\end{itemize}

  We let $V_{H}=V_{K}\cup V_{G_{1}}\cup\dots\cup V_{G_{k}}$,
  $E_{H}=E_{K}\cup E_{G_{1}}\cup\dots\cup E_{G_{k}}
  -\{e_{1},\dots,e_{k}\}$ and $s_{i}(H)=s_{i}(K)$ for $i=1,2$. If the
  graphs are not disjoint as required, one takes disjoint copies and
  one fuses the sources of the graphs $G_{i}$ with the end vertices of
  the edges $e_{i}$ of $K$. In this case, the result of the
  substitution is well-defined up to isomorphism.

  The \emph{parallel composition} of two 2-graphs $G$ and $H$ is the
  graph $G//H$ defined as $K[G/e,H/f]$ where $K$ consists of two
  parallel edges, $e$$,f$: $s_{1}(K)\longrightarrow s_{2}(K)$. This
  operation is associative and commutative so that the expression
  $G_{1}//\dots//G_{k}$ is well-defined, and the ordering of the
  arguments is irrelevant. We define similarly the \emph{series
  composition}: $G\bullet H=K[G/e,H/f]$ where $K$ consists of two
  edges $e:s_{1}(K)\longrightarrow u$ and $f:u\longrightarrow
  s_{2}(K)$ for some (arbitrary) $u$. This operation is associative,
  so that the expression $G_{1}\bullet \dots\bullet G_{k}$ is
  well-defined, but the order of arguments matters. In order to have a
  shorter notation we will use $\theta _{K}(G_{1},\dots,G_{k})$ for
  $K[G_{1}/e_{1},\dots,G_{k}/e_{k}]$ where $e_{1},\dots,e_{k}$ is an
  enumeration of the set of edges of $K$ (it is not made explicit in
  the notation $\theta _{K}$).

  We will also use the constant \textbf{e} denoting the 2-dag
  consisting of the single edge $e:x\longrightarrow y$, with
  $s_{1}(\mathbf{e})=x,s_{2}(\mathbf{e})=y$, and the constant
  $\overline{\mathbf{e}}$ defined similarly, with
  $s_{1}(\overline{\mathbf{e}})=y,s_{2}(\overline{\mathbf{e}})=x$.
\end{defi}

  Terms built with these operations and constants denote 2-graphs. In
  some proofs, we will require that the arguments of each operation in
  a term are graphs with disjoint sets of edges (hence not graphs up
  to isomorphism).  In this case, if $t$ is a term denoting a graph
  with edges $e_{1},\dots,e_{k}$, it has $k$ occurrences of
  constants, which are $\mathbf{e}_{i}$ or $\overline{\mathbf{e}_{i}}$
  for $i=1,\dots,k$.

  We now consider the case of 2-dags.

\begin{prop}\label{P:2.4}
  Let $G$ be a 2-dag. An internal node $N$ of its decomposition tree
  $T_{\mSFE(G)}$ is of one of the following mutually exclusive types:
\begin{itemize}
\item[1)] $N$ is a complete node with sons $N_{1},\dots,N_{k}$,
  $N=N_{1}\cup\dots\cup N_{k}$, we have $G[N]=G[N_{1}]//\dots//G[N_{k}]$
  and none of $N_{1},\dots,N_{k}$ is a complete node.

\item[2)] $N$ is a prime node,
  $G[N]=\theta_{K}(G[N_{1}],\dots,G[N_{k}])$ where $K$ is a 2-dag that
  cannot be written $L\bullet M$ or $L//M$ or
  $K'[M_{1}/f_{1},\dots,M_{p}/f_{p}]$ except in a trivial way,
  with either $K'$ or all $M_{1},\dots,M_{p}$ reduced to single
  edges.

\item[3)] $N$ is a linear node with sons $N_{1},\dots,N_{k}$, we have
  $N=N_{1}\cup\dots\cup N_{k}$, none of $N_{1},\dots,N_{k}$ is linear
  and, either $G[N]=G[N_{1}]\bullet\dots\bullet G[N_{k}]$ or
  $G[N]=G[N_{k}]\bullet\dots\bullet G[N_{1}]$.
\end{itemize}

  A leaf of this tree is an edge $e:s_{1}(\mathbf{e})\longrightarrow
  s_{2}(\mathbf{e})$.
\end{prop}

\proof This follows from Proposition \ref{P:2.2} and Theorem
  \ref{T:2.1}. The three types of nodes correspond respectively to
  properties T1, T2 and T3 of Theorem \ref{T:2.1}. In all three cases,
  the graphs $G[N_{1}],\dots,G[N_{k}]$ have disjoint sets of edges and
  we need not make isomorphic copies. In the second case, the vertices
  of $K$ are vertices of $G$. Its edges are not edges of $G$: they
  mark positions where the subgraphs $G[N_{1}],\dots,G[N_{k}]$ must be
  substituted. The graph $K$ is simple because otherwise it can
  expressed as $\mathbf{e}//M$ or as
  $K'[M_{1}/f_{1},\dots,M_{p}/f_{p}]$ in a nontrivial way with
  some $M_{i}$ consisting of two parallel edges.\qed

  In Case 2, $K$ has $k\geq3$ edges because otherwise, it is of the
  form $\mathbf{e}\bullet\mathbf{f}$ or $\mathbf{e}//\mathbf{f}$, and
  $\theta_{K}$ is $\bullet$ or $//$. A 2-dag $K$ satisfying the
  conditions of Case 2 will be called \emph{prime}. In Case 3, the
  sons of a node will be numbered so that
  $G[N]=G[N_{1}]\bullet\dots\bullet G[N_{k}]$. For building a term $t$
  denoting a 2-dag, we need only the operations $\bullet$, // and
  $\theta_{K} $ where $K$ is a prime 2-dag, and the constants
  \textbf{e}.

  We now examine how such a term can be constructed in MS logic.  MS\
  formulas analogous to the formulas $\varphi_{\oplus}$,
  $\varphi_{\otimes}$, $\varphi_{\overrightarrow{\otimes}}$,
  $\varphi_{\Pr}$ used in Subsection \ref{SS:2.2} for the modular
  decomposition, can recognize which case applies to a given module
  $N$, and can specify its sons. According to the general method
  sketched at the end of Subsection \ref{SS:2.1}, we transform the
  structure constructed by the transduction of Proposition \ref{P:2.3}
  into a \emph{graph representation} of the canonical decomposition of
  the considered 2-dag, intended to be as space-efficient as
  possible. We let $\Rep(G)$ be the structure:
\[\<V_{G}\cup E_{G}\cup
  N_{T_{\mSFE(G)}},inc_{G},son_{T_{\mSFE(G)}},src_{1\mSFE(G)},src_{2\mSFE(G)},leaf_{T_{\mSFE(G)}}\>,
\]  
  where:
\begin{itemize}
\item[(1)] $src_{i\mSFE(G)}=\{(x,s_{i}(G(x)))\mid x\in N_{T_{\mSFE(G)}}\}$ and
\item[(2)] $leaf_{T_{\mSFE(G)}}=\{(x,e)\mid x$ is a leaf of
  $T_{\mSFE(G)}$ and $e$ is the unique edge of $G(x)\}$.
\end{itemize}
  We replace thus the relation \emph{mod}$_{\mSFE(G)}$ of $\Dec(G)$ by
  three functional relations, and we avoid a certain amount of
  redundancy.  The relation \emph{mod}$_{\mSFE(G)}$ can be defined by
  an MS\ formula in the structure $\Rep(G)$. The relation
  $leaf_{T_{\mSFE(G)}}$ is useful to establish the bijection between
  the leaves of the tree and the edges of the considered graph.  For
  simple graphs, we can use the simpler structure:
\[Rep'(G)=\<V_{G}\cup N_{T_{\mSFE(G)}},\edg_{G},son_{T_{\mSFE(G)}},src_{1\mSFE(G)},src_{2\mSFE(G)}\>\]
  because there is no need to relate a leaf of the tree to the
  corresponding edge which no longer exists as an element of the
  domain.

\begin{defi}\label{D:2.7}
  \emph{Separated representations.} As explained above, to every
  internal node of the tree $T_{\mSFE(G)}$ corresponds an edge
  substitution operation $\theta_{K}$, and this tree can be considered
  as the syntax tree of a term $t$ that denotes the 2-dag $G$ and is
  written with operations $\theta_{K}$ and constants \textbf{e} \
  denoting the different edges. In a proof in the next subsection, we
  will transform such a term $t$ denoting a 2-dag into another one
  $t'$, intended to denote a 2-graph $G'$, by
  replacing at certain occurrences in $t$, some operations
  $\theta_{K}$ by operations $\theta_{K'}$ of same arity.
  Then the evaluation of $t'$ giving $G'$ will be done
  by an MS transduction. For this purpose, we introduce a variant of
  the structure $\Rep(G)$, called a \emph{separated representation},
  where $G$ is a 2-dag.

  We let 
\[\eqalign{
 &\Rep^{sep}(G)=\cr
 &\<V_{H}\cup E_{G}\cup N_{T_{SF\mathcal{E}(G)}},inc_{H},
\varepsilon-\edg_{H},son_{T_{\mSFE(G)}},ssrc_{1\mSFE(G)},ssrc_{2\mSFE(G)},
leaf_{T_{\mSFE(G)}}\>,}
\]
  where:
\begin{itemize}
\item[(1)] $ V_{H}$ is the set of pairs $(x,i)$ for $x\in
  N_{T_{\mSFE(G)}} $ and $i=1,2$,
\item[(2)] $inc_{H}$ is the set of triples $(e,(x,1),(x,2))$ such that
  $e\in E_{G} $, $x$ is the corresponding leaf of $T_{\mSFE(G)}$,
\item[(3)] $\varepsilon-\edg_{H}((x,i),(y,j))$ is defined as holding if
  and only if $s_{i}(G(x))=s_{j}(G(y))$ and, either $x$ and $y$ are
  adjacent (one is the father of the other) or $x$ and $y$ are sons of
  some node $z$, and $s_{i}(G(x))\neq s_{k}(G(z))$ for $k=1,2$.
\item[(4)] $ssrc_{i\mSFE(G)}=\{(x,(x,i))\mid x\in N_{T_{\mSFE(G)}}\} $,
\item[(5)] the other sets and relations are as in $\Rep(G)$.
\end{itemize}
\end{defi}

  These sets and relations define a graph $H$.  Its vertices are pairs
  $(x,1)$ and $(x,2)$ denoting the two sources of the factors
  associated with the nodes $x$ of the tree.  Each pair represents a
  vertex of $G$. Since a vertex of $G$ belongs to several factors, it
  has several representations by vertices of $H$. For an example, if
  $z$ is a node with sons $x$ and $y$ such that $G(z)=G(x)\bullet
  G(y)$, then $s_{1}(G(z))=s_{1}(G(x))$, $s_{2}(G(z))=s_{2}(G(y))$,
  $s_{2}(G(x))=s_{1}(G(y))$.  The undirected
  $\varepsilon$-\emph{edges}, defined by the symmetric relation
  $\varepsilon-\edg_{H}$ materialize such equalities.  In this case, we
  have the following $\varepsilon$-edges: $(z,1)-(x,1)$, $(z,2)-(y,2)$
  and $(x,2)-(y,1)$. In the case where $G(z)=G(x)//G(y)$, we have the
  $\varepsilon$-edges: $(z,i)-(x,i)$, $(z,i)-(y,i)$, for $i=1,2$,
  which represent the equalities $s_{i}(G(z))=s_{i}(G(x))$,
  $s_{i}(G(z))=s_{i}(G(y))$, and the equalities
  $s_{i}(G(x))=s_{i}(G(y))$ follow by transitivity. The graph $H$ has
  also edges which correspond to those of $G$, however, they are not
  adjacent in $H$. They are "separated" by $\varepsilon$-edges. Since
  $t$ is a term denoting a 2-dag, its leaves correspond to factors
  with an edge directed from the first source to the second one. This
  justifies condition (2).

  This graph $H$, denoted by $\Sep(G)$, can be "extracted" from
  $\Rep^{sep}(G)$ which contains also $T_{\mSFE(G)}$ as it is defined
  from $\<V_{H}\cup E_{G},inc_{H},\varepsilon-\edg_{H}\>$. Figure 1
  below shows a graph $G$, and Figure 2 shows the corresponding graph
  $\Sep(G)$.  Edge directions are omitted for the purpose of
  readability. Dotted lines represent the pairs in
  $\varepsilon$-$\edg_{H}$. The following fact, which shows how one can
  reconstruct $G$ from $\Sep(G)$, is clear from the definition.

\begin{lem}\label{L:2.2}
  The graph $G$ is obtained from $\Sep(G)$ by the contraction of all
  $\varepsilon$-edges. There exist MS transductions transforming
  $\Rep(G)$ and $\Rep^{sep}(G)$ into each other, and $\Sep(G)$ into
  $\Inc(G)$.\qed
\end{lem}

\medskip\noindent\textbf{Remark.}
%\begin{rem}\label{R:rep}
  The structures $\Rep(G)$ and $\Rep^{sep}(G)$ use fixed finite
  signatures.  They encode terms written with the operations $//$ and
  $\bullet $ of variable arity, and the infinitely many operations
  associated with the prime graphs $K$. They are interesting from the
  point of view of the study of graph structure, but they are not
  space efficient as can be the graph representations of modular
  decompositions.
\medskip%\end{rem}

  We now give an application related to matroids.

\subsection{Whitney's 2-isomorphism theorem}\label{SS:2.4}

  We consider directed graphs without loops or isolated vertices and
  possibly with multiple edges. The \emph{cycle matroid} of a graph
  $G$ is the pair $M(G)=\<E_{G},indep_{G}\>$ where, for $X\subseteq
  E_{G}$, $indep_{G}(X)$ holds if and only if $G[X]$ has no undirected
  cycle. The matroid $M(G)$ does not depend on the directions of
  edges, however, the definitions are given for directed graphs
  because edge directions will be useful for some constructions. For
  matroids in general we refer the reader to the books by White
  \cite{white86} and Oxley \cite{oxley92}. Actually, we will need no
  more than this definition.

  We say that two graphs $G$ and $H$ are \emph{equivalent} if
  $E_{G}=E_{H}$ and $M(G)=M(H)$.  We require the equality of the sets
  of edges but nothing on the sets of vertices.  The vertices and the
  incidencies may be different in the two graphs.  In particular, any
  two forests with the same sets of edges have the same (trivial)
  cycle matroids, independently of how their edges are incident with
  vertices.  A theorem by Whitney characterizes the equivalence of
  2-connected graphs.

\begin{defi}\label{D:2.8}
  \emph{Twisting
.}
% and the 2-isomorphism theorem.}
  For a 2-graph $G$, we let $\widetilde{G}$ be the 2-graph with same
  underlying graph as $G$ except that its sources are swapped:
  $s_{1}(\widetilde{G})=$ $s_{2}(G)$ and $s_{2}(\widetilde{G})=$
  $s_{1}(G)$. We recall that $G^{0}$ is $G$ with its sources made into
  ordinary vertices. For disjoint 2-graphs $G$ and $H$, we let $G//H$
  be their parallel composition, obtained from the union of $G$ and
  $H$ by the fusion of $s_{1}(G)$ and $s_{1}(H)$, and of$ s_{2}(G)$
  and $s_{2}(H)$. It is important to note here that $E_{G//H}=
  E_{G}\cup E_{H}$.

  A graph is \emph{2-connected} if it is connected and the deletion of
  any vertex yields a connected graph.  A graph with just one edge or
  several parallel edges is 2-connected, and we consider graphs
  without loops.  A 2-graph $G$ is \emph{2-connected} if
  $(\mathbf{e}//G)^{0}$ is 2-connected.  Edge directions do not matter
  in these definitions. A 2-dag is a 2-connected 2-graph.

  If $G=(L//M)^{0}$ and $H=(L//\widetilde{M})^{0}$ where $L$ and $M$
  are connected 2-graphs, we say that $H$ is obtained from $G$ by a
  \emph{twisting}. Note that $E_{H}=E_{G}$.  Reversing an edge
  direction is a twisting.
\end{defi}

  The equivalence of two graphs $G$ and $H$ without isolated vertices
  is characterized by Whitney's \emph{2-isomorphism Theorem} as the
  existence of a transformation of $G$ into $H$ by a finite sequence
  of twistings and of transformations of two other types called
  \emph{vertex splitting} and \emph{vertex identification}.  See the
  chapter by J.~Oxley in the book edited by N.~White \cite{white86},
  or \cite{truemper80}.  However, these latter transformations do not
  apply to 2-connected graphs.  Hence, this theorem yields the
  following:

\begin{prop}\label{P:2.5}
  Two 2-connected graphs are equivalent if and only if one can be
  transformed into the other by a finite sequence of twistings.\qed
\end{prop}

\begin{figure}
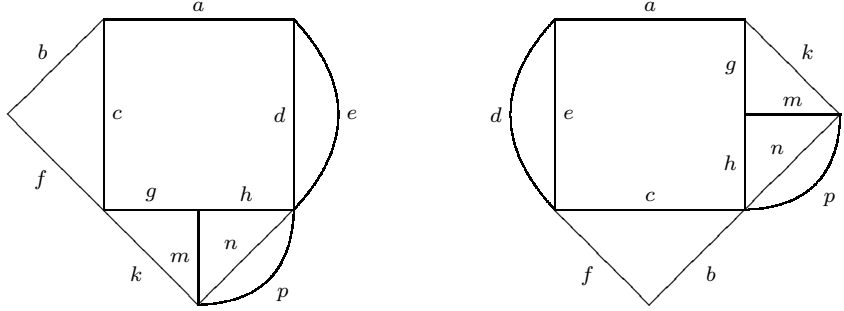
\label{F:Whitney1a}
\[\xy;<36 pt,0 pt>:
  (1,3)="A",
  (3,3)="B",
  (0,2)="C",
  \ar@{-}^a"A";"B"
  \ar@{-}_d"B";(3,1)
  \ar@{-}_(.25)h_(.75)g(3,1);(1,1)
  \ar@{-}_c(1,1);"A"
  \ar@{-}_b"A";"C"
  \ar@{-}_(.25)f_(.75)k"C";(2,0)
  \ar@{-}^m(2,0);(2,1)
  \ar@{-}^n(2,0);(3,1)
  \ar@{-}@/_12 pt/_p(2,0);(3,1)
  \ar@{-}@/_16.8 pt/_e(3,1);"B"
\endxy
\qquad\qquad
\xy;<36 pt,0 pt>:
  (0,3)="A",
  (2,3)="B",
  (3,2)="C",
  (2,2)="D",
  \ar@{-}^a"A";"B"
  \ar@{-}_(.25)g_(.75)h"B";(2,1)
  \ar@{-}_e(0,1);"A"
  \ar@{-}^c(0,1);(2,1)
  \ar@{-}_f(0,1);(1,0)
  \ar@{-}_b(1,0);(2,1)
  \ar@{-}^k"B";"C"
  \ar@{-}^m"D";"C"
  \ar@{-}^n(2,1);"C"
  \ar@{-}@/_12 pt/_p(2,1);"C"
  \ar@{-}@/^16.8 pt/^d(0,1);"A"
\endxy
\]
\caption{Two 2-isomorphic graphs \emph{G} (to the left) and \emph{H}.}
\end{figure}

  Figure 1 shows two graphs which are 2-isomorphic. Our aim is to
  prove the following theorem:

\begin{thm}\label{T:2.2}
  There exists an MS transduction that associates with $\<V_{G}\cup
  E_{G},inc_{G},\preccurlyeq\>$ where $G$ is a 2-connected graph and
  $\preccurlyeq$ ranges over all linear orders on $V_{G}$, the set of
  graphs having the same cycle matroid as $G$.
\end{thm}

\begin{lem}\label{L:2.3}
  A 2-connected 2-graph $G$ is either:
\begin{itemize}
\item[(o)] \textbf{e} or $\overline{\mathbf{e}}$ or
\item[(i)] $G_{1}//\dots//G_{k}$ for $k\geq2$, and some 2-connected
  2-graphs $G_{1},\dots,G_{k}$ not of this form, or
\item[(ii)] $G_{1}\bullet\dots\bullet G_{k}$ for $k\geq2$ and some
  2-connected 2-graphs $G_{1},\dots,G_{k}$ not of this form, or
\item[(iii)] $\theta_{K}(G_{1},\dots,G_{k})$ where $K$ is a prime
  2-dag and $G_{1},\dots,G_{k}$ are 2-connected 2-graphs.
\end{itemize}
\end{lem}

  Prime 2-dags are defined after Proposition \ref{P:2.4}. These
  expressions are unique except for the directions of the edges of $K$
  and the ordering of $G_{1},\dots,G_{k}$ in (i), but we will not need
  this fact.

\proof For a 2-dag $G$, this result is Proposition \ref{P:2.4}.
  Otherwise $G$ can be made into a 2-dag $H$ by reversing some edge
  directions. The result holds for $H$, whence also for $G$ by
  reversing again the same edges.  This corresponds to changing
  certain constants \textbf{e} into $\overline {\mathbf{e}}$.\qed

  It follows that every 2-connected 2-graph $G$ can be expressed as
  the value of a term $t$ belonging to the set $\mathcal{T}$ of finite
  terms defined recursively as follows:
\begin{itemize}
\item[]
\begin{itemize}
\item[either] $t=\mathbf{e}$, 
\item[or] $t=\overline{\mathbf{e}}$,
\item[or] $t=//(t_{1},\dots,t_{k})$,
\item[or] $t=\bullet(t_{1},\dots,t_{k})$,
\item[or] $t=\theta_{K}(t_{1},\dots,t_{k})$,
\end{itemize}
\end{itemize}
  where $t_{1},\dots,t_{k}$  are in $\mathcal{T}$.

  We recall that we denote by $\theta_{K}$ the substitution operation
  associated with $K$ and the list $e_{1},\dots,e_{k}$ of its edges:
  $\theta_{K}(G_{1},\dots,G_{k})=K[G_{1}/e_{1},\dots,G_{k}/e_{k}]$. (The
  list $e_{1},\dots,e_{k}$ is implicit in the notation $\theta_{K}).$
  The operations // and $\bullet$ have a variable arity. A term in
  $\mathcal{T}$ obtained by using recursively Lemma \ref{L:2.3} will
  be called a \emph{canonical term for G}.  It is unique up to the
  ordering of the arguments of the operations // and up to the
  directions of edges in $K$. If the set of edges of $G$ is
  $\{f_{1},\dots,f_{n}\}$, then the constants occurring in a canonical
  term for $G$ are $\mathbf{f}_{1}$ or
  $\overline{\mathbf{f}_{1}},\dots,\mathbf{f}_{n}$ or
  $\overline{\mathbf{f}_{n}}$.

\begin{defi}\label{D:2.9}
  \emph{Twistings of 2-graphs.}  Twisting for graphs is defined above.
  In order to characterize the twistings of 2-connected graphs in
  terms of their decompositions in 2-graphs, we extend the notion of
  twisting to 2-graphs. A \emph{twisting of a 2-graph} $G$ is a
  2-graph $H$ such that either $H=\widetilde{G}$, or $H^{0} $ is a
  twisting of $G^{0}$, $s_{i}(H)=s_{i}(G)$ for $i=1,2$, and
  $G^{0}=(L//M)^{0}$, $H^{0}=(L//\widetilde{M})^{0}$ in such a way
  that the two sources of $G$ are two vertices of $L$, of $M$, or of
  both (we may have $G=L//M$). These conditions imply that for every
  2-graph $K$, $K//H$ is a twisting of $K//G$. They also imply that if
  $G$ is a 2-connected 2-graph, then so are $H$ and, $L$ and $M$ when
  the second case of the definition is used.

  For every 2-graph $G$, we denote by $\triangledown(G)$ the least set
  of 2-graphs containing $G$ and closed under twisting, hence of
  2-graphs obtained from $G$ by a finite sequence of twistings.  For
  every graph $H$ in $\triangledown(G)$, $E_{H}=E_{G}$, and either
  $s_{i}(H)=s_{i}(G)$ for $i=1,2, $ or $s_{i}(H)=s_{3-i}(G)$ for
  $i=1,2.$
\end{defi}

\begin{lem}\label{L:2.4}

  Let $G$ be a 2-connected 2-graph and $H$ be a twisting of $G$.
\begin{itemize}
\item[(i)] If $G=G_{1}//\dots//G_{k}$ for $k\geq2$, and some
  2-connected 2-graphs $G_{1},\dots,G_{k}$ not of this form, then
  $H=G_{1}//\dots//G_{i-1}//H_{i}//G_{i+1}\dots//G_{k}$ where $H_{i}$ is
  a twisting of $G_{i}$, or $H=L_{1}//\dots//L_{k}$ where each $L_{i}$
  is either $G_{i}$ or $\widetilde {G_{i}}$,

\item[(ii)] if $G=G_{1}\bullet\dots\bullet G_{k}$ for $k\geq2$ and
  some 2-connected 2-graphs $G_{1},\dots,G_{k}$ not of this form, then
  $H=G_{1}\bullet\dots\bullet G_{i-1}\bullet H_{i}\bullet
  G_{i+1}\bullet\dots\bullet G_{k}$ where $H_{i}$ is a twisting of
  $G_{i}$, or $H=G_{1}\bullet\dots\bullet G_{i-1}\bullet
  \widetilde{G_{j}}\bullet\widetilde{G_{j-1}}\bullet\dots\bullet\widetilde
  {G_{i+1}}\bullet\widetilde{G_{i}}\bullet G_{j+1}\bullet\dots\bullet
  G_{k}$ \ for $1\leq i<j\leq k$,

\item[(iii)] $G=K[G_{1}/e_{1},\dots,G_{k}/e_{k}]$ where $K$ is a prime
  2-dag and $G_{1},\dots,G_{k}$ are 2-connected 2-graphs, then
  $H=K[G_{1}/e_{1},\dots,H_{i}/e_{i},\dots,G_{k}/e_{k}]$ where $H_{i}$
  is a twisting of $G_{i}$, or $
  H=\widetilde{K}[G_{1}/e_{1},\dots,G_{k}/e_{k}]=\widetilde{G}.$
\end{itemize}
  Conversely, in all cases, every graph $H$ of the above forms is
  either $G$ or a twisting of $G$.
\end{lem}

  Note the special cases of 
\begin{itemize}
\item[(i)] $H=\widetilde{G}=\widetilde{G_{1}}//\dots//\widetilde{G_{k}}$, and
\item[(ii)]$H=\widetilde{G}=\widetilde{G_{k}}\bullet\widetilde{G_{k-1}}
  \bullet\dots\bullet\widetilde{G_{2}}\bullet\widetilde{G_{1}}$.
\end{itemize}

\proof Let $G$ be defined from $G_{1},\dots,G_{k}$ by one of cases
  (i)-(iii) and $H$ be a twisting of $G.  $If $H=\widetilde{G}$, then
  the conclusions hold in all three cases.  Let us now assume that $\
  G^{0}=(L//M)^{0}$ and $H^{0}=(L//\widetilde{M})^{0}$ and, without
  loss of generality, that the two sources of $G$ are vertices of $L$.

  We will prove that we have one of the following three cases:

\begin{itemize}
\item[\(a] $M^{0}$ is a subgraph of some $G_{i}$ in any of cases
  (i)-(iii), then the replacement of $M$ by $\widetilde{M}$ yields a
  2-graph $H_{i}$, and by replacing $G_{i}$ by $H_{i}$, we obtain $H$
  from $G$, as required.

\item[\(b] $G$ satisfies case (i) and the two sources of $M$ are those
  of $G$: then $M$ is the parallel composition of some of the $G_{i}$'s,
  and we obtain $H$ from $G$ by replacing each of these $G_{i}$'s by
  $\widetilde{G_{i}}$, this is the second possibility of case (i).

\item[\(c] $G$ satisfies case (ii) and one source of $M$ is a source
  of some factor $G_{i}$.  Then the other one is also a source of some
  factor $G_{j}$ (with $i\neq j$, otherwise case (a) applies). Then
  $M=G_{i'}\bullet \dots\bullet G_{j'}$ for some $1\leq
  i'<j'\leq k$, and $H$ is defined by the second
  possibility of case (iii).
\end{itemize}
  To complete the proof, we need only verify that there are no other cases.

  As in the proof of Lemma \ref{L:2.3}, we make $G$ into a 2-dag
  $G'$ by reversing if necessary some edge directions. We
  denote by $G_{i}'$, $L'$ and $M'$ the
  2-graphs obtained from $G_{i}$, $L$ and $M$ by these reversals. An
  internal vertex of $M'$ is on a directed path from
  $s_{1}(G')$ to $s_{2}(G')$. This path goes through
  the two sources of $M'$.  By changing if necessary the
  source numbers of $M'$ we may assume that this path
  traverses $M'$ from $s_{1}(M')$ to
  $s_{2}(M')$. All paths associated in this way with the
  internal vertices of $M'$ do the same.  They must traverse
  $M'$ from $s_{1}(M')$ to $s_{2}(M')$
  otherwise $M'$ whence $G'$ has a circuit.  Hence
  $M'$ is a 2-dag and a factor of $G'$. Clearly, the
  $G_{i}$'s are also factors of $G'$.  Consider its
  decomposition tree $T_{\mSFE}(G')$: the $G_{i}$'s are the
  sons of its root. We apply Theorem \ref{T:2.1}(4) to $M'$:
  if it is a strong module (with respect to $\mSFE(G')$), it
  corresponds to a node of this tree, and thus is a factor of
  (possibly equal to) some $G_{i}$. If it is not strong it is a union
  of sons of a strong module $N$, satisfying T1 or T3. If $N$ is the
  root, we are in the above cases (b) or (c) for
  $M',G_{1}',\dots,G_{k}'$ instead of
  $M,G_{1},\dots,G_{k}$. Otherwise, $M'$ is a factor of some
  $G_{i}'$.  By resestablishing the original edge directions,
  we see that $M$ satisfies one of \(a, \(b, \(c\ as required.

  This completes the proof of the ``only if'' direction. The other one
  is easy to verify.\qed

  A \emph{$k$-permutation} is a permutation of $\{1,\dots,k\}$.  For
  every $k$-permutation $\pi$, we denote by $\bullet_{\pi}$ the
  operation of arity $k$ such that:
\[\bullet_{\pi}(G_{1},\dots,G_{k})=\bullet(G_{\pi(1)},\dots,G_{\pi(k)}).\]
  For every canonical term $t$, we denote by $\triangledown(t)$ the
  set of terms defined inductively as follows:
\[\eqalign{
  \triangledown(\mathbf{e})
&=\triangledown(\overline{\mathbf{e}})
 =\{\mathbf{e,}\overline{\mathbf{e}}\}\cr
  \triangledown(//(t_{1},\dots,t_{k}))
&=\{//(s_{1},\dots,s_{k})\mid s_{i}\in\triangledown(t_{i})\}\cr
  \triangledown(\bullet(t_{1},\dots,t_{k}))
&=\{\bullet_{\pi}(s_{1},\dots,s_{k})\mid s_{i}\in\triangledown(t_{i}),
    \pi\,\hbox{\ is a $k$-permutation}\,\}\cr
  \triangledown(\theta_{K}(t_{1},\dots,t_{k}))
&=\{\theta_{K}(s_{1},\dots,s_{k})\mid s_{i}\in\triangledown(t_{i})\}\cup\{\theta_{\widetilde{K}}(s_{1},\dots,s_{k})\mid s_{i}\in\triangledown(t_{i})\}.\cr
  }
\]

\begin{lem}\label{L:2.5}
  For every 2-connected 2-graph $G$ with canonical term $t$, the set
  of 2-graphs $\triangledown(G)$ is the set of values of the terms in
  $\triangledown(t)$.
\end{lem}

\proof For every 2-connected 2-graph $G$ we have in the four cases of
  Lemma \ref{L:2.3}\:
\begin{itemize}
\item[(o)] $\triangledown(\mathbf{e})=\triangledown(\overline{\mathbf{e}})
  =\{\mathbf{e,}\overline{\mathbf{e}}\}$,

\item[(i)] $\triangledown(G_{1}//\dots//G_{k})
  =\triangledown(G_{1})//\dots//\triangledown (G_{k})$,

\item[(ii)] $\triangledown(G_{1}\bullet\dots\bullet G_{k})
  =\bigcup\{\triangledown(G_{\pi(1)})\bullet\dots\bullet
  \triangledown(G_{\pi(k)})\mid\pi\,\hbox{\ is a $k$-permutation}\,\}$,

\item[(iii)] $\triangledown(K[G_{1}/e_{1},...,G_{k}/e_{k}])=
  K[\triangledown(G_{1})/e_{1},...,\triangledown(G_{k})/e_{k}]\cup
  \widetilde {K}[\triangledown(G_{1})/e_{1},...,\triangledown(G_{k})/e_{k}]$,
\end{itemize}
  where in all cases the operations on 2-graphs extend to sets of
  2-graphs in the natural way.

  The result for cases (o),(i),(iii) follows from Lemmas \ref{L:2.3}
  and \ref{L:2.4}. For case (ii), the inclusion $\subseteq$ follows
  from Lemma \ref{L:2.4} (ii), and the inclusion $\supseteq$ follows
  also from the facts that every permutation is a composition of
  transpositions and that
\[G_{1}\bullet\dots\bullet G_{i-1}\bullet G_{i+1}\bullet G_{i}
  \bullet\dots\bullet G_{k}
\]
  is a twisting of
\begin{center}
\hfill$G_{1}\bullet\dots\bullet G_{i-1}\bullet\widetilde{G_{i}}\bullet
\widetilde{G_{i+1}}\bullet G_{i+1}\dots\bullet G_{k}\;.$\hfill\qEd
\end{center}

  The idea of the proof of Theorem \ref{T:2.2} is illustrated by Figures 1
  to 3.  By reversing some edge directions if necessary, we make the
  given graph into a 2-dag $G$ with its two sources the ends of some
  edge. From the decomposition tree $T_{\mSFE(G)}$ \ we construct the
  canonical term $t$ of $G$ and the graph $\Sep(G)$ from which $G$ is
  obtained by contraction of the $\varepsilon$-edges. The
  $\varepsilon$-edges of $\Sep(G)$ represent the graph operations with
  which $t$ is built.  In order to produce a term $t'$ in
  $\triangledown(t)$ yielding $G'$ equivalent to $G$, it suffices to
  modify some operations in $t$ according to Lemma \ref{L:2.5}.  These
  modifications are reflected by modifications of the
  $\varepsilon$-edges of $\Sep(G)$ giving a graph $M(\Sep(G))$ (where
  $M$ means "modification") from which $G'$ is obtained by contracting
  the $\varepsilon$-edges.  All these manipulations can be done by MS
  transductions.

\begin{exa}\label{E:1-3}
  These definitions are illustrated in Figures 1--3.  Figure 1 shows a
  graph $G$ and a graph $H$ that is 2-isomorphic to $G$.  We make $G$
  into a 2-dag, the two sources of which are the ends of edge $a$.
  The corresponding canonical term is:
\[t=//(\mathbf{a},\bullet\lbrack//(\mathbf{c},\bullet(\mathbf{b},\mathbf{f})),
  \theta_{K}(\mathbf{g},\mathbf{k},\mathbf{m},\mathbf{h},//(\mathbf{n},
  \mathbf{p})),//(\mathbf{d},\mathbf{e})])
\]
  where $K$ is the graph $K_{4}^{-}$ (defined as $K_{4}$ minus one
  edge).  Figure 2 shows the corresponding graph $\Sep(G)$. The
  $\varepsilon$-edges are represented by dotted lines.  The graph
  $M(\Sep(G))$ on Figure 3 is obtained by modifying certain
  $\varepsilon$-edges, and the modified edges are represented by
  broken lines.  These modifications correspond to replacing in $t$
  the subterm $\bullet(\mathbf{b},\mathbf{f)}$ by
  $\bullet(\mathbf{f},\mathbf{b)}$, $\theta_{K}$ by
  $\theta_{\widetilde{K}}$ and the operation $\bullet$ occurring first
  in the subterm $\bullet\lbrack t_{1},t_{2},t_{3}]$ by
  $\bullet_{\pi}$ where $\pi(1)=3$, $\pi(2)=1$, $\pi(3)=2$. For the
  purpose of readability, the edges of $G$ are undirected.
\begin{figure}
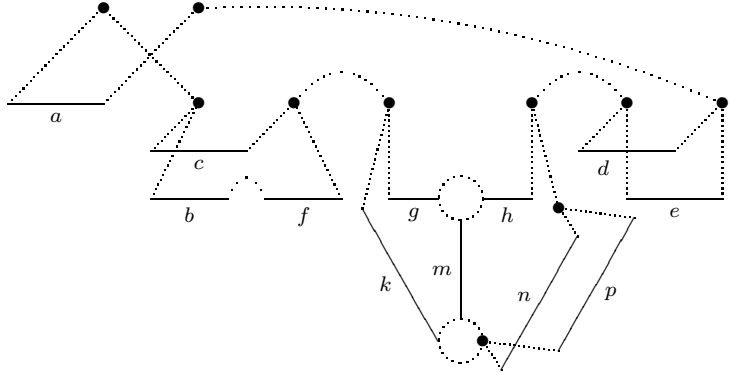
\label{F:w1a}
\[\xy;<36 pt,0 pt>:
  (2,3)*{\bullet}="A",
  (3,3)*{\bullet}="B",
  (3,2)*{\bullet}="C",
  (4,2)*{\bullet}="D",
  (5,2)*{\bullet}="E",
  (6.5,2)*{\bullet}="F",
  (7.5,2)*{\bullet}="G",
  (8.5,2)*{\bullet}="H",
  (5.75,1)*++{\ }="I"*\frm{.o},
  (5.75,-.5)*++{\ }="J"*\frm{.o},
  (3.5,1)="K",
  "I"!R+(.8,-.1)*{\bullet}="L",
  "J"!R*{\bullet}="M",
  \ar@{.}"E";(5,1)
  \ar@{.}"F";(6.5,1)
  \ar@{.}"G";(7,1.5)
  \ar@{.}"H";(8,1.5)
  \ar@{.}"G";(7.5,1)
  \ar@{.}"H";(8.5,1)
  \ar@{-}_-g(5,1);"I"
  \ar@{-}^-h(6.5,1);"I"
  \ar@{-}_m"I"!D-(0,.05);"J"!U+(0,.05)
  \ar@{-}_(.25)d(7,1.5);(8,1.5)
  \ar@{-}_e(7.5,1);(8.5,1)
  \ar@{-}^b"K"-(.2,0);"K"-(1,0)
  \ar@{-}_f"K"+(.2,0);"K"+(1,0)
  \ar@{.}"C";"K"-(1,0)
  \ar@{.}"D";"K"+(1,0)
  \ar@{.}@/^8 pt/"K"-(.2,0);"K"+(.2,0)
  \ar@{.}@/^12 pt/"D";"E"
  \ar@{.}@/^12 pt/"F";"G"
  \ar@{.}"A";"C"
  \ar@{.}"B";(2,2)
  \ar@{.}"A";(1,2)
  \ar@{-}_a(1,2);(2,2)
  \ar@{.}"C";(2.5,1.5)
  \ar@{.}"D";(3.5,1.5)
  \ar@{-}_c(2.5,1.5);(3.5,1.5)
  \ar@{.}@/^12 pt/"B";"H"
  \ar@{-}_k"I"!L-(.8,.1);"J"!L
  \ar@{.}"I"!L-(.8,.1);"E"
  \ar@{.}"L";"L"+(.8,-.1)
  \ar@{.}"M";"M"+(.8,-.1)
  \ar@{-}^p"L"+(.8,-.1);"M"+(.8,-.1)
  \ar@{.}"L";"L"+(.2,-.3)
  \ar@{.}"M";"M"+(.2,-.3)
  \ar@{-}_n"L"+(.2,-.3);"M"+(.2,-.3)
  \ar@{.}"F";"L"
\endxy
\]
\caption{The separated graph \protect{$\Sep(G)$}.}
\end{figure}
\end{exa}

  We now detail the proof more formally.

\medskip\noindent\emph{Proof of Theorem \ref{T:2.2}.}
  We choose in the given graph $G$ two adjacent vertices, we make them
  into sources $s_{1}(G)$ and $s_{2}(G)$, and we change some edge
  directions to make $G$ into a 2-dag. This is possible by Lemma 3.1
  in \cite{courcelle99} since $G$ is 2-connected.  Furthermore this
  can be done by an MS transduction taking $\Inc(G)$ as input (by the
  reorientation technique of \cite{courcelle95}). Hence, we obtain a
  2-dag from $G$ by a finite sequence of twistings if $G$ is not a
  2-dag, because reversing an edge is a twisting.  Without loss of
  generality we now consider that the given graph $G$ is a
  2-dag. Using Lemma \ref{L:2.2}, we can construct the structure
  $\Rep^{sep}(G)$, and from it the structure $\Sep(G)$.

  The tree $T=T_{\mSFE(G)}$ in the structure $\Rep^{sep}(G)$ is the
  syntactic tree of the canonical term $t$ for $G$.

  The terms $t'$ in $\triangledown(t)$ are obtained by selecting:
\begin{itemize}
\item[-] a set $X$ of prime nodes of $T$ corresponding to an operation
  $\theta_{K}$ to be replaced by $\theta_{\widetilde{K}}$,

\item[-] for each linear node $x$ of arity $k$ a $k$-permutation $\pi$
such that the operation $\bullet$ at $x$ is to be replaced by
$\bullet_{\pi}$,

\item[-] a set $Y$ of leaves corresponding to reversals of edge
directions.
\end{itemize}
  The sets $X,Y$ are straightforward to specify as parameters
  $X,Y\subseteq N_{T}$ of the MS transduction we are constructing.

  The permutations associated with the linear nodes are obtained from
  a linear order $\preccurlyeq$ on $V_{G}$ as follows. We call
  $s_{2}(G(x))$ the \emph{leading vertex of the factor }$G(x)$ of $G$,
  for $x$ in $N_{T}$. (See after Proposition \ref{P:2.2} for the
  notation $G(x)$). A vertex may be leading for several factors. Let
  $x$ be a linear node with sequence of sons $y_{1},\dots,y_{k}$. The
  linear order $\preccurlyeq$ on $V_{G}$ will be used here to permute
  this sequence. The leading vertices of $G(y_{1})$, \dots, $G(y_{k})$
  are pairwise distinct. There exists a unique permutation $\pi $\
  such that $ s_{2}(G(y_{\pi(1)}))\prec \dots$ $\prec
  s_{2}(G(y_{\pi(k)}))$. We obtain in this way a (possibly identity)
  permutation of the list of sons of $x$. It is clear that the new
  ordering of the sons of $x$ is MS\ definable from $\preccurlyeq$ and
  the other relations of the structure $\Rep^{sep}(G)$.
\begin{figure}
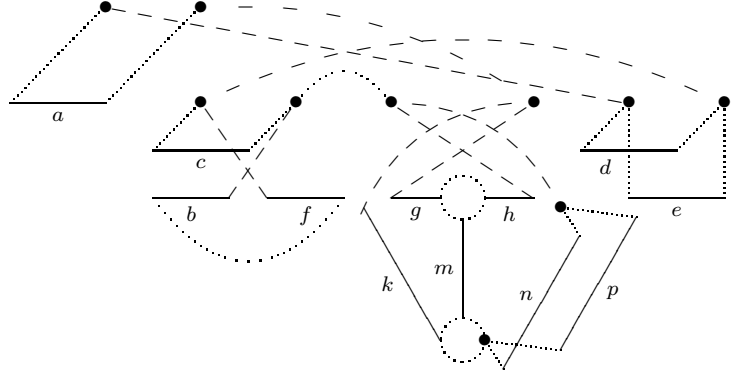
\label{F:w2b}
\[\xy;<36 pt,0 pt>:
  (2,3)*{\bullet}="A",
  (3,3)*{\bullet}="B",
  (3,2)*{\bullet}="C",
  (4,2)*{\bullet}="D",
  (5,2)*{\bullet}="E",
  (6.5,2)*{\bullet}="F",
  (7.5,2)*{\bullet}="G",
  (8.5,2)*{\bullet}="H",
  (5.75,1)*++{\ }="I"*\frm{.o},
  (5.75,-.5)*++{\ }="J"*\frm{.o},
  (3.5,1)="K",
  "I"!R+(.8,-.1)*{\bullet}="L",
  "J"!R*{\bullet}="M",
  \ar@{--}"F";(5,1)
  \ar@{--}"E";(6.5,1)
  \ar@{.}"G";(7,1.5)
  \ar@{.}"H";(8,1.5)
  \ar@{.}"G";(7.5,1)
  \ar@{.}"H";(8.5,1)
  \ar@{-}_-g(5,1);"I"
  \ar@{-}^-h(6.5,1);"I"
  \ar@{-}_m"I"!D-(0,.05);"J"!U+(0,.05)
  \ar@{-}_(.25)d(7,1.5);(8,1.5)
  \ar@{-}_e(7.5,1);(8.5,1)
  \ar@{-}^b"K"-(.2,0);"K"-(1,0)
  \ar@{-}_f"K"+(.2,0);"K"+(1,0)
  \ar@{--}"C";"K"+(.2,0)
  \ar@{--}"D";"K"-(.2,0)
  \ar@{.}@/^24 pt/"K"+(1,0);"K"-(1,0)
  \ar@{.}@/^12 pt/"D";"E"
  \ar@{--}@/^24 pt/"C";"H"
  \ar@{--}"A";"G"
  \ar@{.}"B";(2,2)
  \ar@{.}"A";(1,2)
  \ar@{-}_a(1,2);(2,2)
  \ar@{.}"C";(2.5,1.5)
  \ar@{.}"D";(3.5,1.5)
  \ar@{-}_c(2.5,1.5);(3.5,1.5)
  \ar@{--}@/^12 pt/"B";"F"
  \ar@{-}_k"I"!L-(.8,.1);"J"!L
  \ar@{--}@/^12 pt/"I"!L-(.8,.1);"F"
  \ar@{.}"L";"L"+(.8,-.1)
  \ar@{.}"M";"M"+(.8,-.1)
  \ar@{-}^p"L"+(.8,-.1);"M"+(.8,-.1)
  \ar@{.}"L";"L"+(.2,-.3)
  \ar@{.}"M";"M"+(.2,-.3)
  \ar@{-}_n"L"+(.2,-.3);"M"+(.2,-.3)
  \ar@{--}@/^12 pt/"E";"L"
\endxy
\]
\caption{The graph \protect{$M(\Sep(G))$}.}
\end{figure}
  The variable $\Pi$ will denote families of permutations of
  appropriate types associated with linear nodes (a $k$-permutation
  for a node with $k$ sons) and $\Pi(\preccurlyeq)$ will denote the
  one induced as defined above by a linear order $\preccurlyeq$ on
  $V_{G}$. We denote by $t_{X,Y,\Pi}$ the term obtained from $t$ by
  the modifications described above, based on $X,Y$ and $\Pi$. The set
  $\triangledown(t)$ is thus the set of all terms $t_{X,Y,\Pi}$.

\medskip\noindent\textbf{Claim 1}: There exists an MS transduction that
  associates the graph defined by the term $t_{X,Y,\Pi(\preccurlyeq)}$
  with the structure $(\Rep^{sep}(G),X,Y,\preccurlyeq)$, where $X$ is a
  set of prime nodes of the decomposition tree $T$, $Y$ a set of leaves
  and $\preccurlyeq$ is a linear order on $V_{G}$.

\medskip\noindent\emph{Proof of the claim.}
  By using $X,Y$ and $\preccurlyeq$, we transform $\Rep^{sep}(G)$
  into a graph $H'$ from which the graph $G'$ defined
  by the term $t_{X,Y,\Pi(\preccurlyeq)}$ can be obtained by the MS
  transduction that contracts the $\varepsilon$-edges. The
  construction consists in modifying the $\varepsilon$-edges in
  $\Sep(G)$, so as to represent the replacements of $\theta_{K}$ by
  $\theta_{\widetilde{K}}$ at the nodes in $X$, those of the operation
  $\bullet$ by $\bullet_{\pi}$ at every linear node where the
  corresponding permutation $\pi$ is specified by $\Pi(\preccurlyeq)$,
  and the reversal of edges at the leaves of $Y$.

  For this purpose we modify in $\Rep^{sep}(G)$ the relation
  $\varepsilon$-$\edg_{H}$ into $\varepsilon$-$\edg_{H^{\prime }}$ as
  follows:
\begin{itemize}
\item[1)] For every $x$ in $X$ and every son $y$ of $x$, we replace
  the pairs $((x,i),(y,j))$ and $((y,j),(x,i))$ in $\varepsilon$-$\edg_{H}$
  by $((x,3-i),(y,j))$ and $((y,j),(x,3-i))$.

\item[2)] For every linear node with ordered list of sons
  $y_{1},y_{2},\dots,y_{k}$ we have in $\varepsilon$-$\edg_{H}$ the
  following pairs, together with their inverses:
\[((x,1),(y_{1},1)),((y_{1},2),(y_{2},1)),\dots,
  ((y_{k-1},2),(y_{k},1)),((y_{k},2),(x,2))\;. 
\]
  We replace them by the following ones, together with their inverses:
\[((x,1),(y_{\pi(1)},1)),((y_{\pi(1)},2),(y_{\pi(2)},1)),\dots
((y_{\pi(k-1)},2),(y_{\pi(k)},1)),((y_{\pi(k)},2),(x,2))
\]
  where $\pi$ is the $k$-permutation of the family $\Pi(\preccurlyeq)$
  associated with $x$,

\item[3)] For every $y$ in $Y$, we replace a triple $(e,(y,1),(y,2))$
  in $inc_{H}$ by $(e,(y,2),(y,1))$.
\end{itemize}
  This modification of $\Rep^{sep}(G)$ can be done by an MS
  transduction using $X,Y$ and $\preccurlyeq$. We obtain in this way
  the graph $M(\Sep(G))$ from which can be defined by edge contractions
  the value $G'$ of the term $t_{X,\Pi(\preccurlyeq)}$.  It
  follows from Lemma \ref{L:2.2} that $G^{\prime }$ can be obtained
  from $M(\Sep(G))$, whence also from $(\Rep^{sep}(G),X,Y,\preccurlyeq)$
  by an MS transduction (we use here Proposition A.1.2).  This proves
  Claim 1.\qed

  Since $\Rep^{sep}(G)$ can be constructed from $\Inc(G)$ by an MS
  transduction, and by composing this transduction with the one just
  constructed, we get an MS transduction $\tau$ that defines
  \emph{some} graph 2-isomorphic to $G$ from $\Inc(G),X,Y$ and any
  linear order $\preccurlyeq$ on $V_{G}$.  To complete the proof, it
  remains to establish that \emph{every} family $\Pi$ of permutations
  (of appropriate types) is $\Pi(\preccurlyeq)$ for some linear order
  $\preccurlyeq$ on $V_{G}$. This will give us that every graph
  2-isomorphic to $G$ is obtained by $\tau$ from $\Inc(G)$ for some
  sets $X$ and $Y$ and \emph{some} linear order $\preccurlyeq$ on
  $V_{G}$.

\medskip\noindent\textbf{Claim 2}: Every family $\Pi$ of permutations
  associated with the linear nodes is $\Pi(\preccurlyeq)$ for some
  linear order $\preccurlyeq$ on $V_{G}$.

\medskip\noindent\emph{Proof of the claim.}
  Let a family $\Pi$ be given.  By using bottom up induction on
  $T$, we construct, for every node $x$ of $T$, an appropriate linear
  order on the vertices of $V_{G(x)}-\{s_{1}(G(x))\}$ handled as the
  increasing sequence $Seq(x)$ of its elements.

  Let us note that if $x$ and $y$ are incomparable nodes in the tree
  $T$ (\emph{incomparable} means that no one is an ancestor of the
  other) then the only vertices that can be common to $G(x)$ and
  $G(y)$ are among the source vertices.

  We can construct the sequences $Seq(x)$ by bottom up induction as follows:

  If $x$ is a leaf (it corresponds to an edge of $G)$ then $Seq(x)$ is
  the sequence with single element $s_{2}(G(x)).$

  If $x=//(x_{1},\dots,x_{k})$ we have sequences $Seq(x_{i})$ by
  induction.\ Since two distinct factors $G(x_{i})$ and $G(x_{j})$
  have only in common the sources
  $s_{a}(G(x))=s_{a}(G(x_{i}))=s_{a}(G(x_{j}))$ for $a=1,2$, \ the
  only vertex common to two sequences $Seq(x_{i})$ and $Seq(x_{j})$ is
  $s_{2}(G(x_{i}))=s_{2}(G(x_{j}))=s_{2}(G(x))$, and all sequences
  $Seq(x_{i})$ can be merged into a single one, that we can take as
  $Seq(x).$

  If $x=\bullet(x_{1},\dots,x_{k})$ and $\pi$ is the permutation of
  $\{1,\dots,k\}$ associated with $x$ (by the family $\Pi$), then,
  since the sequences $Seq(x_{i})$ obtained by induction are pairwise
  disjoint, we can take $Seq(x)=Seq(x_{\pi(1)})\dots
  Seq(x_{\pi(k)})$. By this construction, the permutation associated
  with $x$ by any linear order for which $Seq(x)$ is increasing is
  actually $\pi$.

  If $x=\theta_{K}(x_{1},\dots,x_{k})$ we have sequences $Seq(x_{i})$
  by induction. Two sequences $Seq(x_{i})$ and $Seq(x_{j})$ either are
  disjoint or share the only vertex
  $s_{2}(G(x_{i}))=s_{2}(G(x_{j}))=v$, in the case where the two edges
  $x_{i},x_{j}$ of $K$ have the same target $v$. Hence we can merge
  all sequences $Seq(x_{i})$ such that $x_{i}$ has target $v$ into a
  sequence $L(v)$. Then we concatenate the sequences $L(v)$ for all
  vertices of $v$ of $K$ (except for $v=s_{1}(K)$), which gives
  $Seq(x)$.

  Every sequence $Seq(x)$ is a subsequence of $Seq(y)$ if $y$ is an
  ancestor of $x$ by this construction.  From the choice made for
  linear nodes $x$ it follows that the sequence $Seq(root_{T})$ of the
  root of $T$ yields the appropriate permutation of $\Pi$ at each
  $x$. To have a linear order on $V_{G}$, we add to $Seq(root_{T})$
  the vertex $s_{1}(G)$ as very first element, proving Claim 2.\qed

  This concludes the proof of the theorem.\qed

  By a simple counting argument, one can see that it is impossible to
  specify all permutations of arbitrarily large sets $X$ with
  $k$-tuples of subsets of $X$ for fixed $k$.  For this reason, we
  specify the permutations associated with the linear nodes by linear
  orders on the vertices.

\section{Partitive families of bipartitions}\label{S:3}

  The general framework for split decomposition is defined by
  Cunnigham and Edmonds in \cite{CuEd80}.  It applies to other cases,
  in particular to hypergraphs and matroids.  Our presentation owes a
  lot to the dissertation of Montgolfier \cite{montgolfier03}.

\subsection{Definitions and general properties}\label{SS:3.1}

  We define families of bipartitions of a set $V$ associated with a
  partition of this set, the blocks of which are organized into an
  unrooted tree.  These definitions generalize two important examples:
  the decomposition defined by Tutte of a 2-connected graph in
  3-connected components and the \emph{split decomposition} defined by
  Cunnigham \cite{cunningham82}.

\begin{defi}\label{D:3.1}
  \emph{Bipartitions, overlapping bipartitions}.  A \emph{bipartition}
  of a nonempty set $V$ is an unordered pair of subsets, $P=\{A,B\}$
  such that $V=A\cup B$, $A\cap B =\varnothing$, $\{A,B\}\neq
  \{\varnothing,V\}$.  The sets $A,B$ will be called the \emph{blocks}
  of $P$.  We denote by $\mathcal{B(}V)$ the set of bipartitions of
  $V$. Two bipartitions $P$ and $Q$ \emph{overlap} if $A\perp B$ for
  all $A\in P$ and $B\in Q$. Hence $P$ and $Q$ do not overlap if and
  only if $A\cap B=\varnothing$ for some $A\in P$ and $B\in Q$.  A
  bipartition of the form $\{\{v\},V-\{v\}\}$ does not overlap any
  bipartition.

  By an \emph{unrooted tree} we mean a simple undirected connected
  graph without cycles (and without loops).  It has no root, and the
  \emph{leaves} are the nodes of degree 1. Its other nodes are the
  \emph{internal} nodes.\ The sets of nodes and of internal nodes of a
  tree $T$ are denoted by $N_{T}$ and $N_{T}^{int}$ respectively. For
  each edge $e:x-y$ of $T$, we denote by $T(x,y)$ the set of nodes,
  including $x$, that are reachable from $x$ by a path that \emph{does
  not} use edge $e$.

  Let $T$ be an unrooted tree with at least two nodes and
  $\mV=(V(x))_{x\in N_{T}}$ be a partition of a set $V$ such that
  $V(x)$ is not empty if $x$ is a node of degree 1 or 2.  We call
  $(T,\mV)$ a \emph{tree-partition} of $V$. For each edge $e:x-y$ of
  $T$, we let $P_{e}=\{P_{x},P_{y}\}$ where $P_{x}$ is the union of
  the sets $V(z)$ for $z\in T(x,y)$, and similarly for $P_{y}$ with
  $T(y,x)$. The family $\mathcal{B=B(}T,\mathcal{V)}$ of bipartitions
  $P_{e}$ is not empty and satisfies the following property:
\begin{itemize}
\item[B1:] no two bipartitions of  $\mB$  overlap.
\end{itemize}
  If $V(x)$ is empty for every internal node $x$ and is singleton for
  each leaf $x$, then $\mathcal{B=B(}T,\mathcal{V)}$ satisfies in
  addition the property:
\begin{itemize}
\item[B0:] $\{\{v\},V-\{v\}\}\in\mB$ for every $v\in V$.
\end{itemize}
  For a tree-partition $(T,\mV)$ we define $box_{T}(v,x)$ to hold if
  and only if $x\in N_{T}$ and $v\in V(x)$. For every nonempty family
  $\mB$ of bipartitions, we let
  $\mB^{+}=\mathcal{B\cup\{}\{\{v\},V-\{v\}\}\mid v\in V\}.  $Since
  $\{\{v\},V-\{v\}\}$ does not overlap any bipartition, $ \mB^{+}$
  satisfies B1 if and only if $\mB$ satisfies B1. A block of a
  bipartition of $\mB$ is called a $\mB$-\emph{block}.
\end{defi}

\begin{lem}\label{L:3.1}
  For every nonempty family $\mB\subseteq\mB(V)$ satisfying B1, there
  exists a tree-partition $(T,\mV) $ such that $\mB(T,\mV)=\mB$. It is
  unique up to isomorphism.
\end{lem}

\proof We first make some observations concerning $\mB(T,\mV)$ where
  $(T,\mV)$ is a tree-partition, by using the notation of the
  definition.

\medskip\noindent\textbf{Claim}: A $\mB$-block is minimal if and only
  if it is $V(x)$ for some leaf $x$ of $T$.

\medskip\noindent\emph{Proof of the claim.}
  It is clear that $V(x)$ is a minimal $\mB$-block if $x$ is a
  leaf. For the other direction, assume that $A$ is a minimal
  $\mB$-block and $A=P_{x}$ where $P_{e}=\{P_{x},P_{y}\},e:x-y$. If
  $x$ is not a leaf there is an edge $x-z$, $z\neq y$, and $P_{z}$ is
  a proper subset of $P_{x}$ by the condition that $V(u)$ is not empty
  if $u$ is a node of degree 1 or 2.  Hence, this cannot happen, $x$
  is a leaf and $A=P_{x}=V(x)$.\qed

  We first prove the unicity property.  Assume $\mB(T,\mV)=\mB(T',\mV')$
  for two tree-partitions.

  Let $R$ be a minimal block. We have $R=V(r)$ for some leaf $r$ of
  $T$ by the claim. We make $T$ into a rooted (directed) tree with
  root $r$ and we let $s$ be adjacent to $r$ in $T$.  Let
  $V_{1}(u)=V(u)$ for every node $u\neq r$.  By Definition
  \ref{D:2.1}, $T(s)$ is the subtree of $T$ with root $s$; its nodes
  are those of $T$ except $r$. Clearly, $\mF(T(s),\mV_{1})$ is the set
  of $\mB$-blocks that do not include $R$.

  Similarly for $\mB(T',\mV')$ we have $R=V'(r')$ for some leaf $r'$
  of $T'$, we let $s'$ be adjacent to $r'$ in $T'$, and we have
  $\mF(T'(s'),\mV_{1}')=\mF(T(s),\mV_{1})$.  For a node $u$ of $T(s)$
  of outdegree 1, hence of degree 2 in $T$, the set $V_{1}(u)$ is not
  empty, and the same holds for $(T'(s'),\mV_{1}')$.  Hence, by an
  observation made in Definition \ref{D:2.1}, $(T(s),\mV_{1})$ and
  $(T'(s'),\mV_{1}')$ are isomorphic.  So are $(T,\mV)$ and
  $(T',\mV')$, as was to be proved.

  We now prove the existence $(T,\mV)$ such that $\mB(T,\mV)=\mB$
  where $\mB\subseteq\mB(V)$ satisfies B1. We let $R$ be a minimal
  block and $\mF$ be the set of $\mB$-blocks that do not include
  $R$. Hence $\mF$ is a family of subsets of $V-R$ that satisfies P0.
  It satisfies Condition P1 because if $A$ and $A'$ in $\mF$ overlap,
  then $\{A,B\}$ and $\{A',B'\}$ overlap since $R\subseteq B\cap B'$
  and this cannot happen since $\mB$ satisfies B1.

  Let $(T_{\mF},V_{\mF})$ be as in Definition \ref{D:2.1}.  Hence
  $\mF(T_{\mF},V_{\mF})=\mF$. We recall that
  $V_{\mF}(N)=N-\bigcup\{M\mid\hbox{\ $M$ is a son of $N$ in
  $T_{\mF}$}\}$. We add a new node $r$ linked to the root $s$ of
  $T_{\mF}$, we denote by $T$ the undirected tree obtained in this
  way, and we let $V(r)=R$, $V(N)=V_{\mF}(N)$ if $N$ is a node of
  $T_{\mF}$.  Then $(T,V) $ is a tree-partition of $V$.  In particular
  if $N$ has degree 2, then $V(N) $ is not empty because $N$ has
  outdegree 1 in $T_{\mF}$ and $V_{\mF}(N)$ is not empty. We claim
  that $\mathcal{B(}T,\mathcal{V)=B}$.

  Let $P_{e}$ be the bipartition of $V$ associated with an edge $e$ of
  $T$.\ If $e:r-s$, then $P_{e}=\{R,V-R\}$ which belongs to
  $\mB$. Otherwise let $e$ be directed $x\longrightarrow y$ in
  $T_{\mF}$.  Then $P_{e}=\{P_{x},P_{y}\}$ .  The set $P_{y}$ is the
  $\mF$-module associated with the node $y$ of $T_{\mF}$.  Hence
  $\{P_{y},V-P_{y}\}\in\mB$.  But $V-P_{y}=P_{x}$.  Hence
  $P_{e}\in\mB$.

  Conversely, let $P=\{A,B\}\in\mB$.  If $P=\{R,V-R\}$ it is in
  $\mathcal{B(}T,\mathcal{V)}$, corresponding to the edge
  $r-s$. Otherwise it does not overlap $\{R,V-R\}$ and since $R$ is
  minimal, we have $R\subset A$ or $R\subset B$. Assume the
  first. Then $B\in\mF$, hence is a node $y $ of $T_{\mF}$, its father
  is some $x$ and we have $\{B,V-B\}\in \mB(T,\mV)$.  But $A=V-B$,
  hence $P\in\mB(T,\mV)$. This completes the proof.\qed

  We denote by $(T_{\mB},\mV_{\mB})$ the tree-partition associated
  with $\mB$ by Lemma \ref{L:3.1}. It does not depend on the choice of
  $r$ by the unicity property. It will be useful to extend this
  definition to the case where $\mB$ is empty: we let then $T_{\mB}$
  consist of a single node $r$ and $V_{\mB}(r)=V$.

\begin{lem}\label{L:3.2}
  Let $\mB\subseteq\mB(V)$ satisfy B1.  For a node $x\in N_{T_{\mB}}$
  of degree $k$ with incident edges
  $e_{1}:x-y_{_{1}},\dots,e_{k}:x-y_{k}$, the sets
  $P_{y_{_{1}}},\dots,P_{y_{k}}$ (where
  $P_{e_{i}}=\{P_{x}^{i},P_{y_{i}}\}$) are pairwise disjoint and we
  have:
\[V_{\mB}(x)=V-(P_{y_{_{1}}}\cup\dots  \cup
P_{y_{k}})=\bigcap\{P_{x}^{i}\mid i=1,\dots,k\}\;.
\]
  If $x$ is a leaf, then $k=1$ and $V_{\mB}(x)=$ $P_{x}^{1}$.
\end{lem}

\proof This is clear from the construction of Lemma \ref{L:3.1}, and the
  definition of $\overline{V_{\mF}}$ in Definition \ref{D:2.1}.\qed

  Let $\mC$ be a class of relational structures as in Section
  \ref{S:2}. For each $S$ in $\mC$, we let $\mB(S)$ be a family of
  bipartitions of the domain $D_{S}$ of $S$ satisfying condition
  B1. We say that $\mB$ is \emph{MS-definable} if there exists an MS
  formula $\varphi(X)$ such that for every $S$ in $\mC$, $\{A\mid
  \{A,B\}\in\mB(S)$ for some $B\}=\{A\subseteq D_{S}\mid S\models
  \varphi(A)\}$. With these definitions:

\begin{prop}\label{P:3.1}
  Let $\mC$ be a class of $\mR$-structures and $\mB$ be an MS
  definable family of bipartitions of\ the domains of the structures
  in $\mC$ which satisfies conditions B1. There exists a domain
  extending MS-transduction that associates with $(S,\preccurlyeq)$
  where $S=\<D_{S},(R_{S})_{R\in\mR}\>$ $\in\mC$ and $D_{S}$ is
  linearly ordered by $\preccurlyeq$, the structure
\[\Dec(S)=\<D_{S}\cup
  N_{T_{\mB(S)}},(R_{S})_{R\in\mR},\edg_{T_{\mB(S)}},box_{T_{\mB(S)}}\>
\]
  such that $\<N_{T_{\mB(S)}},\edg_{T_{\mB(S)}}\>=T_{\mB(S)}$.
\end{prop}

\proof Lemma \ref{L:3.1} reduces the construction of the tree $T_{\mB(S)}$ to
  that of a tree associated with a family $\mF$ of subsets of
  $D_{S}$. Since the structure $S$ is linearly ordered, one can take
  for $R $ the unique minimal $\mB(S)$-block that contains the
  $\preccurlyeq$-smallest element of $D_{S}$.  The corresponding
  family $\mF$ is thus MS definable. Using Proposition \ref{P:2.1}, an
  MS\ transduction can construct the corresponding rooted tree
  $T_{\mF}$, modified so as to yield the tree $T$ (cf.\ the proof of
  Lemma \ref{L:3.1}). One gets the desired unrooted tree
  $T_{\mB(S)}=\<N_{T_{\mB(S)}},\edg_{T_{\mB(S)}}\>$.  The definition of
  the relation $box_{T_{\mB(S)}}$ is easy to write in MS
  logic.\qed

  The constructed structure is, up to isomorphism, independent on
  $\preccurlyeq $, by the unicity result of Lemma \ref{L:3.1}. This
  proposition has a corollary fully similar to Corollary \ref{C:2.1}
  of Proposition \ref{P:2.1}.

\begin{defi}\label{D:3.2}
  \textit{Partitive families of bipartitions.} Let $V$ be a nonempty
  set. A family $\mB$ of bipartitions of $V$ is \emph{weakly
  partitive} if it satisfies the following property:
\begin{itemize}
\item[B2:] For every two overlapping elements $P$ and $Q$ of $\mB$, we
  have $\{A\cap B,A'\cup B'\}$ $\in\mB$, whenever $\{A,A'\}=P$ and
  $\{B,B'\}=Q.$
\end{itemize}
  It is \emph{partitive} if, in addition, it satisfies the following property:
\begin{itemize}
\item[B3:] For every two overlapping elements $P$ and $Q$ of $\mB$, we
  have $\{A\Delta B,A\Delta B'\}$ $\in\mB$, whenever $\{A,A'\}=P$ and
  $\{B,B'\}=Q$.
\end{itemize}
  Note that in B3, we have  $A\Delta B'=A'\Delta B$.   

  The bipartitions of $\mB$ will be called the $\mB$-\emph{splits} of
  $V$ (or of the structure $S$, if $\mB=\mB(S)$).\ Those which do not
  overlap any other bipartition of $\mB$ are called the \emph{good}
  $\mB$-splits. (We keep our terminology close to that of
  \cite{CuEd80} and \cite{cunningham82} which are the fundamental
  articles for these notions). If $\mB$ is weakly partitive, the
  family $\Good(\mB)$ of good $\mB$-splits is nonempty: let $A$ be a
  minimal $\mB$-block among those containing an element $v$; if
  $\{A,V-A\}$ overlaps $\{B,C\}$ where $B$ contains $v$, then, by B2,
  $\{A\cap B,(V-A)\cup C\}$ $\in\mB$, and $A$ is not a minimal block
  containing $v$; hence $\{A,V-A\}$ is a good $\mathcal{B-}$split.
  Clearly, $\Good(\mB)$ satisfies B1.  The corresponding unrooted tree
  is $T_{\Good(\mB)}$. If we transform a family $\mB$ into $\mB^{+}$ so
  as to insure Property B0, then $\mB^{+}$ is weakly partitive or
  partitive if $\mB$ is weakly partitive or partitive respectively.
\end{defi}

  If $\{A,B\}$ is a split of a structure $S$, we consider $S$ as a
  composition of the smaller induced substructures $S[A]$ and
  $S[B]$. By iterating the splitting, one reaches a decomposition of
  $S$ into unsplittable pieces. The objective is to obtain in this way
  a canonical decomposition.

  As in Theorem \ref{T:2.1}, the conditions of partitivity and weak
  partitivity on a family $\mB$ imply some particular structure
  associated with the nodes of $T_{\Good(\mathcal{B)}}$, and we also
  express this structural property for the tree
  $T_{\Good(\mB^{+}\mathcal{)}}$. We recall that the leaves of this
  tree are the singletons $\{v\}$ for $v$ in $V$.  If $N$ and $M$ are
  adjacent nodes of $T_{\Good(\mB^{+}\mathcal{)}}$, we also recall that
  $T_{\Good(\mB^{+})}(N,M)$ denote the set of nodes of
  $T_{\Good(\mB^{+})}$ (including $N$) that are reachable from $N$ by a
  path not containing $M$.  For $\mB$ defined by the context, we
  denote by $V(N,M)$ the set of elements of $V$ at the leaves
  belonging to $T_{\Good(\mB^{+})}(N,M)$.

\begin{thm}\label{T:3.1} \emph{(\cite{CuEd80}, \cite{montgolfier03})}
  Let $\mB\subseteq\mB(V)$ be partitive.
%\begin{itemize}
%\item[1)]

  \noindent\emph{(1)}\
  Every internal node $N$ of the tree $T_{\mB^{+}}$ satisfies
  one and only one of the following two properties:
\begin{itemize}
\item[S1:] $N$ has $k$ neighbours, $N_{1},\dots,N_{k},k\geq3$, and for
  every nonempty proper subset $I$ of $\{1,\dots,k\}$, the pair
\[\qquad\qquad\mB(N,I):=\Bigl\{\,\bigcup\{V(N_{i},N)\mid i\in
I\},\bigcup\{V(N_{i},N)\mid i\in\{1,\dots,k\}-I\}\,\Bigr\}
\]
  belongs to $\mB$.
\item[S2:]
  \noindent\emph{(1)}\
  $N$ has $k$ neighbours, $N_{1},\dots,N_{k},k\geq3$, and for
  every subset $I$ of $\{1,\dots,k\}$, the pair $\mB(N,I)$ (as defined
  above) belongs to $\mB$ if and only if $I$ or $\{1,\dots,k\}-I$ is
  singleton.
\end{itemize}

%\item[2)]
  \noindent\emph{(2)}\
  If a $\mB$-split is not good, it is of the form $\mB(N,I)$
  for some node $N$ satisfying T1 and a non singleton set $I\subset
  \{1,\dots,k\}$.
%\end{itemize}

  Let $\mB\subseteq\mB(V)$ be weakly partitive.
%\begin{itemize}
%\item[3)]

  \noindent\emph{(3)}\ Every internal node $N$ of the tree
  $T_{\Good(\mB^{+})}$ satisfies one and only one of properties
  \emph{S1}, \emph{S2} or the following
\begin{itemize}
\item[S3:] $N$ has at least 3 neighbours that can be numbered as
  $N_{1},\dots,N_{k}$ in such a way that for every subset $I$ of
  $\{1,\dots,k\}$, the pair $\mB(N,I)$ belongs to $\mB$ if and only if
  $I$ is an interval $[m,n]$ or its complement for some $m,n$ with
  $1\leq m\leq n\leq k$ (and $\{1,\dots,k\}\neq\lbrack m,n]$).
\end{itemize}

%\item[4)]

  \noindent\emph{(4)}\
  If a $\mB$-split is not good, it is of the form $\mB(N,I)$
  for some node $N$ satisfying \emph{S1} or \emph{S3} (with $m<n$).\qed
%\end{itemize}
\end{thm}

  The nodes satisfying S1, S2, S3 are said to be, respectively,
  \emph{complete}, \emph{prime}, and \emph{circular}. Montgolfier's
  dissertation \cite{montgolfier03} reviews several applications from
  \cite{CuEd80}, together with other ones that we do not discuss here.

\subsection{The Tutte decomposition of 2-connected graphs}\label{SS:3.2}

  We review briefly the Tutte decomposition of 2-connected graphs used
  in \cite{CuEd80} to introduce the theory of graph
  decomposition. This notion does not depend on edge directions, hence
  graphs will be \emph{undirected} in this section.  They are
  loop-free, without isolated vertices, they may have multiple
  edges. The notation and definitions of Subsection \ref{SS:2.3} for
  2-graphs are used here with obvious adaptations to undirected
  graphs.

\begin{defi}\label{D:3.3}
  \emph{2-separations.}  A \emph{2-separation} of a graph $G$ is a
  bipartition $\{A,B\}$ of its set of edges $E_{G}$ such that $A$ and
  $B$ have at least two elements and there are exactly two vertices,
  $u$ and $v$, which are incident with edges from both blocks of the
  bipartition. (For example $\{\{a,b,c,f\},\{d,e,g,h,k,m,n$, $p\}\}$
  is a 2-separation of the graph $G$ of Figure 1.) Hence,
  $G=(G_{uv}[A]//G_{uv}[B])^{0}$, where
  $s_{1}(G_{uv}[A])=s_{1}(G_{uv}[B])=u$,
  $s_{2}(G_{uv}[A])=s_{2}(G_{uv}[B])=v$, $G_{uv}[A]^{0}=G[A]$ and
  similarly for $B$. For the purpose of iterating the decomposition
  process, it is convenient to consider that the two graphs resulting
  from this decomposition step are $G^{+}[A]$ and $G^{+}[B]$, obtained
  from $G[A]$ and $G[B]$ by the addition of a new undirected edge
  $u-v$, labelled in a particular way, and called a \emph{marker}. The
  graphs $G^{+}[A]$ and $G^{+}[B]$ have in common the marker edge, its
  two ends and nothing else. They have no distinguished vertices. The
  decomposition process is applied to them recursively.
\end{defi}

  If $G$ is 2-connected, then $G[A]$ and $G[B]$ are connected, and
  furthermore $G^{+}[A]$ and $G^{+}[B]$ are 2-connected
  (\cite{CuEd80}, Lemma 1). A graph without any 2-separation is
  3-connected. We denote by $2\mS(G)$ the set of 2-separations of a
  graph $G$.

  A \emph{decomposition of a graph H} is a set of graphs (called the
  \emph{components} of the decomposition) which is either $\{H\}$ or
  the set obtained from a decomposition by replacing one of its
  components, say $G$,\ by $G^{+}[A]$ and $G^{+}[B]$ defined from a
  2-separation $\{A,B\}$ of $G$. This process is applied recursively
  to a 2-connected graph $H$ and each component of a decomposition is
  2-connected. The graphs in a decomposition are not disjoint, they
  form a single connected graph. If we delete from this graph the
  marker edges, we obtain $H$. To every decomposition corresponds a
  tree, the nodes of which are the components of the decomposition.
  Two nodes are adjacent if they share a marker edge.

  It is proved in \cite{CuEd80} that the family $2\mS(G)$ for a
  2-connected graph $G$ is weakly partitive. If we decompose a
  2-connected graph by using only good bipartitions at each step, we
  obtain at the end a canonical (unique up to isomorphism)
  decomposition (Theorem 1 of \cite{CuEd80}). This canonical
  decomposition is the one defined by Tutte and proved unique in
  \cite{tutte66}, chapter 11: every 2-connected graph has a unique
  decomposition in terms of \emph{bonds} (i.e., graphs consisting of
  several parallel edges between two vertices), cycles and 3-connected
  graphs such that no two bonds and no two cycles share an edge.

  Proposition \ref{P:3.1} is applicable and shows that the tree of the
  Tutte decomposition can be constructed by an MS transduction using
  an auxiliary order on the set of edges.  However, another
  construction, not using any linear order is given in
  \cite{courcelle99}, Theorem 4.7.  It uses the detour through 2-dags,
  as in the proof of Lemma \ref{L:2.3}.  We do not discuss any longer
  this construction.

\textbf{Question}: In the case of a simple graph $G$, one might hope, by
  using Corollary \ref{C:2.2} to be able to construct the structure
  $\<V_{G}\cup N_{T},\edg_{G},\edg_{T},box_{T}\>$ from
  $\<V_{G},\edg_{G}\>$ by an MS transduction.  However this is not
  immediate from the above results because the proof of this corollary
  is valid for directed graphs, and in order to orient the edges of a
  graph, edge set quantifications are necessary (see
  \cite{courcelle95}). However, an alternative construction might be
  possible, giving a statement analogous to Corollary \ref{C:2.2}. We
  leave this as an open question.

\section{The split decomposition}\label{S:4}

  In this section we apply the results of Section \ref{S:3} to the
  \emph{split decomposition} of graphs defined by Cunnigham in
  \cite{cunningham82} and used as a preliminary step in several
  algorithms: for the polynomial time recognition of circle graphs in
  \cite{bouchet87}, for the recognition of \emph{parity graphs} in
  \cite{CiSt99} and for the construction of \emph{distance labellings}
  in \cite{GaPa03}.

\subsection{Splitting a graph}\label{SS:4.1}

  As in Subsection \ref{SS:2.2}, graphs are simple, directed and
  loop-free. The split decomposition will be applied to connected
  graphs.  Hence, most definitions are restricted to connected graphs,
  which permits to avoid some technical difficulties. A directed graph
  is \emph{strongly connected} if for any two vertices $u,v$, there
  are directed paths from $u$ to $v$ and $v$ to $u$. An undirected
  graph is connected if and only if it is strongly connected.

\begin{defi}\label{D:4.1}
  \emph{Splitting a graph.} A \emph{split} of a connected graph $G$ is
  a bipartition $\{A,B\}$ of $V_{G}$ such that $E_{G}=E_{G[A]}\cup
  E_{G[B]}\cup(A_{1}\times B_{1})\cup(B_{2}\times A_{2})$ for some
  $A_{i}\subseteq A$, $B_{i}\subseteq B$, and each of $A$ and $B$ has
  at least 2 elements. If $\{A,B\}$ is a split, then $G$ can be
  expressed as the union of $G[A]$ and $G[B]$ linked by one or two
  directed, complete bipartite graphs. (Since $G$ is connected the set
  $(A_{1}\times B_{1})\cup(B_{2}\times A_{2})$ is not empty).

  The inverse of splitting is the \emph{join operation}, defined as
  follows.  Let $H$ and $K$ be two disjoint graphs with distinguished
  vertices $h$ in $H$ and $k$ in $K$. We define $H\boxtimes_{(h,k)}K$
  as the graph with set of vertices $V_{H}\cup V_{K}-\{h,k\}$ and
  edges $x\longrightarrow y$ such that, either $x\longrightarrow y$ is
  an edge of $H$, or an edge of $K$, or we have $x\longrightarrow h$
  in $H$ and $k\longrightarrow y$ in $K$, or we have $h\longrightarrow
  y$ in $H$ and $x\longrightarrow k$ in $K$. The subscript $(h,k)$ in
  $\boxtimes_{(h,k)}$ will be omitted whenever possible.

  If $\{A,B\}$ is a split, then $G=H\boxtimes_{(h,k)}K$ where $H$ is
  $G[A]$ \ augmented with a new vertex $h$ and edges $x\longrightarrow
  h$ whenever there are in $G$ edges from $x$ to some $u$ in $B$, and
  edges $h\longrightarrow x$ whenever there are edges from some $u$ in
  $B$ to $x$.\ The graph $K$ is defined similarly from $G[B]$, with a
  new vertex $k$.  These new vertices are called \emph{markers} in
  \cite{cunningham82}. We say that $h$ and $k$ are \emph{neighbours}
  if they are created from a same split. Note that the graphs $H$ and
  $K$ have at least 3 vertices and strictly less vertices than $G$.

  A technical variant (used in \cite{cunningham82}) consists in
  letting $h=k$. In this case the graphs $H$ and $K$ have in common
  the marker vertex $h$ and nothing else. We write in this case
  $G=H\boxtimes_{(h,h)}K$.  The advantage is that $H\cup K$ is a
  single connected graph.  However, the marker must be identified in
  some way.  But when one iterates the decomposition process, it is
  easier to handle of the components of the decomposition as disjoint
  graphs.
\end{defi}

\begin{defi}\label{D4.2}
  \emph{Decompositions.}  A \emph{decomposition} of a connected graph
  $G$ is defined inductively as follows: $\{G\}$ is the only
  decomposition of size 1; if $\{G_{1},\dots,G_{n}\}$ is a
  decomposition of size $n$ and $G_{i}=H\boxtimes_{(h,k)}K$, then
  $\{G_{1},\dots,G_{i-1},H,K,G_{i+1},\dots,G_{n}\}$ is a decomposition
  of $G$ of size $n+1$. The graphs $G_{i}$ are called the
  \emph{components} of the decomposition. The graph $G$ can be
  reconstructed without ambiguity provided the marker vertices and
  their matchings are specified.  We say that two components are
  \emph{neighbours} if they have neighbour marker vertices. From the
  inductive definition of decompositions, it is clear that the
  components of a decomposition form an unrooted tree for the
  neighbourhood relation.
\end{defi}

  It will be convenient to handle a decomposition
  $\mD=\{G_{1},\dots,G_{n}\}$ of a graph $G$ as a single graph
  $\Sdg(\mD)$ called a \emph{split decomposition graph}, an
  \emph{SD graph} in short. The components of $\mD$ being
  pairwise disjoint, we let $\Sdg(\mD)$ be their union together
  with particular edges, called $\varepsilon$-edges between any two
  neighbour marker vertices.  Every vertex of $G$ is a vertex of
  $\Sdg(\mD)$.  The graph $G$ can be reconstructed in a unique
  way from $\Sdg(\mD)$.  Two decompositions $\mD$ and
  $\mD'$ of a graph $G$ are \emph{isomorphic} if there exists
  an isomorphism of $\Sdg(\mD)$ onto $\Sdg(\mD')$ which
  is the identity on $V_{G}$.

  The objective is to construct for every connected graph a \emph{canonical
  decomposition} by iterated good splittings.
\begin{figure}
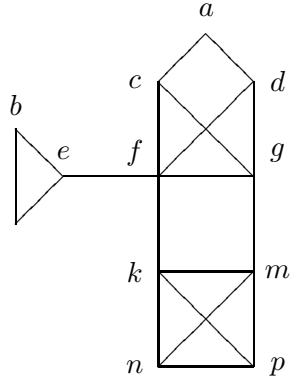
\label{F:splitdec1}
\[\xy;<18 pt,0 pt>:
  (0,5.5)*+{b};
  (1,4.5)*+{e};
  (2.5,4.5)*+{f};
  (2.5,6)*+{c};
  (2.5,2)*+{k};
  (2.5,0)*+{n};
  (5.5,4.5)*+{g};
  (5.5,6)*+{d};
  (5.5,2)*+{m};
  (5.5,0)*+{p};
  (4,7,5)*+{a};
  \ar@{-}( 0,5);( 0,3)
  \ar@{-}( 0,5);( 1,4)
  \ar@{-}( 0,3);( 1,4)
  \ar@{-}( 1,4);( 5,4)
  \ar@{-}( 3,2);( 5,2)
  \ar@{-}( 3,0);( 5,0)
  \ar@{-}( 3,0);( 3,6)
  \ar@{-}( 5,0);( 5,6)
  \ar@{-}( 3,6);( 4,7)
  \ar@{-}( 5,6);( 4,7)
  \ar@{-}( 3,6);( 5,4)
  \ar@{-}( 5,6);( 3,4)
  \ar@{-}( 3,2);( 5,0)
  \ar@{-}( 5,2);( 3,0)
\endxy
\]
\caption{{A graph $G$.}}
\end{figure}
  Figure 4 shows a graph $G$ and Figure 5 shows the graph representing
  its canonical split decomposition.  The dotted lines are the
  $\varepsilon$-edges.
\begin{figure}
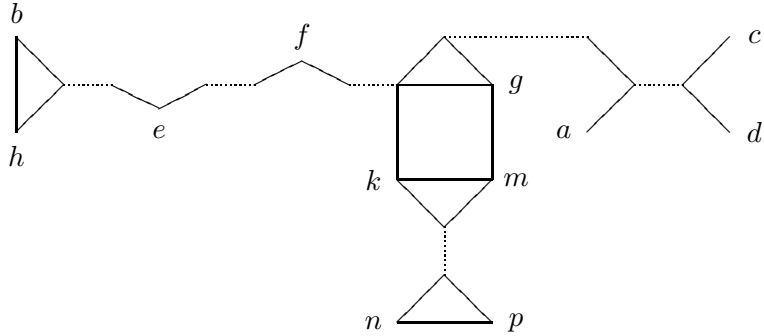
\label{F:splitdec2}
\[\xy;<18 pt,0 pt>:
  (   0,6,5)*+{b};
  (   0,3,5)*+{h};
  (   3,4)*+{e};
  (   6,6)*+{f};
  (10.5,5)*+{g};
  (11.5,4)*+{a};
  (15.5,4)*+{d};
  (15.5,6)*+{c};
  ( 7.5,3)*+{k};
  (10.5,3)*+{m};
  ( 7.5,0)*+{n};
  (10.5,0)*+{p};
  \ar@{-}( 0,6);( 0,4)
  \ar@{-}( 0,6);( 1,5)
  \ar@{-}( 0,4);( 1,5)
  \ar@{.}( 1,5);( 2,5)
  \ar@{-}( 2,5);( 3,4.5)
  \ar@{-}( 4,5);( 3,4.5)
  \ar@{.}( 4,5);( 5,5)
  \ar@{-}( 5,5);( 6,5.5)
  \ar@{-}( 7,5);( 6,5.5)
  \ar@{.}( 7,5);( 8,5)
  \ar@{-}( 8,5);( 9,6)
  \ar@{.}( 9,6);(12,6)
  \ar@{-}(12,6);(13,5)
  \ar@{-}(13,5);(12,4)
  \ar@{.}(13,5);(14,5)
  \ar@{-}(14,5);(15,6)
  \ar@{-}(14,5);(15,4)
  \ar@{-}( 9,6);(10,5)
  \ar@{-}(10,5);(10,3)
  \ar@{-}( 8,5);(10,5)
  \ar@{-}( 8,5);( 8,3)
  \ar@{-}( 8,3);(10,3)
  \ar@{-}( 8,3);( 9,2)
  \ar@{-}(10,3);( 9,2)
  \ar@{.}( 9,2);( 9,1)
  \ar@{-}( 9,1);( 8,0)
  \ar@{-}( 9,1);(10,0)
  \ar@{-}( 8,0);(10,0)
\endxy
\]
\caption{The split decomposition graph \protect{$\Sdg(\Split(G))$}.}
\end{figure}

  In the perspective of obtaining a canonical decomposition, we first
  observe that the graphs $H$ and $K$ associated with a split
  $\{A,B\}$ of a graph $G$ such that $G=H\boxtimes K$, $H$ contains
  $A$ and $K$ contains $B$ are not always uniquely defined. Consider
  $G$: $1\longrightarrow2\longrightarrow 3\longrightarrow4$,
  $A=\{1,2\}$, $B=\{3,4\}$.  One can take
  $H=1\longrightarrow2\longrightarrow h$,
  $K=k\longrightarrow3\longrightarrow4$, but one can also add an edge:
  $h\longrightarrow2$ to $H$, or an edge from $3\longrightarrow k$ to
  $K$ (but not both simultaneously), and we still have
  $G=H\boxtimes_{(h,k)}K$. However, $H$ and $K$ are uniquely defined
  in certain situations, as shows the following lemma.
\vfill\eject

\begin{lem}\label{L:4.1}\hfill
\begin{itemize}
\item[1)] Let $G$ be a strongly connected (resp. undirected and
  connected) graph and $\{A,B\}$ be a split.  There is a unique pair of
  graphs $(H,K)$ such that $H$ contains $A$, $K$ contains $B$ and
  $G=H\boxtimes K$, where unique is meant up to isomorphism.

\item[2)] Furthermore, the graphs $H$ and $K$ are strongly connected
  (resp.\ undirected and connected).  They are isomorphic to induced
  subgraphs of $G$ or to graphs obtained from induced subgraphs of $G$
  by fusing one vertex of indegree 0 and one vertex of outdegree 0.  If
  $G$ is undirected, only the former case occurs.
\end{itemize}
\end{lem}

\proof The verifications are easy.  For assertion 2), if $G$ is
  undirected then $H$ is isomorphic to $G[A\cup\{v\}]$ where $v$ is
  any vertex of $B$ adjacent to some vertex of $A$. Otherwise, $H$ is
  isomorphic to the graph obtained from $G[A\cup\{u,v\}]$ by the
  fusion of $u$ and $v$ where $u$ is any vertex of $B$ such that
  $x\longrightarrow u$ for some vertex $x$ of $A$, and $v$ is any
  vertex of $B$ such that $v\longrightarrow x$ for some vertex $x$ of
  $A.  $ It may happen that $u=v$.\qed

\medskip\noindent\textbf{Remark.}
%\begin{rem}\label{R:third}
  These assertions are not true for nonconnected graphs: the
  undirected graph $I_{4}$ (where $I_{n}$ has $n$ vertices and no
  edge) is equal to $I_{3}\boxtimes H$, where $H$ consist of one
  isolated vertex and an edge, one end of which is the marker vertex
  $h$.  If $G$ is strongly connected, the graphs $H$ and $K$ need not
  be induced subgraphs: consider
  $\overrightarrow{C}_{4}=\overrightarrow{C}_{3}\boxtimes
  \overrightarrow{C}_{3}$ where $\overrightarrow{C}_{n}$ denotes the
  directed circuit with $n$ vertices.
\medskip%\end{rem}

  The decomposition process must terminate because the components are
  getting smaller and smaller, and are thus at the end
  "unsplittable". A graph is \emph{prime} if it has at least 4
  vertices and no split. The graphs with at most 3 vertices have no
  split for the trivial reason that they have not enough vertices.
  They are not called prime. We give easy verifiable examples of
  prime and splittable graphs. For further reference, we put them into
  a lemma.

\begin{lem}\label{L:4.2}\hfill
\begin{itemize}
\item[1)] A prime graph is 2-connected.
\item[2)] There is no prime undirected graph with 4 vertices.
\item[3)] For each $n\geq5$, the graph $C_{n}$ is prime, and the
  graphs $P_{n},K_{n},S_{n-1},L_{n}$, all with $n$ vertices, are not.\qed
\end{itemize}
\end{lem}

  As usual, we denote by $K_{n}$ the \emph{n-clique,} i.e., the
  complete undirected graph with $n$ vertices, by $S_{n}$ the
  $n$-\emph{star} consisting of one vertex, the \emph{center},
  adjacent to $n$ vertices by undirected edges, by $P_{n}$ the
  undirected path with $n-1$ edges and $n$ vertices, by $L_{n}$ the
  transitive (acyclic) tournament on $n$ vertices (the directed graph
  of a strict linear order), by $C_{n}$ the undirected cycle with $n$
  vertices. The graphs $K_{n},S_{n-1}$ for $n\geq4$ are "highly
  decomposable", or \emph{brittle} in the terminology of
  \cite{CuEd80,cunningham82}: every bipartition, each block of which
  has at least 2 elements is a split. They are the only undirected
  graphs with this property. The highly decomposable directed graphs
  have a more complex structure that we will review later.

  The 2-connected undirected graphs having 4 vertices are
  $K_{4},C_{4}$, and $K_{4}^{-}$ (i.e., $K_{4}$ minus one edge).  None
  of them is prime.  The directed, 2-connected graph with 4 vertices
  defined as the union of the paths $a\longrightarrow b\longrightarrow
  c\longrightarrow d$ and $a\longrightarrow d\longrightarrow c$ is
  prime, as one checks by trying the three possibilities to split it.

\begin{defi}\label{D:4.3}
  \emph{Canonical decompositions.}  A decomposition of a connected
  undirected graph $G$ is \emph{canonical} if and only if:
\begin{itemize}
\item[(1)] each component is either prime or is isomorphic to $K_{n}$
  or to $S_{n-1}$ for $n$ at least 3,
\item[(2)] no two clique components are neighbour,
\item[(3)] two neighbour vertices in star components are both centers
  or both not centers.
\end{itemize}
  If $G$ has one or two vertices, we define $\{G\}$ as its canonical
  decomposition.
\end{defi}

  Restrictions (2) and (3) can be justified as follows: if two clique
  components, isomorphic to $K_{n}$ and $K_{m}$ are neighbour they can
  be merged into a single one isomorphic to $K_{n+m-2}$, by using the
  elimination of $\varepsilon$-edges described below and $K_{n+m-2}$
  has several overlapping splits ($n+m-2\geq4$). Similarly, if two
  star components, isomorphic to $S_{n}$ and $S_{m}$ are neighbours,
  and the center of one is linked by an $\varepsilon$-edge to a
  non-center vertex of the other, they can be merged into a single
  star isomorphic to $S_{n+m-1}$, $n+m-1\geq3$,\ and $S_{n+m-1}$ has
  several overlapping splits. It is thus necessary to assume (2) and
  (3) in order to obtain a \emph{unique decomposition theorem} because
  stars and cliques have several overlapping, hence "incompatible"
  splits. Note that the connected undirected graphs with 3 vertices
  are $K_{3}$ and $S_{2}$, hence are among the possible types of
  nonprime components.

  As in Section \ref{S:3}, a split is \emph{good} if it does not
  overlap other splits.  Starting from a graph $G$ and the
  decomposition $\{G\}$, one can refine it by iteratively splitting
  its components \emph{with respect to good splits only}.  Since a
  graph breaks into two strictly smaller graphs, one reaches a
  decomposition that cannot be refined by any split. Since one only
  applies good splits, one cannot generate neighbour components that
  are cliques, or that are stars with a center neighbour to a
  non-center vertex. It is thus canonical.

  The following theorem concerns \emph{connected graphs}.  By using
  the obvious decomposition of a graph into connected components, we
  get thus a canonical decomposition for every undirected graph.  The
  isomorphism of decompositions is defined in Definition 4.2.

\begin{thm}\label{T:4.1}\cite[Theorem 3]{cunningham82} 
  A connected undirected graph has a canonical decomposition.  It is
  unique up to isomorphism.  It can be obtained by iterated splitting
  relative to good splits.\qed
\end{thm}

  For directed graphs, there exists a similar notion of canonical
  decomposition, for which one needs another notion of "highly
  decomposable" graph, called a \emph{circle of transitive
  tournaments, }a CTT in short. A \emph{CTT} is a graph with $n\geq3$
  vertices $v_{0},\dots,v_{n-1}$, such that its edges are described in
  terms of a sequence of integers
  $0=p_{1}<p_{2}<\dots<p_{k}<p_{k+1}=n$ as the pairs
  $v_{i}\longrightarrow v_{j}$ such that $p_{m}\leq i<j\leq p_{m+1}$
  for some $m,1\leq m\leq k$. (We let $v_{n}=v_{0}$.) In the special
  case $k=1$, the loop $v_{0}\longrightarrow v_{0}$ is excluded. The
  vertices $v_{p_{1}},\dots,v_{p_{k}}$ are called the
  \emph{hinges}. We write that this graph is a \emph{k-CTT} to specify
  the number $k$ of hinges.

  A CTT is strongly connected and is not undirected. Each of its
  splits has a block of the form $\{v_{i},\dots,v_{j}\}$ for some
  $i,j$ with $0\leq i<j\leq n-1$.

  Here are some examples: For $k=n$, one gets a directed circuit. For
  $k=2,n=4$, $p_{1}=0,p_{2}=2,p_{3}=4$ one gets the graph
  $0\longrightarrow 1\longrightarrow2\longrightarrow3\longrightarrow0$
  with additional edges $0\longrightarrow2$ and
  $2\longrightarrow0$. For $k=1,n=3$, one gets the graph
  $0\longrightarrow1\longrightarrow2\longrightarrow0$ with additional
  edges $1\longrightarrow0$ and $0\longrightarrow2$. A 1-CTT with $n$
  vertices has all its vertices of degree $n$, except the hinge which
  has\ degree $2n-2$. (Since graphs are defined as directed, a vertex
  in a loop-free directed graph with $n$ vertices has maximum degree
  $2n-2$).

  A decomposition of a strongly connected graph $G$ is
  \emph{canonical} if and only if:
\begin{itemize}
\item[(1)] each component is either prime, or is isomorphic to $K_{n}$
  or to $S_{n-1}$ for $n$ at least 3, or is a CTT,

\item[(2)] and (3) hold as for undirected graphs,

\item[(4)] if two neighbour components are CTTs and each of them has
  at least two hinges, then the neighbour vertices are not two hinges.
\end{itemize}

  If $G$ has one or two vertices, we define $\{G\}$ as its canonical
  decomposition.

  To justify the roles of cliques and stars, we recall that an
  undirected edge is defined as a pair of opposite directed edges. If
  two neighbour components in a decomposition are CTTs with
  respectively $k$ and $m$ vertices and $k'+1$ and $m'+1$ hinges, and
  two hinges are linked by an $\varepsilon$-edge $e$, then they can be
  merged (by what we will call in the next subsection the
  \emph{elimination} of $e$) into a single $(k^{\prime }+m')$-CTT
  with $k+m-2$ vertices.  This is shown on Figures 6 and 7 : two
  3-CTTs are merged into a single 4-CTT. In all other cases where two
  CTTs are neighbour, the elimination of the $\varepsilon$-edge
  linking them does not yield a CTT, a star or a clique.

\begin{figure}
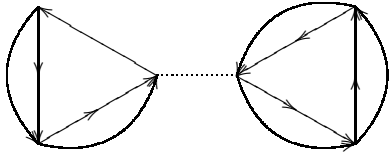
\label{F:ctt1}
\[\xy;<30 pt,0 pt>:
  a(  0)="A",
  a(120)="B",
  a(240)="C",
  (-.5,0)="BC",
  {"C";"A":0;a(300),x}="CA",
  (3,0)="Q",
  "Q"+a(180)="a",
  "Q"+a(300)="b",
  "A"+(2.5,\halfrootthree)="c",
  "CA"+(2.5,\halfrootthree)="ca",
  "C"+(2.5,\halfrootthree)="a",
  {"a";"b":"Q";a(180)+"Q",x}="ab",
  (3.5,0)="bc",
  \ar "A";"B"
  \ar "B";"BC"
  \ar "BC";"C"
  \ar "C";"CA"
  \ar "CA";"A"
  \ar@/_12 pt/"B";"C"
  \ar@/_12 pt/"C";"A"
  \ar@{.}"A";"a"
  \ar "a";"ab"
  \ar "ab";"b"
  \ar "b";"bc"
  \ar "bc";"c"
  \ar "c";"ca"
  \ar "ca";"a"
  \ar@/_12 pt/"c";"a"
  \ar@/_12 pt/"a";"b"
  \ar@/_12 pt/"b";"c"
\endxy
\]
\caption{Two 3-CTT's linked by an $\protect\varepsilon$-edge.}
\end{figure}

\begin{figure}
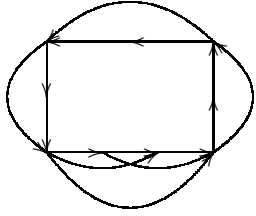
\label{F:ctt2}
\[\xy;<21 pt,0 pt>:
  \ar (3,2);(1.5,2)
  \ar (1.5,2);(0,2)
  \ar (0,2);(0,1)
  \ar (0,1);(0,0)
  \ar (0,0);(1,0)
  \ar (1,0);(2,0)
  \ar (2,0);(3,0)
  \ar (3,0);(3,1)
  \ar (3,1);(3,2)
  \ar@/_15 pt/(3,2);(0,2)
  \ar@/_15 pt/(0,2);(0,0)
  \ar@/_21 pt/(0,0);(3,0)
  \ar@/_15 pt/(3,0);(3,2)
  \ar@/_6 pt/(0,0);(2,0)
  \ar@/_6 pt/(1,0);(3,0)
\endxy
\]
\caption{Two CTT's merged into a 4-CTT.}
\end{figure}

\begin{thm}\label{T:4.2}\cite[Theorem 2]{cunningham82}
  A strongly connected graph has a canonical decomposition.  It is
  unique up to isomorphism.  It can be obtained by iterated splitting
  relative to good splits.\qed
\end{thm}

  The \emph{split decomposition} of a strongly connected graph $G$
  (which includes the case of a connected undirected graph) denoted by
  $\Split(G)$ is the canonical decomposition of Theorems \ref{T:4.1}
  and \ref{T:4.2}. In the next subsections we define its representation by a
  graph, and its construction by an MS\ transduction.

\subsection{Graph representations of decompositions}\label{SS:4.2}

  We have defined a single graph $\Sdg(\mD)$ linking all components of
  a decomposition $\mD$. We obtain in this way a binary relational
  structure on a fixed finite signature, actually an edge-labelled
  graph, from which the decomposed graph can be reconstructed by an MS
  transduction, as we will see. We get something similar to the
  \emph{graph representation of modular decompositions} defined in
  Section \ref{S:2}.

\begin{defi}\label{D:4.4}
  \emph{Split decomposition graphs.}  A \emph{split decomposition
  graph}, (an \emph{SD graph} in short) is a graph $H$ with two types
  of edges, defined as a triple
  $\<V_{H},\edg_{H},\varepsilon$-$\edg_{H}\>$ satisfying the following
  conditions
\begin{itemize}
\item[(i)] the \emph{solid edges} are represented by a binary relation
  $\edg_{H}$, and the (undirected) $\varepsilon$-\emph{edges} are
  represented by a symmetric binary relation $\varepsilon$-$\edg_{H}$;
\item[(ii)] every vertex is incident to a solid edge;
\item[(iii)] no two $\varepsilon$-edges are adjacent;
\item[(iv)] the graph obtained from $H$ by contracting the solid edges
  is an undirected tree.
\end{itemize}
\end{defi}

  Condition (iv) implies that a cycle in $H$ can only consist of solid
  edges, and that $H$ is connected.  The connected components of the
  subgraph $H[E_{H}^{sol}]$ where $E_{H}^{sol}$ is the set of solid
  edges, are called the \emph{components} of $H$.  They are not
  isolated vertices and they are linked to one another by
  $\varepsilon$-edges, in the global shape of a tree.

\begin{lem}\label{L:4.3}
  The graph $\Sdg(\mD)$ associated with a decomposition $\mD$ of a
  connected graph $G$ is an SD graph. Its vertices incident to no
  $\varepsilon$-edge are the vertices of $G$.\qed
\end{lem}

  We now explain how a graph can be reconstructed from the SD graph
  $\Sdg(\mD)$ representing one of its decompositions.

\begin{defi}\label{D:4.5}
  \emph{Evaluating SD graphs} If $e$ is an $\varepsilon$-edge $u-v$ of
  an SD graph $H$, we define an SD\ graph $H'=\emph{Elim}_{e}(H)$ as
  follows:
\begin{itemize}
\item[\(a] $V_{H'}=V_{H} -\{u,v\}$,
\item[\(b] the edges of $H'$ are those of $H$ not incident to $u$ or
  $v$, and the edges $x\longrightarrow y$ if $x\longrightarrow
  u-v\longrightarrow y$ or $x\longrightarrow v-u\longrightarrow y$ in
  $H$.  (The edges $x\longrightarrow u$, $v\longrightarrow y$,
  $x\longrightarrow v$\ and $u\longrightarrow y$ are necessarily solid
  edges).
\end{itemize}
  We will say that this operation \emph{eliminates} the edge $e$.
\end{defi}

\begin{lem}\label{L:4.4}
  If $H$ and $K$ are two disjoint graphs and $G$ is their union with
  an $\varepsilon$-edge linking $h$ in $H$ and $k$ in $K$, then
  $Elim_{e}(G)=H \boxtimes_{(h,k)}K$. If $e$ and $f$ are two
  $\varepsilon$-edges of an SD graph $H$, we have
  $\Elim_f(\Elim_e(H))=\Elim_e(\Elim_f(H))$.
\end{lem}

  Hence one can eliminate simultaneously (or in any order) the
  $\varepsilon$-edges of a given set. We let $\Eval(H)$ be obtained
  by eliminating all $\varepsilon$-edges of an SD-graph $H$.  We use
  the notation $\Eval$ because we consider this mapping as the
  evaluation of a kind of algebraic expression, the operations of
  which are defined by the components of $H$.

\begin{lem}\label{L:4.5}
  For an SD graph $H$, the graph $G=\Eval(H)$ can be defined as
  follows
\begin{itemize}
\item[(a')] $V_{G}$ is the set of vertices of $H$ incident to no
  $\varepsilon$-edge,
\item[(b')] the edges of $G$ are the solid edges of $H$ not adjacent
  to any $\varepsilon$-edge and the edges $x\longrightarrow y$ such that
  there is in $H$ a path
\[x\longrightarrow u_{1}-v_{1}\longrightarrow
u_{2}-v_{2}\longrightarrow\dots\longrightarrow
u_{k}-v_{k}\longrightarrow y
\]
  where the edges $u_{i}-v_{i}$ are $\varepsilon$-edges and alternate
  with solid edges.\qed
\end{itemize}
\end{lem}

\begin{exa}\label{E:two}
  The following graph $H$ is an SD graph:
\[a\longrightarrow b\longrightarrow u-v\longrightarrow c\longleftarrow
u'-v'\longleftarrow d\longleftarrow u"-v"\longrightarrow e
\]
  and $\Eval(H)$ is the non connected graph: $a\longrightarrow
  b\longrightarrow c\longleftarrow d$ $ e$. This example shows that
  not every SD graph is associated with a decomposition of a connected
  graph.
\end{exa}

\begin{prop}\label{P:4.1}
  If $\mD$ is a decomposition of a connected graph $G$, then
  $\Eval(\Sdg(\mD))$ $=G$.
\end{prop}

\proof By induction on the size $k$ of $\mD$.  If $k=1$, then we have
  $\Sdg(\mD)=G$. For the induction step, we let
  $\mD'=\{G_{1},\dots,G_{k}\}$ be a decomposition with corresponding
  graph $\Sdg(\mD')$ such that $\Eval(\Sdg(\mD'))=G$. We prove the
  assertion for $\mD=\{G_{1},\dots,G_{k-1},M,M'\}$, obtained by
  splitting one component, say $G_{k}$ without loss of generality.
  The graph $\Sdg(\mD)$ is obtained from $\Sdg(\mD')$ by the
  replacement of the subgraph $G_{k}$ by the union of $M$ and $M'$
  linked by an $\varepsilon$-edge, say $e$, and
  $\Elim_{e}(\Sdg(\mD))=\Sdg(\mD')$.  We have
  $\Eval(\Sdg(\mD))=\Eval(\Elim_{e}(\Sdg(\mD)))=\Eval(\Sdg(\mD'))$. Since
  $\Eval(\Sdg(\mD'))=G$ by the induction hypothesis, we obtain
  $\Eval(\Sdg(\mD))=G$.\qed

  The notion of \emph{clique-width} of a directed or undirected graph
  $G$, denoted by $\cwd(G)$, and a few results about it, are recalled
  in Appendix 2.  It is defined for graphs with labelled edges, hence
  is applicable to SD\ graphs.

\begin{prop}\label{P:4.2}
  The mapping $\Eval$ from SD graphs to graphs is an MS transduction.
  There exists a function $f$ such that $\Eval(H)$ has clique-width
  $\leq f(k)$ if each component of an SD graph $H$ has clique-width
  $\leq k$.
\end{prop}

\proof That the mapping $\Eval$ is an MS transduction is clear from
  its definition and the fact that the transitive closure of an MS
  definable binary relation is MS definable.

  For the second assertion, we use the fact for every MS transduction
  $\tau$, there exists a function $f$ such that $\cwd(G')\leq
  f(\cwd(G))$ whenever $G'$ is obtained from $G$ by $\tau$. (See Lemma
  A.2.2).  Hence assuming that each connected component of
  $H[E_{H}^{sol}]$ has clique-width $\leq k$, it is enough to prove
  that $\cwd(H)\leq k+2$.

  We need a few technical facts about the algebraic expressions
  defining clique-width.  We recall here that if $C$ is a set of $k$
  labels, a $C$-expression defining a graph $G$ witnesses that $G$ has
  clique-width at most $k$ (full definitions in Appendix 2). Let $C$
  and $D$ be disjoint sets of labels. Let $M$ be a graph with pairwise
  distinct vertices $v_{1},\dots,v_{m}$.  Let $N_{1},\dots,N_{m}$ be
  pairwise disjoint graphs such that $N_{i}$ has in common with $M$
  the single vertex $v_{i}$. We assume that $M$ is defined by a
  $C$-expression, and that each $N_{i}$ is defined by a $(C\cup
  D)$-expression, its vertices are labelled in $D$, and $v_{i}$ has a
  label $r_{i}$ that is different from those of the other vertices of
  $N_{i} $.

\medskip\noindent\textbf{Claim 1}: The graph $L=M\cup
  N_{1}\cup\dots\cup N_{m}$ can be defined by a $(C\cup D)$-expression.

\medskip\noindent\emph{Proof of the claim.}
  Let $E$ be an expression defining $M$. It has occurrences of
  constants $\mathbf{p}_{1},\dots.,\mathbf{p}_{m}$ which define
  respectively the vertices $v_{1},\dots,v_{m}$. Let
  $F_{1},\dots,F_{m}$ be $(C\cup D)$-expressions defining respectively
  $N_{1},\dots,N_{m}$. The expressions $F_{i}^{\prime
  }=\ren_{r_{i}\rightarrow p_{i}}(F_{i})$ define the graphs $N_{i} $
  with $v_{i}$ now labelled by $p_{i}$. The desired $(C\cup
  D)$-expression for $L$ is obtained by substituting in $E$ the
  expressions $F_{1}',..,F_{m}'$ for the occurrences of
  $\mathbf{p}_{1},\dots.,\mathbf{p}_{m}$ defining $v_{1},\dots,v_{m}$,
  giving an expression $E'$. Since $E$ does not contain operations
  involving labels in $D$, the substitution of the expressions
  $F_{1}',..,F_{m}'$ \ in $E$ does not result in edge creations
  between the vertices of the graphs $N_{i}$ other than $v_{i}$ and
  the vertices not in $N_{i}$. Hence, $E' $ is a $(C\cup
  D)$-expression defining $L$.\qed

  We now continue the proof of the proposition. Let $H$ be an SD graph
  with components of clique-width $\leq k$. We wish to prove that
  $\cwd(H)\leq k+2$.  The case where $H$ has a single component is
  obvious.

  We let $D =\{\top,\bot\}$, and $C$ be a set of $k$ labels. For every
  $\varepsilon$-edge $e:v-u$, we let $H_{v,e}$ be the subgraph of $H$
  consisting of $v,e$ and the connected component of $u$ in the graph
  $H$ minus the edge $e$.  (We recall that the $\varepsilon$-edges
  link the components of $H$ in the global shape of a tree). We label
  $v$ by $\top$ and all other vertices of $H_{v,e}$ by $\bot$.

\medskip\noindent\textbf{Claim 2}: Each graph $H_{v,e}$ labelled in
  this way is definable by a $(C\cup D)$-expression.

\medskip\noindent\emph{Proof of the claim.}
  The proof is by induction on the number of $\varepsilon$-edges of
  $H_{v,e}$. We let $M$ be the component of $H$ containing $u$, the
  other end of $e$. It is a subgraph of $H_{v,e}$. By the hypothesis,
  $M$ is defined by a $C$-expression. We let $v_{1},\dots,v_{n}$ be
  the other vertices of $M$ incident with $\varepsilon$-edges,
  respectively $e_{1},\dots,e_{n}$, which are the $\varepsilon$-edges
  linking $M$ at vertices $v_{1},\dots,v_{n}$ to other components of
  $H$. Using induction, we obtain that each graph $H_{v_{i},e_{i}}$ is
  definable by a $(C\cup D)$-expression.  We let $N$ be the edge $e$,
  with $v$ labelled by $\top$ and $u$ labelled by $\bot$. Claim 1 is
  applicable to the graph $L=M\cup H_{v_{1},e_{1}}\cup\dots\cup
  H_{v_{n},e_{n}}\cup N$, which is equal to $H_{v,e}$.  Hence
  $H_{v,e}$ is definable by a $(C\cup D)$-expression.

  This argument applies for $n=0$ which is the basis of the
  induction.\qed

  The graph $H$ is itself is expressible as $M\cup
  H_{v_{1},e_{1}}\cup\dots\cup H_{v_{n},e_{n}}$ where $M$ is any
  component, using the notation of Claim 2.  Its proof yields the
  desired result since the cardinality of $C\cup D$ is $k+2$. This
  completes the proof of the proposition.\qed

  We leave as an open question to determine a good bounding function $f$.

\begin{prop}\label{P:4.3}
  A set of strongly connected graphs has bounded clique-width if and
  only if the prime components of their split decompositions have
  bounded clique-width.
\end{prop}

\proof We first consider undirected graphs (for them strong
  connectedness is just connectedness). The "only if" direction is
  clear because the prime components of the split decomposition of an
  undirected graph are isomorphic to induced subgraphs of this graph,
  and clique-width is monotone with respect to induced subgraph
  inclusion. (See Lemma A.2.1).

  For the other direction, it suffices to apply Theorem \ref{T:4.1}
  and Proposition \ref{P:4.2} knowing that the cliques $K_{n}$ and the
  stars $S_{n}$ have clique-width 2.

  We now consider directed graphs.  We will use Theorem
  \ref{T:4.2}. For the "only if" direction, we note that, by Lemma
  \ref{L:4.1}, a prime component $M$ of the split decomposition of a
  strongly connected graph $G$ is either an induced subgraph of $G$ or
  is obtained from an induced subgraph $N$ by the fusion of a vertex
  of indegree 0 and a vertex of indegree 1. In this case, $\cwd(N)\leq
  k$ implies $\cwd(M)\leq4k$ by Lemma A.2.3.

  For the "if" direction, we argue as above, and it remains to prove
  that CTTs have bounded clique-width.  Actually they have
  clique-width at most 4. Let $G $ be a $k$-CTT with vertices
  $v_{0},\dots,v_{n-1}$ , $n\geq3$, and edges $v_{i}\longrightarrow
  v_{j}$ such that $p_{m}\leq i<j\leq p_{m+1}$ for some $m,1\leq m\leq
  k$, where $0=p_{1}<p_{2}<\dots<p_{k}<p_{k+1}=n$ and
  $v_{n}=v_{0}$. For every $i=0,\dots,n-1$, we let $G_{i}$ be the
  subgraph of $G$ defined as follows
\begin{itemize}
\item[\(a] its vertices are $v_{0},\dots,v_{i}$ ,
\item[\(b] its edges are those of $G$ of the form
  $v_{j}\longrightarrow v_{k}$ for $0\leq j<k\leq i$ (hence $G_{n-1}$ is
  $G$ minus the edges towards $v_{0}$);
\item[\(c] its vertices are labelled as follows: we label $v_{0}$ by
  1; letting $m $ be such that $p_{m}\leq i<p_{m+1}$, we label $v_{j}$
  by $\bot$ if $0<j<p_{m}$ and we label $v_{j}$ by $2$ if $p_{m}\leq
  j\leq i$.
\end{itemize}
  The graphs $G_{i}$ are defined by the following expressions:
\begin{itemize}
\item[-] $G_{1}=\add_{1,2}(\mathbf{1}\oplus\mathbf{2})$;
\item[-] if $2\leq i<p_{2}$, then
  $G_{i}=\ren_{3\rightarrow2}(\add_{2,3}(\add_{1,3}(G_{i-1}\oplus\mathbf{3})))$;
\item[-] if $p_{m}<i<p_{m+1}$ and $m\geq2$, then
  $G_{i}=\ren_{3\rightarrow2}(\add_{2,3}(G_{i-1}\oplus\mathbf{3}))$;
\item[-] if $i= p_{2}>1$, then
  $G_{i}=\ren_{3\rightarrow2}(\ren_{2\rightarrow\bot}
  (\add_{2,3}(\add_{1,3}(G_{i-1}\oplus\mathbf{3}))))$,
\end{itemize}
  and finally
\begin{itemize}
\item[-] if $i= p_{m}$ and $m>2$, then
  $G_{i}=\ren_{3\rightarrow2}(\ren_{2\rightarrow\bot}(\add_{2,3}(G_{i-1}%
  \oplus\mathbf{3})))$.
\end{itemize}
  Then $ G=\add_{2,1}(G_{n-1})$. This shows that $G$ can be constructed
  with the 4 labels $1,2,3,\bot$ hence has clique-width at most 4. If
  $G$ has a single hinge, then $n-1<p_{2}=n$ and labels 1,2,3 suffice.
  Thus 1-CTTs have clique-width at most 3.\qed

\medskip\noindent\textbf{Remark.}
%\begin{rem}\label{R:four}
  The clique-width of a graph may be strictly larger than the maximum
  clique-width of the components of its split decomposition. For an
  example the clique-width of $P_{4}$ is 3, $P_{4}=P_{3}\boxtimes
  P_{3}$ and the clique-width of $P_{3}$ is $2$. By contrast, the
  clique-width of a graph is the maximum clique-width of its prime
  components for the modular decomposition (by Lemma A.2.1).
\medskip%\end{rem}

  Another complexity measure for undirected graphs called
  \emph{rank-width} is defined by Oum and Seymour (see
  \cite{CoOu04,OuSe04}). It is equivalent to clique-width in the sense
  that the same sets of undirected graphs have bounded clique-width
  and bounded rank-width (because $\rwd(G)\leq
  \cwd(G)\leq2^{\rwd(G)+1}-1$ where $\rwd(G)$ denotes the rank-width of
  $G$). The rank-width of a graph is the maximal rank-width of its
  prime components for the split decomposition.

\subsection{Monadic Second-Order definition of the split decomposition}
\label{SS:4.3}

  The following theorem is actually the basis for Theorems \ref{T:4.1}
  and \ref{T:4.2}.

\begin{thm}\label{T:4.3}\cite[Theorem 9]{cunningham82}: The family of
  splits of a strongly connected graph is weakly partitive.  The
  family of splits of a connected undirected graph is partitive.\qed
\end{thm}

\medskip\noindent\textbf{Remark.}
%\begin{rem}\label{R:five}
  This result may not hold for a graph that is not strongly
  connected. Take for example: $1\longleftarrow2\longrightarrow
  3\longrightarrow4\longleftarrow5\longrightarrow6$ with additional
  edge $6\longrightarrow1$.  The two splits $\{\{1,2,3\},\{4,5,6\}\}$
  and $\{\{2,3,4\},\{5,6,1\}\}$ overlap but $\{\{2,3\},\{4,5,6,1\}\}$
  is not a split. Hence, the family of splits of this graph is not
  weakly partitive.
\medskip%\end{rem}

  We denote by $\mBS(G)$ the family of splits of a graph $G$, and by
  $\mBS_{g}(G)$ the family of good ones. The tree $T_{\mBS_{g}(G)}$
  (defined in Section \ref{S:3}) is the tree of the split
  decomposition $\Split(G)$. To simplify the notation, we will denote
  it by $T_{\mBS(G)}$, remembering that it is based on good splits.
  Proposition \ref{P:3.1} yields the following:

\begin{prop}\label{P:4.4}
  There exists an MS transduction that associates with a strongly
  connected graph $G$ and a linear ordering $\preccurlyeq$ of its set
  of vertices the structure 
\[\<V_{G}\cup
  N_{T_{\mBS(G)}},\edg_{G},\edg_{T_{\mBS(G)}},box_{T_{\mBS(G)}}\>
\]
  such that $T_{\mBS(G)}=\<N_{T_{\mBS(G)}},\edg_{T_{\mBS(G)}}\>$.\qed
\end{prop}

  From the tree $T_{\mBS(G)}$, we build an SD graph $H_{\mBS(G)} $ and
  we will prove that it represents $\Split(G)$, i.e. that $H_{\mBS(G)}
  =\Sdg(\Split(G)).$

\begin{defi}\label{D:4.6} 
  \emph{The SD graph $H_{\mBS(G)}$ constructed from $T_{\mBS(G)}$.}
  To avoid special cases, we assume that $G$ has at least 3 vertices.
  The tree-partition $(T_{\mBS(G)},\mV_{\mBS(G)})$ is defined by Lemma
  \ref{L:3.1} from the family $\mBS_{g}(G)$ (the set of good splits,
  which do not overlap any other). We let $N$ be the set of nodes of
  the unrooted tree $T_{\mBS(G)}$. An edge $e:x-y$ of $T_{\mBS(G)}$
  corresponds to a bipartition $\{P_{x},P_{y}\}\in\mBS_{g}(G)$.

  For each such edge, we create two new vertices $(e,x)$ and $(e,y)$:
  they will be the marker vertices of Definition \ref{D:4.1}. More
  precisely, the nodes of $T_{\mBS(G)}$ correspond to the
  components of the split decomposition, and the markers of the
  component at a node $x$ will be the vertices $(e,x)$ for all edges
  $e$ of $T$ incident with $x$.

  For a node $x\in N$ with neighbours $y_{_{1}},\dots,y_{k}$ we let
  $P_{y_{_{1}}},\dots,P_{y_{k}}$ be the sets associated with the edges
  $e_{1}:x-y_{_{1}},\dots,e_{k}:x-y_{k}$ (we use the notation of Lemma
  \ref{L:3.2}). They are pairwise disjoint. By this lemma,
  $V_{\mBS(G)}(x)=V_{G}-(P_{y_{_{1}}}\cup\dots \cup P_{y_{k}})$ (this
  set may be empty). We define a graph $H(x)$ as follows:
\begin{itemize}
\item[(i)] $V_{H(x)}=V_{\mBS(G)}(x)\cup\{(e_{i},x)\mid i=1,\dots,k\}$,
\item[(ii)] its edges are of several types:
%\begin{itemize}
\item[-] the edges $u\longrightarrow v$ in $G$, for $u,v\in V_{\mBS(G)}(x) $,
\item[-] the edges $u\longrightarrow(e_{i},x)$ if $u\in
  V_{\mBS(G)}(x)$ and there is in $G$ an edge $u\longrightarrow v$ for
  some $v$ in $P_{y_{i}}$,
\item[-] the edges $u\longleftarrow(e_{i},x)$ if $u\in V_{\mBS(G)}(x)$
  and there is in $G$ an edge $u\longleftarrow v$ for some $v$ in
  $P_{y_{i}}$,
\item[-] the edges $(e_{i},x)$ $\longrightarrow(e_{j},x),i\neq j$ if
  there is in $G$ an edge $u\longrightarrow v$ for some $u$ in
  $P_{y_{i}}$ and some $v$ in $P_{y_{j}}$.
%\end{itemize}
\end{itemize}
\end{defi}

  As we will prove, these graphs are the components of the split
  decomposition.  In order to obtain an SD graph $H_{\mBS(G)}$,
  we take their union and we link them by undirected
  $\varepsilon$-edges between $(e,x)$ and $(e,y)$ for every edge
  $e:x-y$ of $T_{\mBS(G)}$. This completes the definition of
  $H_{\mBS(G)}$. If $G$ has no good split, then $\mBS_{g}(G)$
  is empty, the tree $T_{\mBS(G)}$ has one node and no edge,
  and $H_{\mBS(G)}=G$.

\begin{prop}\label{P:4.5}
  If a graph $G$ is strongly connected with at least 3 vertices, we
  have $H_{\mBS(G)}=\Sdg(\Split(G))$ and $\Eval(H_{\mBS(G)})=G$.
\end{prop}

\proof The proof is by induction on the number of vertices of $G$.
\begin{itemize}
\item[1)] The case of graphs with 3 vertices is checked directly: each
  graph is a clique, a star or a CTT, hence is necessarily a component,
  $\mBS(G)$ is empty, and $H_{\mBS(G)}=G$.
\item[2)] If $G$ has no good split, then it follows from
  \cite{cunningham82}, Theorems 10 and 11, that $G$ is either $S_{n}$,
  or $K_{n}$, or a CTT, or is prime.  In all cases we have
  $H_{\mBS(G)}=G$.
\item[3)] If none of these cases hold, then $G$ has a good split
  $\{A,B\}$ and $G$ can be written as $H\boxtimes K$ in a unique way
  (Lemma \ref{L:4.1}) with $V_{H}\supseteq A$, $V_{K}\supseteq B$. We
  let $h$ and $k$ be their marker vertices (cf.\ Definition
  \ref{D:4.1}).
\end{itemize}

\medskip\noindent\textbf{Claim 1}: The tree $T_{\mBS(G)}$ is the union
  of the trees $T_{\mBS(H)}$ and $T_{\mBS(K)}$ linked by an edge
  between $x$ and $y$, where $x$ is the node of $T_{\mBS(H)}$ such
  that $h\in V_{\mBS(H)}(x)$ and $y$ is the node of $T_{\mBS(K)}$ such
  that $k\in V_{\mBS(K)}(y)$.

\medskip\noindent\emph{Proof of Claim 1.} 
  Property F3 of Theorem 8 of \cite{cunningham82}, states that for a
  split $\{A,B\}$, if $A'\subset A$, then $\{A',B\cup A-A^{\prime }\}$
  is a split of $G$ if and only if $\{A',\{h\}\cup A-A^{\prime }\}$ is
  a split of $H$. It follows that if $\{A,B\}$ is a good split, then,
  with $H$ and $K$ associated with it as above:
\[\eqalign{
  \mBS_{g}(G)=\{\{A,B\}\}
 &\cup\{\{A',C\cup B\}\mid
  \{A',C\cup\{h\}\}\in\mBS_{g}(H)\}\cr
 &\cup\{\{B',C\cup A\}\mid\{B',C\cup\{k\}\}\in\mBS_{g}(K)\}\;.}
\]
  This fact gives the bijection between $T_{\mBS(G)}$ and the union of
  the trees $T_{\mBS(H)}$ and $T_{\mBS(K)}$ linked by an edge as in
  the statement. The edge $x-y$ corresponds to $\{A,B\}$. $\square$

\medskip\noindent\textbf{Claim 2}: The graph $H_{\mBS(G)}$ is
  isomorphic to the union of the graphs $H_{\mBS(H)}$ and $H_{\mBS(K)}$
  linked by an $\varepsilon$-edge between $h$ and $k$.

\medskip\noindent\emph{Proof of Claim 2.}
  Let $e$ be the $\varepsilon$-edge linking $h$ and $k$.  Let $x_{h}$
  and $x_{k}$ be the nodes of $T_{\mBS(H)}$ and $T_{\mBS(K)}$ such
  that $h\in V_{\mBS(H)}(x_{h}),k\in V_{\mBS(K)}(x_{k})$.

  We denote by $H_{\mBS(H)}+H_{\mBS(K)}$ the union of the graphs
  $H_{\mBS(H)}$ and $H_{\mBS(K)}$ together with $e$ where $h$ is
  replaced by $(e,x_{h})$ and $k$ by $(e,x_{k})$.

  Our goal is to prove that $H_{\mBS(G)}=H_{\mBS(H)}+H_{\mBS(K)}$. By
  Claim 1 and the definitions, the vertices of the graph $H_{\mBS(G)}$
  are those of $H_{\mBS(H)}+H_{\mBS(K)}$. It remains to prove that the
  edges are the same in both.

  This is clear for the $\varepsilon$-edges as an immediate
  consequence of Claim 1. We now consider the various types of solid
  edges.
\begin{itemize}
\item[a)] A solid edge of the form $u\longrightarrow v$, $u,v\in
  V_{\mBS(G)}(x)$, where none of $u$ and $v$ is a vertex $(f,y)$, is in
  $H_{\mBS(G)}$ if and only if it is in $H_{\mBS(H)}+H_{\mBS(K)}$
  because $V_{\mBS(G)}(x)=V_{\mBS(H)}(x)\cap V_{G}$ for $x$ a node of
  $T_{\mBS(H)}$ and similarly for $K$.
\item[b)] Consider a solid edge $(e_{i},x)\longrightarrow(e_{j},x)$.
  Without loss of generality, we assume that $x$ is a node of
  $T_{\mBS(H)}.$

  \noindent Consider such an edge in $H_{\mBS(G)}$: there is in $G$ an edge
  $u\longrightarrow v$ for some $u$ in $P_{y_{i}}$ and some $v$ in
  $P_{y_{j}}$, where $y_{1},\dots,y_{n}$ are the neighbours of $x$ in
  $T_{\mBS(G)}$ as in Definition \ref{D:4.6}.

  \medskip\noindent\emph{Subcase 1}: One of $(e_{i},x)$ or $(e_{j},x)$, say
  $(e_{j},x)$, is $(e,x_{h})$.

  \noindent Then we have $u\longrightarrow v$ in $G$ with $v\in P_{y_{j}}=B$.
  Hence, we have an edge $(e_{i},x)\longrightarrow h$ in $H$, hence
  the edge $(e_{i},x)\longrightarrow(e_{j},x)$ in
  $H_{\mBS(H)}+H_{\mBS(K)}$ since $(e,x_{h})=(e_{j},x)$ replaces $h$.

  \medskip\noindent\emph{Subcase 2}: None of $(e_{i},x),(e_{j},x)$ is $(e,x_{h})$ or
  $(e,x_{k})$, $u$ and $v$ are both in $H$, and they are not $h$
  (because $u\longrightarrow v$ is an edge of $G$).

  \noindent Then the edge $(e_{i},x)\longrightarrow(e_{j},x)$ is also in
  $H_{\mBS(H)}$, because if we denote by $P_{y_{i}}'$ the blocks like
  $P_{y_{i}}$ relative to $H$, then we have either
  $P_{y_{i}}'=P_{y_{i}}$ or $P_{y_{i}}'=P_{y_{i}}-V_{K}\cup\{h\}$, by
  the result recalled in the proof of Claim 1.

  \medskip\noindent\emph{Subcase 3}: As in the previous subcase except that one of
  $u$$,v$, say $u$ is in $H$, and the other is in $K$.

  \noindent Then the edge $u\longrightarrow h$ is in $H$, and we also have the
  edge $(e_{i},x)\longrightarrow(e_{j},x)$ in $H_{\mBS(H)}$ because
  $h$ $\in P_{y_{j}}'$, since $P_{y_{j}}'=P_{y_{j}}-V_{K}\cup\{h\}$,
  with the notation of the previous subcase.

  \noindent Conversely, let us assume that $(e_{i},x)$
  $\longrightarrow(e_{j},x)$ in $H_{\mBS(H)}$. We have in $H$ an edge
  $u\longrightarrow v$ for some $u$ in $P_{y_{i}}'$ and some $v$ in
  $P_{y_{j}}'$.

  \medskip\noindent\emph{Subcase 1}: None of $u,v$ is $h$, then we have also
  $(e_{i},x)$ $\longrightarrow(e_{j},x)$ in $H_{\mBS(G)}$, using the
  observation on the blocks $P_{y_{i}},P_{y_{i}}' $ made above in
  Subcase 2.

  \medskip\noindent\emph{Subcase 2}: If $u=h$, then we have $w\longrightarrow v$ in
  $G$ for some $w$ in $K$. Hence $(e_{i},x)$
  $\longrightarrow(e_{j},x)$ is in $H_{\mBS(G)}$.

  \noindent The arguments are of course the same with $K$ in place of $H$.

\item[c)] Consider a solid edge $u\longrightarrow(e_{i},x)$ in
  $H_{\mBS(G)}$, $u\in V_{\mBS(G)}(x)$. There is in $G$ an edge
  $u\longrightarrow v$ for some $v$ in $P_{y_{i}}$, where
  $e_{1}:x-y_{1},\dots,e_{n}:x-y_{n}$ are the edges of $T_{\mBS(G)} $
  incident to $x$, as in Definition \ref{D:4.6}.  There are several
  subcases:

  \medskip\noindent\emph{Subcase 1}:  $u\in V_{H}$, $(e_{i},x)=(e,x_{h}).$

  \noindent Then $v\in V_{K}$, but we have $u\longrightarrow h$ in
  $H_{\mBS(H)}$.  Hence the edge $u\longrightarrow(e,x_{h})$ is in
  $H_{\mBS(H)}+H_{\mBS(K)}$.

  \medskip\noindent\emph{Subcase 2}:  $u\in V_{H}$, $(e_{i},x)\neq(e,x_{h}).$

  \noindent Then $(e_{i},x)$ is in $H_{\mBS(H)}$. Either $v\in V_{H}$, and then
  the edge $u\longrightarrow(e_{i},x)$ is also in $H_{\mBS(H)}$ or
  $v\in V_{K}$, so the edge $u\longrightarrow h$ is in $H$ and the
  edge $u\longrightarrow(e_{i},x)$ is also in $H_{\mBS(H)}$ because
  $h\in P_{y_{i}}'.$

  \noindent The argument is similar if $u\in V_{K}$ and for the edges
  $u\longleftarrow (e_{i},x)$.

  \noindent Consider conversely a solid edge $u\longrightarrow(e_{i},x)$ in
  $H_{\mBS(H)}$, $u\in V(x)$, $u\neq h$.  There is in $H$ an edge
  $u\longrightarrow v$, where $v$ in $P_{y_{i}}'$ (a block relative to
  $H$, same notation as in case b).

  \noindent If $v=h$, we have $u\longrightarrow w$ for some $w\in V_{K}$, hence
  $u\longrightarrow(e_{i},x)$ in $H_{\mBS(G)}$. If $v\neq h$, we have
  $u\longrightarrow v$ in $G$, hence also $u\longrightarrow(e_{i},x)$
  in $H_{\mBS(G)}$.

  \noindent Again the argument is similar for a solid edge
  $u\longrightarrow(e_{i},x)$ in $H_{\mBS(K)}$, and for the edges
  $u\longleftarrow(e_{i},x)$.\qed
\end{itemize}

  We can now complete the Proof.  We have $G=H\boxtimes K$.
  By induction, we can assume that $H_{\mBS(H)}=\Sdg(\Split(H))$ and
  $H_{\mBS(K)}=\Sdg(\Split(K))$. Using the notation of Claim 2, the SD
  graph $\Sdg(\Split(G))$ is, by its definition, equal to
  $\Sdg(\Split(H))+\Sdg(\Split(K))$.  Hence, by Claim 2 and these
  equalities following from induction, it is isomorphic to
  $H_{\mBS(G)}$.  This completes the proof.\qed

\begin{thm}\label{T:4.4}
  There exists an MS transduction that associates with a linearly
  ordered strongly connected graph the SD graph representing its
  canonical split decomposition.
\end{thm}

\proof By Proposition \ref{P:4.4}, we have an MS transduction
  associating with $(G,\preccurlyeq)$ the structure $\<V_{G}\cup
  N_{T},\edg_{G},\edg_{T},box_{T}\>$ where $T$ is the tree of the
  canonical decomposition, i.e, $T=T_{\mBS(G)}$.

  The next task is to specify the pairs $(e,x)$ for the edges $e$ of
  $T$ and their nodes $x$ as pairs $(u,i)$ for $u$ in $V_{G}\cup
  N_{T}$ and integers $i$ in a fixed finite set. By using the ordering
  of $V_{G}$ one can select the leaf $r$ of $T$ which contains a
  smallest vertex of $G$.  We make $T$ into a directed tree with root
  $r$.  This orientation is MS definable. For an edge $e$ of $T$,
  directed, say $:x\longrightarrow y$, we can define $(e,x) $ as the
  pair $(y,1)$ and $(e,y)$ as the pair $(y,2)$. Since $T$ is a
  directed tree, each edge is specified in a unique way by its
  target. Hence, the vertex $y$ refers to a single edge $e$.

  Hence the set of vertices of $H_{\mBS(G)}$ is defined as
  $V_{G}\times\{1\}\cup(N_{T}-\{r\})\times\{1,2\}$. The conditions
  defining the edges of the graph $H_{\mBS(G)}$ are straightforward to
  express in MS logic, provided for each edge of $T$ one can determine
  the corresponding good split.  This is possible using the relation
  $box_{T}$.\qed

  Hence we have proved that the split decomposition of a strongly
  connected graph is definable by an MS transduction from the graph
  and a linear order of its vertices. It follows from Proposition
  A.1.1 (in Appendix 1) that a property of graphs expressed as an MS
  property of their prime components and/or of the underlying trees of
  their split decompositions is an order-invariant MS property.

\section{Conclusion}\label{S:5}

  In this article, we have applied Monadic Second-Order logic to the
  graph decompositions which follow the pattern of modular
  decomposition and to those defined in the framework of Cunnigham and
  Edmonds \cite{CuEd80}.  We have established general definability
  results in Monadic Second-Order logic, and we have applied them to
  the canonical decompositions of 2-connected graphs. We have obtained
  as new results a logical expression of Whitney's 2-isomorphism
  Theorem and the definability in Monadic Second-Order logic of the
  split decomposition of Cunnigham \cite{cunningham82}.  The article
  \cite{cour05} applies this result to \emph{circle graphs} studied in
  the framework of Monadic Second-order logic. This application is
  presented in the Introduction.

  Here are some open questions (a few others are presented also in the
  main text).

\textbf{Question 1}: The \emph{split decomposition} works well for
  undirected graphs and for strongly connected directed graphs,
  because these graphs have canonical decompositions.  What about
  \emph{connected directed graphs ?} The strongly connected components
  of a graph $G$ form a \emph{directed acyclic graph} $D$.  Directed
  acyclic graphs have unique modular decompositions.  However, it is
  not clear how to combine the modular decomposition of $D$ and the
  split decompositions of the strongly connected components of $G$ in
  order to obtain a notion of canonical decomposition subsuming these
  cases. Although directed graphs have no \emph{canonical} split
  decomposition, it may be useful to construct non canonical ones for
  algorithmic purposes or for investigations on the structure of
  graphs.

\textbf{Question 2}: Our logical formalization of decompositions,
  based on families of sets and on families of bipartitions can be
  applied to hypergraphs (along the lines of \cite{ChHaMa81}), to
  \emph{k}-structures which are also hypergraphs (see \cite{EhMc94}),
  to matroids (the MS logic of matroids has been studied by
  Hlin\u{e}ny \cite{hlineny03}). These applications should be
  developped.

\textbf{Question 3}: Another topic for future research is the
  extension of split decomposition to countable graphs, generalizing
  what is done in \cite{CoDe05} for modular decomposition.

\section*{Acknowledgement}

  The dissertation of F.~de Montgolfier \cite{montgolfier03} contains
  a very good introduction to the articles by W.~Cunnigham and
  J.~Edmonds \cite{CuEd80,cunningham82}.  Many thanks to
  A.~Blumensath, S.~Oum and the referees for their numerous useful
  comments.  I also thank J.~Koslowski, layout editor, for his
  important editing work on this article.

\bibliography{co}

\begin{thebibliography}{10}

\bibitem{BeSe05}
M.~Benedikt and L.~Segoufin.
\newblock Towards a characterization of order-invariant queries over tame
  structures.
\newblock In {\em Computer Science Logic 2005}, volume 3634 of {\em Oxford,
  Lec. Notes Comput. Sci. \textbf{3634}}, pages 276--291. S-V, 2005.

\bibitem{bouchet87}
A.~Bouchet.
\newblock Reducing prime graphs and recognizing circle graphs.
\newblock {\em Combinatorica}, 7:243--254, 1987.

\bibitem{capelle97}
C.~Capelle.
\newblock Block decomposition of inheritance hierarchies.
\newblock In R.~M{\"o}hring, editor, {\em Proceedings of WG'97}, volume 1335 of
  {\em LNCS}, pages 118--131. S-V, 1997.

\bibitem{ChHaMa81}
M.~Chein, M.~Habib, and M.~Maurer.
\newblock Partitive hypergraphs.
\newblock {\em Discrete mathematics}, 37:35--50, 1981.

\bibitem{CiSt99}
S.~Cicerone and G.~Stefano.
\newblock On the extension of bipartite to parity graphs.
\newblock {\em Discrete Appl. Math.}, 95:181--195, 1999.

\bibitem{cour05}
B.~Courcelle.
\newblock
  \href{http://www.labri.fr/Perso/~courcell/ArticlesEnCours/CircleGraphsSubmit%
ted.pdf}% {Circle graphs and monadic second-order logic}.
\newblock submitted.

\bibitem{courcelle94}
B.~Courcelle.
\newblock Monadic second-order graph transductions: A survey.
\newblock {\em Theoret. Comput. Sci.}, 126:53--75, 1994.

\bibitem{courcelle95}
B.~Courcelle.
\newblock The monadic second-order logic of graphs
  \uppercase\expandafter{\romannumeral8}: Orientations.
\newblock {\em Ann. Pure Appl. Logic}, 72:103--143, 1995.

\bibitem{courcelle96}
B.~Courcelle.
\newblock The monadic second-order logic of graphs
  \uppercase\expandafter{\romannumeral10}: Linear orderings.
\newblock {\em Theoret. Comput. Sci.}, 160:87--143, 1996.

\bibitem{courcelle97}
B.~Courcelle.
\newblock The expression of graph properties and graph transformations in
  monadic second-order logic.
\newblock In G.~Rozenberg, editor, {\em Handbook of graph grammars and
  computing by graph transformations}, volume 1: Foundations, pages 313--400.
  World Scientific, 1997.

\bibitem{courcelle99}
B.~Courcelle.
\newblock The monadic second-order logic of graphs
  \uppercase\expandafter{\romannumeral11}\: Hierarchical decompositions of
  connected graphs.
\newblock {\em Theoret. Comput. Sci.}, 224:35--58, 1999.

\bibitem{courcelle04}
B.~Courcelle.
\newblock The monadic second-order logic of graphs
  \uppercase\expandafter{\romannumeral15}: On a {C}onjecture by {D}. {S}eese.
\newblock {\em J. Appl. Logic}, 4:79--114, 2006.

\bibitem{CoDe05}
B.~Courcelle and C.~Delhomm{\'e}.
\newblock The modular decomposition of countable graphs: Constructions in
  {M}onadic {S}econd-{O}rder {L}ogic.
\newblock In {\em Computer Science Logic 2005}, volume 3634 of {\em Oxford,
  Lec. Notes Comput. Sci.}, pages 325--338. S-V, 2005.

\bibitem{CoMaRo00}
B.~Courcelle, J.~Makowsky, and U.~Rotics.
\newblock Linear time solvable optimization problems on graphs of bounded
  clique-width.
\newblock {\em Theory of Computer Systems}, 33:125--150, 2000.

\bibitem{CoOl00}
B.~Courcelle and S.~Olariu.
\newblock Upper bounds to the clique-width of graphs.
\newblock {\em Discrete Appl. Math.}, 101:77--114, 2000.

\bibitem{CoOu04}
B.~Courcelle and S.~Oum.
\newblock
  \href{http://www.labri.fr/Perso/~courcell/Textes/BC-OumSubmitted(2004).pdf}%
  {Vertex-minors, monadic second-order logic and a conjecture by {S}eese}.
\newblock To appear in J. of Combinatorial Theory B.

\bibitem{cunningham82}
W.~Cunnigham.
\newblock Decomposition of directed graphs.
\newblock {\em SIAM J. Algebraic Discrete Methods}, 3:214--228, 1982.

\bibitem{CuEd80}
W.~Cunnigham and J.~Edmonds.
\newblock A combinatorial decomposition theory.
\newblock {\em Canad. J. Math}, 32:734--765, 1980.

\bibitem{montgolfier03}
F.~de~Montgolfier.
\newblock {\em D{\'e}composition modulaire des graphes, Th{\'e}orie, extensions
  et algorithmes}.
\newblock PhD thesis, Montpellier 2 University, 2003.

\bibitem{EhMc94}
A.~Ehrenfeucht and R.~McConnell.
\newblock A $k$-structure generalization of the theory of 2-structures.
\newblock {\em Theoretical Computer Science}, 132:209--227, 1994.

\bibitem{GaPa03}
C.~Gavoille and C.~Paul.
\newblock Distance labeling scheme and split decomposition.
\newblock {\em Discrete Mathematics}, 273:115--130, 2003.

\bibitem{habib81}
M.~Habib.
\newblock {\em Substitution des structures combinatoires, th{\'e}orie et
  algorithmes}.
\newblock PhD thesis, Universit{\'e} Paris-6, 1981.

\bibitem{HaHuSp95}
M.~Habib, M.~Huchard, and J.~Spinrad.
\newblock A linear algorithm to decompose inheritance graphs into modules.
\newblock {\em Algorithmica}, 13:573--591, 1995.

\bibitem{hlineny03}
P.~Hlin{\u e}ny.
\newblock On matroid properties definable in the {MSO} logic.
\newblock In {\em Mathematical Foundations of Computer Science 2003}, volume
  2747 of {\em LNCS}, pages 470--479. S-V, 2003.

\bibitem{HoTa72}
J.~Hopcroft and R.~Tarjan.
\newblock Isomorphism of planar graphs.
\newblock In {\em Complexity of computer computations}. Plenum Press, New York,
  1972.

\bibitem{MoRa84}
R.~M\"{o}hring and F.~Radermacher.
\newblock Substitution decomposition for discrete structures and connections
  with combinatorial optimization.
\newblock {\em Ann. Discrete Math.}, 19:257--356, 1984.

\bibitem{OuSe04}
S.~Oum and P.~Seymour.
\newblock Appoximating clique-width and branch-width.
\newblock 2004, To appear in J. of Combinatorial Theory B.

\bibitem{oxley92}
J.~Oxley.
\newblock {\em Matroid theory}.
\newblock Oxford University Press, 1992.

\bibitem{spinrad03}
J.~Spinrad.
\newblock {\em Efficient graph representations}, volume~19 of {\em Fields
  Institute Monographs}.
\newblock A.M.S., Providence, 2003.

\bibitem{truemper80}
K.~Truemper.
\newblock On {W}hitney's 2-isomorphism theorem for graphs.
\newblock {\em J. Graph Theory}, 4:43--49, 1980.

\bibitem{tutte66}
W.~Tutte.
\newblock {\em Connectivity in graphs}.
\newblock University of Toronto Press, 1966.

\bibitem{white86}
N.~White.
\newblock {\em Theory of matroids}.
\newblock Cambridge University Press, 1986.

\end{thebibliography}
\bibliographystyle{plain}

\section*{Appendix 1: Monadic second-order logic}

  We review Monadic Second-Order (MS) logic and transformations of
  structures expressed in this language, called \emph{MS
  transductions}.  The reader is refered to the book chapter
  \cite{courcelle97}, or to the preliminary sections of the articles
  \cite{courcelle94,courcelle96,courcelle04} for more detailed
  expositions. However all necessary definitions are given in full in
  the present section.

\subsection*{Relational structures and monadic second-order logic}

  Let $R=\{A,B,C,\dots\}$ be a finite set of \emph{relation symbols}
  each of them given with a nonnegative integer \textit{ }$\rho(A)$
  called its \textit{arity}. We denote by $\mSTR(R)$ the set
  of \emph{finite} $R$-structures $S=\<D_{S}$$,(A_{S})_{A\in R}\>$
  where $A_{S}\subseteq D_{S}^{\rho(A)}$ if $A\in R$ is a relation
  symbol.  If \emph{R} consist of relation symbols of arity one or
  two, then we say that the structures in $\mSTR(R)$ are
  \emph{binary}.

  A simple graph $G$ can be defined as an $\{edg\}$-structure
  $G=\<V_{G},\edg_{G}\>$ where $V_{G}$ is the set of vertices of $G$
  and $\edg_{G}$ $\subseteq V_{G}\times V_{G}$ is a binary relation
  representing the edges.  For undirected graphs, the relation
  $\edg_{G}$ is symmetric. If in addition we need vertex labels, we
  will represent them by unary relations.  Binary structures can be
  seen as vertex- and edge- labelled graphs. If we have several binary
  relations say $A,B,C$, the corresponding graphs have edges of types
  $A,B,C$.

  We recall that \emph{Monadic Second-order logic} (\emph{MS logic}
  for short) is the extension of \emph{First-Order logic} (\emph{FO
  logic} for short) by variables denoting subsets of the domains of the
  considered structures, and new atomic formulas of the form $x\in X$
  expressing the membership of $x$ in a set $X$. (Uppercase letters
  will denote set variables, lowercase letters will denote first-order
  variables).

  We denote by $FO(R,W)$ (resp.\ by $MS(R,W)$) the set of
  $\emph{First-order}$ (resp. \emph{Monadic Second-order}) formulas
  written with the set $R$ of relation symbols and having their free
  variables in a set $W$ consisting of \emph{first-order as well as of
  set variables}. Hence, we allow first-order formulas with free set
  variables and written with atomic formulas of the form $x\in X$.  In
  first-order formulas, only first-order variables can be quantified.

  As a typical and useful example of MS formula, we give a formula
  with free variables $x$ and $y$ expressing that $(x,y)$ belongs to
  the reflexive and transitive closure of a binary relation $A$:
\[\forall X(x\in X\wedge\forall u,v[(u\in X\wedge A(u,v))\Longrightarrow v\in
  X]\Longrightarrow y\in X)\;.
\]
  If the relation $A$ is not given in the structure but defined by an
  MS\ formula, then one replaces $A(u,v)$ by this formula with
  appropriate substitutions of variables.

  A \emph{monadic second-order (MS) property} of the structures $S$ of
  a class $\mC\subseteq\mSTR(R)$ is a property $\mP$
  such that for $S\in\mC$:
\[\mP(S)\quad\hbox{holds if and only if}\quad S\vDash\varphi\ ,\]
  for some fixed formula $\varphi$ in $MS(R,\varnothing)$. Let $\leq$
  be a binary relation symbol not in $R$.  A formula $\varphi$ in
  $MS(R\cup \{\leq\},\varnothing)$ is \emph{order-invariant on a
  class} $\mC$, if for every $S\in\mC$, for every two linear orders
  $\preccurlyeq$ and $\preccurlyeq'$ on the domain $D_{S}$
\[(S,\preccurlyeq)\vDash\varphi
  \quad\hbox{if and only if}\quad
  (S,\preccurlyeq')\vDash\varphi\ ,
\]
  where $\preccurlyeq$ and $\preccurlyeq'$ interpret $\leq$.  We say
  that $\mP$ is an \emph{order-invariant} MS property of the
  structures of a class $\mC\subseteq\mSTR(R)$ if and only if 
\[\mP(S)
  \quad\hbox{holds if and only if}\quad
  (S,\preccurlyeq)\vDash\varphi
  \quad\hbox{for some linear order $\preccurlyeq$ on $D_{S}$}\ ,
\]
  where $\varphi$ is a fixed order-invariant MS
  formula. Order-invariant MS properties are investigated in
  \cite{BeSe05,courcelle96}. A difficulty with this definition is that
  the set of order-invariant MS formulas is undecidable. However, we
  will use formulas that are order-invariant by construction.

\subsection*{Monadic Second-order transductions}

  We will also use FO and MS formulas to define certain graph
  transformations.  As in Language Theory, a binary relation
  $\mR\subseteq\mA\times\mB$ where $\mA$ and $\mB$ are sets of
  relational structures will be called a \emph{transduction}:
  $\mA\rightarrow\mB$.

  An \emph{MS transduction} is a transduction specified by MS
  formulas.  It transforms a structure $S$, given with an $n$-tuple of
  subsets of its domain called the \emph{parameters}, into a structure
  $T$, the domain of which is a subset of
  $D_{S}\times\{1,\dots,k\}$. Furthermore, each such transduction, has
  an associated \emph{backwards translation}, a mapping that
  transforms effectively every MS formula$ \varphi$ relative to $T$,
  possibly with free variables, into one, say $\varphi^{\#}$, relative
  to $S$ having free variables corresponding to those of $ \varphi$
  ($k$ times as many actually) together with those denoting the
  parameters.  This new formula expresses in $S$ the property of $T$
  defined by $\varphi$. \bigskip We now give some details.  More can
  be found in \cite{courcelle94,courcelle97}.

  We let $R$ and $Q$ be two finite sets of relation symbols. Let $W$
  be a finite set of set variables, called \emph{parameters}. A
  $(Q,R)$-\emph{definition scheme} is a tuple of formulas of the form
\[\Delta=(\varphi,\psi_{1},\cdots,\psi_{k},(\theta_{w})_{w\in Q^{\ast}k})  
  \quad\hbox{where $k>0$ and $Q^{\ast}k:=\{(q,\vec{j})\mid q\in
  Q,\vec{j}\in\lbrack k]^{\rho (q)}\}$}\ ,
\]
\[\varphi\in MS(R,W),\psi_{i}\in MS(R,W\cup\{x_{1}\})\quad\hbox{for
  $i=1,\cdots,k$, and}
\]
\[\theta_{w}\in MS(R,W\cup\{x_{1},\cdots,x_{\rho(q)}\})
  \quad\hbox{for $w=(q,\vec
{j})\in Q^{\ast}k$}\;.
\]
  These formulas are intended to define a structure $T$ in
  $\mSTR(Q\mathcal{)}$ from a structure $S$ in
  $\mSTR(R\mathcal{)}$. Let $S\in\mSTR(R\mathcal{)}$, let $\gamma$ be
  a $W$-assignment in $S$. A $Q$-structure $T$ with domain
  $D_{T}\subseteq D_{S}\times\lbrack k]$ is \emph{defined in}
  $(S,\gamma)$ by $\Delta$ if
\begin{itemize}
\item[(i)] $(S,\gamma)\models\varphi$,
\item[(ii)] $D_{T}=\{(d,i)\mid d\in D_{S},i\in\lbrack
  k],(S,\gamma,d)\models\psi _{i}\}$,
\item[(iii)] for each $q$ in $Q$
\[q_{T}=\{((d_{1},i_{1}),\cdots,(d_{t},i_{t}))\in
D_{T}^{t}\mid(S,\gamma,d_{1},\cdots,d_{t})\models\theta_{(q,\vec{j})}\}\ ,
\]
  where $\vec{j}=(i_{1},\cdots,i_{t})$ and $t=\rho(q).$
\end{itemize}

  The notation $S\models\psi$ means that the logical formula $\psi$
  holds true in the structure $S$.  By
  $(S,\gamma,d_{1},\cdots,d_{t})\models \theta_{(q,\vec{j})}$, we mean
  $(S,\gamma')\models\theta_{(q,\vec{j})}$, where $\gamma'$ is the
  assignment extending $\gamma$, such that $\gamma'(x_{i})=d_{i}$ for
  all $i=1,\cdots,t$; a similar convention is used for
  $(S,\gamma,d)\models\psi_{i})$.

  Since $T$ is associated in a unique way with $S,\gamma$ and $\Delta$
  whenever it is defined, i.e., whenever $(S,\gamma)\models\varphi$,
  we can use the functional notation $\dEf_{\Delta}(S,\gamma)$ for
  $T$. The \emph{transduction defined by} $\Delta$ is the binary
  relation
\[\mD_{\Delta}:=\{(S,T)\mid T=\dEf_{\Delta}(S,\gamma)
  \hbox{\ for some $W$-assignment $\gamma$ in\ }S\}\;.
\]
  Hence $\mD_{\Delta}\subseteq\mSTR(R\mathcal{)}$$\times
  \mSTR(Q\mathcal{)}$. A transduction
  $f\subseteq\mSTR(R\mathcal{)}\times\mSTR(Q)$ is an \emph{MS
  transduction } if it is equal, up to isomorphism of structures, to
  $\mD_{\Delta}$ for some $(Q,R)$-definition scheme $\Delta$.

  An MS-transduction is defined as a binary relation. Hence it can be
  seen as a "nondeterministic" partial function associating with an
  $R$-structure one or more $Q$-structures.  However, it is not really
  nondeterministic because the different outputs come from different
  choices of parameters.  In the case where $W$ = $\emptyset$, we say
  that the transduction is \emph{\ parameterless }; it defines a
  partial function. It may also happen that different choices of
  parameters yield isomorphic output structures. This is the case in
  the example of edge contraction detailed below.

  We will refer to the integer $k$ by saying that $\Delta$ and
  $\mD_{\Delta}$ are $k$\emph{-copying}; if $k=1$ we will say that
  they are \emph{noncopying.  }A noncopying definition scheme can be
  written more simply: $\Delta=(\varphi,\psi,(\theta_{q})_{q\in Q})$.
  We will say that an MS\ transduction is \emph{domain extending}, if
  the formula $\psi_{1}$ of its definition scheme $\Delta$ is the
  Boolean constant$ True$. In this case, if
  $T=\dEf_{\Delta}(S,\gamma)$, then $D_{T}$ contains
  $D_{S}\times\{1\}$, an isomorphic copy of $D_{S}$. This transduction
  defines the domain of $T$ as an extension of that of $S$. If in the
  definition scheme $\Delta$ we only use FO formulas, then we will say
  that $\mD_{\Delta}$ is an \emph{FO transduction}.

\medskip\textbf{Example.} \emph{Edge contraction.}
  We consider a graph $G$ with two types of edges, the ordinary edges
  and the $\varepsilon$-edges. It is represented by a structure
  $\<V_{G},\edg_{G},\varepsilon-\edg_{G}\>$ where the binary relation
  $\varepsilon-\edg_{G}$ represents the $\varepsilon$-edges. We want to
  define from $G$ the graph $H$ obtained by the contraction of all
  $\varepsilon$-edges.

  It is formally defined as $\<V_{H},\edg_{H}\>$ where $V_{H}=V_{G}
  /\sim$, $\sim$ is the equivalence relation such that $x\sim y$ if
  and only if $x$ and $y$ are linked by an undirected path made of
  $\varepsilon$-edges, and $\edg_{H}([u],[v])$ holds if and only if
  $x\in\lbrack u]$, $y\in\lbrack v]$ for some $(x,y)$ in $\edg_{G}$
  ($[u]$ denotes the equivalence class of $u$).  The MS formula
  $\xi(x,y)$ defined as
\[\forall X[(x\in X\wedge\forall u,v\{u\in X\wedge(\varepsilon-edg(u,v)\vee
\varepsilon-edg(v,u))\Longrightarrow v\in X\})\Longrightarrow y\in X]
\]
  expresses $x\sim y$. For defining $V_{H}$ we must select a set
  containing one and only one vertex of each equivalence class.  This
  can be done with a set variable $Y$ that will be a parameter of the
  MS transduction, satisfying the formula $\varphi(Y)$ defined as$
  \forall x\exists!y[y\in Y\wedge \xi(x,y)].$

  Edge contraction can be defined by the transduction with noncopying
  definition scheme $\Delta=(\varphi,\psi,\theta_{edg})$ where$
  \psi(Y,x)$\ is $x\in Y$ and $\theta_{edg}(Y,x,y)$ is $\exists
  u,v[x\in Y\wedge y\in Y\wedge edg(u,v)\wedge\xi(x,u)\wedge\xi(y,v)].$

  Notice that the structures associated with all values of the
  parameter $Y$ satisfying $\varphi (Y)$ are isomorphic.  They only
  differ regarding the concrete subsets $Y$ of $V_{G}$ used as sets of
  vertices of $H$.

\bigskip\noindent\textbf{Lemma A.1.1.}
  Let $\tau\,:\,\mSTR(R)\longrightarrow\mSTR(Q)$ be an MS (or FO)
  transduction.  Let $\leq$ be a binary relation symbol not in $R\cup
  Q$.  One can transform $\tau$ into an MS (or FO) transduction $\tau$
  $':$ $\mSTR(R\cup\{\leq\})\longrightarrow\mSTR(Q\cup\{\leq\})$ such
  that, for every $S$ in $\mSTR(R)$ and every linear order $\preceq$
  on its domain, $\tau '(S,\preceq)=(\tau(S),\preceq')$ where
  $\preceq'$ is a linear order on the domain of $\tau(S)$.

\proof Let $\tau$ be $k$-copying.  For
  $w=(\leq,\vec{j})\in\{\leq\}^{\ast}k$ it is easy to define FO
  formulas $\theta_{w}$ belonging to
  $MS(R\cup\{\leq\},W\cup\{x_{1},x_{2}\})$ such that, in
  $\tau^{\prime }(S,\preceq)$
\[(d_{1},i)\preceq'(d_{2},j)\quad
 \hbox{if and only if either $d_{1}\prec d_{2}$ 
       or ($d_{1}=d_{2}$ and $i\leq j$)}\;.
\]
  It is clear that $\preceq'$ is a linear order on the domain of
  $\tau(S)$ if $\preceq$ is one on $S$.\qed

\subsection*{The fundamental property of MS transductions}

  The following proposition says that if $T=\dEf_{\Delta}(S,\gamma)$,
  then the monadic second-order properties of $T$ can be expressed as
  monadic second-order properties of $(S,\gamma)$. The usefulness of
  definable transductions is based on this proposition.

  Let $\Delta=(\varphi,\psi_{1},\cdots,\psi_{k},(\theta_{w})_{w\in
  Q^{\ast}k})$ be a $(Q,R)$-definition scheme, written with a set of
  parameters $W $. Let $V$ be a set of set variables disjoint from
  $W$. For every variable $X $ in $V$, for every $i=1,\cdots,k$, we
  let $X_{i}$ be a new variable. We let $V'$:= $\{X_{i}/X\in V$,
  $i=1,\cdots,k\}$. Let $S$ be a structure in $\mSTR$($R$) with domain
  $D$. For every mapping $\eta:V^{\prime }\longrightarrow\mP(D)$, we
  let $\eta^{k}:V\longrightarrow\mP(D\times\lbrack k])$ be defined by
  $\eta^{k}(X)=\eta(X_{1})\times
  \{1\}\cup\cdots\cup\eta(X_{k})\times\{k\}$. With this notation we
  can state

\bigskip\noindent\textbf{Proposition A.1.2.}
  \emph{For every formula $\beta$ in $MS(Q,V)$ one can construct a formula
  $\beta^{\#}$ in $MS(R,V'\cup W)$ such that, for every $S$ in
  $\mSTR(R)$, for every assignment $\gamma\,:\,W\longrightarrow S$,
  for every assignment $\eta:V'\longrightarrow S$ we have
\[\hfill\qquad\qquad\eqalign{
  (S,\eta\cup\gamma)\models \beta^{\#}
  \quad\hbox{if and only if}\quad
 &\hbox{$\dEf_{\Delta}(S,\gamma)$ is defined,}\cr
 &\hbox{$\eta^{k}$ is a $V$-assignment in $\dEf_{\Delta}(S,\gamma)$,}\cr
  \hbox{and}\quad
 &(\dEf_{\Delta}(S,\gamma),\eta^{k})\models\beta\;.\hbox to121.5 pt{\hfill\qEd}
  }
\]}

  If the definition scheme and the formula $\beta$ are FO the formula
  $\beta^{\#}$ is also FO. Note that, even if
  $T=\dEf_{\Delta}(S,\gamma)$ is well-defined, the mapping $\eta^{k}$
  is not necessarily a $V$-assignment in $T$, because $\eta^{k}(X)$
  may not be a subset of the domain of $T$ which is a possibly proper
  subset of $D_{S}\times\{1,\dots,k\}$. We call $\beta^{\#}$ the
  \emph{backwards translation} of $\beta$ relative to the transduction
  $\dEf_{\Delta}$.

  The composition of two transductions is defined as the composition
  of the corresponding binary relations.  If they are both partial
  functions, then one obtains the composition of these functions. The
  composition of two domain extending MS (or FO) transductions is
  domain extending.

\medskip\noindent\textbf{Proposition A.1.3.}

\unskip\noindent(1)\ \emph{The composition of two MS (or FO)
  transductions is an MS (or an FO) transduction.}

\noindent(2)\ \emph{The inverse image of an MS-definable class of structures
  under an MS transduction is MS-definable. A similar statement holds
  with FO instead of MS.}\qed

\section*{Appendix 2: Clique-width}

  \emph{Clique-width} is, like tree-width a graph complexity measure.
  It is defined and studied by Courcelle and Olariu in \cite{CoOl00},
  and also in \cite{courcelle97,OuSe04}.  Graphs are simple, directed
  or not, and loop-free.

  Let $C$ be a set of $k$ labels.  A \emph{C-graph} is a graph $G$
  given with a total mapping from its vertices to $C$, denoted by
  $\lab_{G}$.  Hence $G$ is defined as a triple
  $(V_{G},\edg_{G},\lab_{G})$. We call $\lab_{G}(v)$ the \emph{label} of
  a vertex $v$. The operations on $C$-graphs are the following ones
\begin{itemize}
\item[(i)] For each $i\in C$, we define a constant \textbf{i} for
  denoting an isolated vertex labelled by $i$.
\item[(ii)] For $i,j\in C$ with $i\neq j$, we define a unary function
  $\add_{i,j}$ such that
\[\add_{i,j}(V_{G},\edg_{G},\lab_{G})=(V_{G},\edg_{G}',\lab_{G})\ ,\]
  where $\edg_{G}'$ is $\edg_{G}$ augmented with the set of pairs
  $(u,v)$ such that $\lab_{G}(u)=i$ and $\lab_{G}(v)=j$.

  \noindent In order to add undirected edges, we take:
  $\add_{i,j}(\add_{j,i}(V_{G},\edg_{G},\lab_{G})).$

\item[(iii)] We let also $\ren_{i\rightarrow j}$ be the unary function
  such that
\[\ren_{i\rightarrow j}(V_{G},\edg_{G},\lab_{G})=(V_{G},\edg_{G},\lab_{G}')\ ,\]
  where $\lab_{G}'(v)=j$ if $\lab_{G}(v)=i$, and $\lab_{G}^{\prime
  }(v)=\lab_{G}(v)$, otherwise. This mapping renames into $j$ every
  vertex label $i$.

\item[(iv)] Finally, we use the binary operation $\oplus$ that makes
  the union of disjoint copies of its arguments.  Hence $G\oplus G\neq
  G$ and its size is twice that of $G$.
\end{itemize}

  A well-formed expression $t$ over these symbols will be called a
  \emph{$C$-expression}, or a \emph{$k$-expression} if we are only
  concerned with the size $k$ of $C$.  Its $value$ is a $C$-graph
  $G=\val(t)$. The set of vertices of $\val(t)$ is (or can be defined
  as) the set of occurrences of the constant symbols in $t$. However,
  we will also consider that an expression $t$ designates any graph
  isomorphic to $\val(t)$. The context specifies whether we consider
  concrete graphs or graphs up to isomorphism.

  A graph is considered as a graph all vertices of which are labelled
  in the same way. The \emph{clique-width} of a graph $G$, denoted by
  $\cwd(G)$ is the minimal $k$ such that $G=\val(t)$ for some
  $k$-expression $t$. A graph with at least one edge has clique-width
  at least 2.  The graphs $K_{n},S_{n-1}$ have clique-width 2, for
  $n\geq3$.

  If we need to define graphs with vertex labels from a set $L$, then
  we use constant symbols \textbf{i}$_{a}$ for $i$ in $C$ and $a$ in
  $L$.  The labels from $L$ are not changed, and do not affect the
  other operations.  The clique-width of a graph does not depend on
  the possible labelling of its vertices.  By contrast, it depends
  strongly on edge directions.  Cliques and transitive tournaments
  have clique-width 2 but tournaments have unbounded clique-width
  (\cite{courcelle95}). To build a graph with labelled edges we use
  the operation $\add_{a,i,j}$ to add edges labelled by $a$ from the
  vertices labelled by $i$ to those labelled by $j$.

\medskip\noindent\textbf{Lemma A.2.1.} \cite{CoOl00}\hfill
\noindent(1) \emph{The clique-width of a graph is
  equal to the maximum clique-width of its induced subgraphs.}

\noindent(2) \emph{The clique-width of $G[H/u]$ is equal to the
  maximum of $\cwd(G)$ and $\cwd(H)$.}

\noindent(3) \emph{The clique-width of a graph is equal to
  $Max\{m,2\}$, where $m$ is the maximum clique-width of the prime
  graphs of its modular decomposition.}\qed

\medskip\noindent\textbf{Lemma A.2.2.} \cite{courcelle97} \emph{For every MS
  transduction $\tau$ from graphs to graphs there exists a fonction $f$
  such that $T\in\tau(S)$ implies $\cwd(T)\leq f(\cwd(S))$.}\qed

\medskip\noindent\textbf{Lemma A.2.3.} \emph{Let $G$ be a graph let $u$ be a
  vertex of indegree 0, and $v$ be a vertex of outdegree 0. Let $G'$ be
  obtained from $G$ by fusing $u$ and $v$. Then $\cwd(G')\leq4\cwd(G)$.}

\proof Let $k=\cwd(G)$ and $E$ be a $\{1,\dots,k\}$-expression for
  $G$, considered as a $\{1\}$-graph. For every $x$ in
  $V_{G}-\{u,v\}$, we let its \emph{type} be 1 if $u\longrightarrow x$
  and $x\longrightarrow v$, be 2 if $u\longrightarrow x$ and
  $x\longrightarrow v$ does not hold, be 3 if $x\longrightarrow v$ and
  $u\longrightarrow x$ does not hold, and 0 otherwise.

  We let $H$ be the graph $G[V_{G}-\{u,v\}]$ where every vertex has
  label $(1,i)$ (instead of 1) and $i$ is its type.  We let
  $C=\{1,\dots,k\}\times\{0,1,2,3\}$. From $E$, by deleting the
  constants which define $u$ and $v$, and by modifying the graph
  operations so that every label $a$ of a vertex is replaced by
  $(a,i)$ where $i$ is its type, one can construct a $C$-expression
  $E'$ defining $H$.  Let $p$ be a label, e.g., (2,0), which does not
  label any vertex of $H$.  The graph $G'$ with all its vertices
  labelled by $p$ is the value of %the expression
\[\ren_{(1,0)\rightarrow p}\circ
  \ren_{(1,1)\rightarrow p}\circ
  \ren_{(1,2)\rightarrow p}\circ
  \ren_{(1,3)\rightarrow p}\circ
  \add_{p,(1,1)}\circ
  \add_{p,(1,2)}\circ
  \add_{(1,1),p}\circ
  \add_{(1,3),p}
\]
  at $E'\oplus\mathbf{p}$.  Hence $G'$ has clique-width at most $4k$.\qed

\end{document}